\documentclass{aa}
\usepackage{pifont}
\usepackage{amsmath}
\usepackage{graphicx}
\usepackage{txfonts}
\usepackage{url}
\usepackage{subfigure}
\usepackage{longtable}
\usepackage{lscape}
\usepackage{natbib}
\usepackage{caption}
\usepackage[colorlinks,linkcolor=red,anchorcolor=green,citecolor=blue]{hyperref}

\newcommand{\HII}{H {\small{II}} }
\newcommand{\kms}{{\rm km~s}^{-1}}
\newcommand{\Msun} {M_\sun}

\newcommand{\mjyb}{{\rm mJy~beam}^{-1}}

\newcommand{\mjybkms}{{\rm mJy~beam}^{-1}{\rm km~s}^{-1}}

\begin{document}

   \title{Probing the initial conditions of high-mass star formation}
   \subtitle{IV. Gas dynamics and NH$_2$D chemistry in high-mass precluster and protocluster clumps}
   \author{Chuan-Peng Zhang
          \inst{1,2,3}
          \and
          Guang-Xing Li\inst{4}
          \and
          Thushara Pillai\inst{5}
          \and
          Timea Csengeri\inst{6}
          \and
          Friedrich Wyrowski\inst{6}
          \and
          Karl M. Menten\inst{6}
          \and
          Michele R. Pestalozzi\inst{7}
          }

    \institute{National Astronomical Observatories, Chinese Academy of Sciences, 100101 Beijing, P.R. China\\
    \email{cpzhang@nao.cas.cn}
    \and
    CAS Key Laboratory of FAST, National Astronomical Observatories, Chinese Academy of Sciences, 100101 Beijing, P.R. China
    \and
    Max-Planck-Institut f\"ur Astronomie, K\"onigstuhl 17, D-69117 Heidelberg, Germany
    \and
    South-Western Institute for Astronomy Research, Yunnan University, Kunming, 650500 Yunnan, P.R. China
    \and
    Institute for Astrophysical Research, 725 Commonwealth Ave, Boston University Boston, MA 02215, USA
    \and
    Max-Planck-Institut f\"ur Radioastronomie, Auf dem H\"ugel 69, D-53121 Bonn, Germany
    \and
    Istituto di Fisica dello Spazio Interplanetario - INAF, via Fosso del Cavaliere 100, 00133, Roma, Italy
             }



   \abstract
   {The initial stage of star formation is a complex area of study because of the high densities ($n_{\rm H_2} >$ 10$^6$\,cm$^{-3}$) and low temperatures ($T_{\rm dust} <$ 18\,K) involved. Under such conditions, many molecules become depleted from the gas phase by freezing out onto dust grains. However, the deuterated species could remain gaseous under these extreme conditions, which would indicate that they may serve as ideal tracers. }
   {We investigate the gas dynamics and NH$_2$D chemistry in eight massive precluster and protocluster clumps (G18.17, G18.21, G23.97N, G23.98, G23.44, G23.97S, G25.38, and G25.71).}
   {We present NH$_2$D 1$_{11}$-1$_{01}$ (at 85.926\,GHz), NH$_3$ (1,\,1), and (2,\,2) observations in the eight clumps using the PdBI and the VLA, respectively. We used 3D \texttt{GAUSSCLUMPS} to extract NH$_2$D cores and provide a statistical view of their deuterium chemistry. We used NH$_3$ (1,\,1) and (2,\,2) data to investigate the temperature and dynamics of dense and cold objects.}
   {We find that the distribution between deuterium fractionation and kinetic temperature shows a number density peak at around $T_{\rm kin}=16.1$\,K and the NH$_2$D cores are mainly located at a temperature range of 13.0 to 22.0\,K. The 3.5\,mm continuum cores have a kinetic temperature with a median width of $22.1\pm4.3$\,K, which is obviously higher than the temperature in NH$_2$D cores. We detected seven instances of extremely high deuterium fractionation of $1.0 \leqslant D_{\rm frac} \leqslant 1.41$. We find that the NH$_2$D emission does not appear to coincide exactly with either dust continuum or NH$_3$ peak positions, but it often surrounds the star-formation active regions. This suggests that the NH$_{2}$D has been destroyed by the central young stellar object (YSO) due to heating. The detected NH$_2$D lines are very narrow with a median width of $0.98\pm0.02\,\kms$, which is dominated by non-thermal broadening. The extracted NH$_2$D cores are gravitationally bound ($\alpha_{\rm vir} < 1$), they are likely to be prestellar or starless, and can potentially form intermediate-mass or high-mass stars in future. Using NH$_3$ (1,\,1) as a dynamical tracer, we find evidence of very complicated dynamical movement in all the eight clumps, which can be explained by a combined process with outflow, rotation, convergent flow, collision, large velocity gradient, and rotating toroids.}
   {High deuterium fractionation strongly depends on the temperature condition. Tracing NH$_2$D is a poor evolutionary indicator of high-mass star formation in evolved stages, but it is a useful tracer in starless and prestellar cores.}


   \maketitle
%

\section{Introduction}    
\label{sect_intro}

\begin{table*}
\caption{Properties of selected sample.}
\label{tab_sample} \centering \footnotesize
\begin{tabular}{lcccccc}
\hline \hline
Source\tablefootmark{a} &  Infrared\tablefootmark{b} & \HII\tablefootmark{c} & Maser\tablefootmark{d} & Outflow\tablefootmark{e} & Evolutionary\tablefootmark{f}  &  Distance  \\   & & & Methanol & & Stage & kpc \\
\hline
G18.17    & quiet   &  no   &  no         &  no         & prestellar   & 3.73$^{(1,2)}$   \\
G18.21    & quiet   &  no   &  no         &  no         & prestellar   & 3.60$^{(1,2)}$   \\
G23.97N   & quiet   &  no   &  no         &  no         & prestellar   & 4.68$^{(1,2)}$  \\
G23.98    & quiet   &  no   &  no         &  no         & prestellar   & 4.68$^{(1,2)}$  \\
G23.44-l  & bright  &  yes  & class\,II   &  yes        & protostellar & 5.88$^{(3)}$    \\
G23.44-u  & quiet   &  no   & class\,II   &  no         & protostellar & 5.88$^{(3)}$    \\
G23.97S   & quiet   &  yes  & class\,II   &  yes        & protostellar & 4.70$^{(1,2)}$   \\
G25.38-l  & quiet   &  no   &  no         &  yes        & protostellar & 5.60$^{(4,5,6)}$  \\
G25.38-u  & quiet   &  no   &  no         &  no         & prestellar   & 5.60$^{(4,5,6)}$  \\
G25.71-l  & bright  &  yes  & class\,II   &  yes        & protostellar & 9.50$^{(6,7)}$   \\
G25.71-u  & bright  &  yes  &  no         &  yes        & protostellar & 9.50$^{(6,7)}$    \\
\hline
\end{tabular}
\tablefoot{
\tablefoottext{a}{Source coordinates are listed in Table\,\ref{tab_pdbi}. The ``-l'' and ``-u'' following source names indicate the lower and upper clusters in corresponding clumps. The coordinates of the infrared sources, \HII regions, and masers are listed in Table\,\ref{tab_coordinates}.}
\tablefoottext{b}{Mainly based on a threshold of MIPS 24\,$\mu$m flux $S_{\rm 24\, \mu m}$ = 15.0\,Jy at a distance of 1.7\,kpc \citep{Motte2007}, this flux limit can be rescaled to the distance of the sources in this table \citep[see the MIPS 24\,$\mu$m flux in][]{Paper3}.}
\tablefoottext{c}{Compact \HII region candidate, judged by whether there is a corresponding 1.3\,cm continuum at its sensitivity \citep{Paper3}.}
\tablefoottext{d}{Identified by the 6-GHz methanol multibeam maser catalogs \citep{Green2010,Breen2015}.}
\tablefoottext{e}{Associated with outflows \citep[G23.44-l;][]{Ren2011}, \citep[G23.97S;][]{Cyganowski2008}, \citep[G25.38-l;][]{Liu2011,Zhu2011}, and \citep[G25.71;][]{Villiers2014}.}
\tablefoottext{f}{Identified by the presence or absence of star formation activity toward the center massive core within each source.}\\
\textbf{References for distance.} \tablefoottext{\rm 1}{\citet{Wienen2012}}; \tablefoottext{\rm 2}{\citet{Reid2009}};
\tablefoottext{\rm 3}{\citet{Brunthaler2009}}; \tablefoottext{\rm 4}{\citet{ande2009}};
\tablefoottext{\rm 5}{\citet{Ai2013}}; \tablefoottext{\rm 6}{\citet{Urquhart2013}};
\tablefoottext{\rm 7}{\citet{Lockman1989}}.
}
\end{table*}

High-mass ($\geqslant$\,8\,$\Msun$) stars dominate the Galactic environment and its evolution but many aspects of their formation are still unclear \citep{Shu1987,Churchwell2002,full2005,Bergin2007,Zinnecker2007,Zhang2018}. Firstly, the short time scales of high-mass protostellar objects and their large distances make it difficult to characterize their evolutionary stages. Secondly,  star formation in the early stage takes place at the densest and coldest clumps \citep{Pillai2007,Pillai2011}. At high densities ($n_{\rm H_2} >$ 10$^6$\,cm$^{-3}$) and low temperatures ($T_{\rm dust} <$ 18\,K), characteristic of the interiors of star-forming cores, many molecules are depleted from the gas phase by freezing out onto dust grain surfaces \citep{Walmsley2004,Bergin2007}. Fortunately, the deuterium fractionation of the remaining gas-phase species increases dramatically above the cosmic [D/H] abundance ratio of $\sim$1.5$\times10^{-5}$ \citep{Oliveira2003} due to increased production of H$_2$D$^+$ through the reaction of H$_3^+$ with HD in places where CO is depleted \citep{Roberts2000}. Deuterated molecules, particularly of N-bearing species, can thus serve as selective tracers of the coldest, densest gas in molecular clouds and star-forming cores \citep{Friesen2018}. Therefore, using deuterated species as tracers to probe the initial conditions of high-mass star formation is very useful.

Ammonia and its deuterated isotopologues are formed on grain surfaces through H/D-atom addition reactions to N atoms \citep{Brown1989a,Brown1989b,Fedoseev2015b} and in the gas phase by reactions with deuterated ions convert part of NH$_3$ to NH$_2$D, NHD$_2$, and ND$_3$ \citep{Rodgers2001,Pillai2007}. Substantial deuteration in both phases occurs after the disappearance of CO from the gas phase. It is likely that the relative abundances of the mentioned molecules could give us an idea of the evolutionary stage of a dense core, for example [${\rm N_2D^+}$]/[${\rm N_2H^+}$], [NH$_2$D]/[NH$_3$], and [ND$_3$]/[NHD$_2$] \citep{Crapsi2005,Roueff2005,Flower2006,Pillai2007,Pagani2009,Daniel2016a}. \citet{Harju2017} presented principal reactions forming and destroying NH$_2$D at the deuteration peak. The abundance of NH$_2$D starts to increase gradually, first through the deuteron transfer to ammonia, primarily by HCND$^+$ or DCNH$^+$. The depletion of CO boosts the abundance of H$_3^+$, which, in turn, is efficiently deuterated to H$_2$D$^+$, D$_2$H$^+$, and D$_3^+$ in successive reactions with HD. This stage is characterized by a rapid increase of N-bearing deuterated isotopologues. A detailed deuteration reaction network has been presented in more detail in \citet{Sipila2015b,Sipila2015a}.

Global organized bulk motions are ubiquitously observed in many star-forming regions with different evolutionary stages, for example, stellar wind in an infrared dust bubble \citep{N131,Zhang2016,Zhang2019} and supernova remnant \citep{Zhou2016a,Zhou2016b}, outflow driven by a powerful jet, large scale flow along a filament \citep{Peretto2014,Lu2018,Yuan2018}, and cloud-cloud collisions \citep{Gong2017,Fukui2017}. This suggests that the large-scale ($\gtrsim1$\,pc) dynamics associated with massive star formation could shape the molecular structure. Additionally, the small-scales ($\lesssim1$\,pc) motion could be identified by rotation, inflow, and flow in fiber \citep{Keto2007,Galvan2009,Csengeri2011a,Csengeri2011b,Zhang2014}. The complicated dynamical processes in different scales are linked to the formation of morphological structure and final mass of central star. Therefore, the gas dynamics associated with star formation warrants further study.

In this work, we mainly report gas dynamics using NH$_3$ (1,\,1) and (2,\,2), and we study o-NH$_2$D 1$_{11}$-1$_{01}$ chemistry in eight high-mass star-forming regions (G18.17, G18.21, G23.97N, G23.98, G23.44, G23.97S, G25.38, and G25.71). Four sources are infrared quiet and in a relatively early evolutionary stage, and the other four sources belong to evolved objects with embedded \HII regions. The sample selection and source properties are listed in Table\,\ref{tab_sample} \citep[see also details in][]{Paper3}. In followed Section\,\ref{sect_data}, we describe the Plateau de Bure Interferometer (PdBI) and the Very Large Array (VLA) observations and data reduction. In Section\,\ref{sect_results}, we show observational results of NH$_3$ and NH$_2$D lines. In Section\,\ref{sect_analysis}, source extraction, optical depth, kinetic temperature, density, velocity dispersion, mass, and virial stability are presented and analyzed. In Section\,\ref{sect_discuss}, we discuss gas dynamics and deuterium chemistry. In Section\,\ref{sect summary}, we give a summary.

\begin{table*}
\caption{Parameters of the PdBI and VLA observations: phase center, beam, and rms.}
\label{tab_pdbi} \centering \footnotesize
\begin{tabular}{lcccccccc}
\hline \hline
Source &        \multicolumn{2}{c}{Phase center (J2000)}        &       \multicolumn{4}{c}{NH$_2$D beam \& rms}
&       \multicolumn{2}{c}{NH$_3$ beam \& rms}     \\
&       h~~m~~s~~       &       $^\circ~~'~~''~~$       & \multicolumn{2}{r}{$''\,\times\,''$; $^\circ$ ~~~~ mJy/beam} &       \multicolumn{2}{r}{$''\,\times\,''$; $^\circ$ ~~~~ mJy/beam}  &       \multicolumn{2}{r}{$''\,\times\,''$; $^\circ$ ~~~~ mJy/beam}   \\
\hline
&&&\multicolumn{2}{c}{PdBI CD configurations} & \multicolumn{2}{c}{PdBI BCD configurations} & \multicolumn{2}{c}{VLA D configuration} \\
\object{G18.17} &       18 25 07.534    &       $-1$3 14 32.75  &       4.82$\,\times\,$2.52;   21.22&        $\sim$15        &\multicolumn{2}{c}{---}&       4.61$\,\times\,$3.24;   13.50 &       $\sim$3.3  \\
\object{G18.21} &       18 25 21.558    &       $-1$3 13 39.56  &       5.01$\,\times\,$2.75;   18.20&        $\sim$13        &\multicolumn{2}{c}{---}&       4.64$\,\times\,$3.24;   15.06 &       $\sim$3.6  \\
\object{G23.97N}        &       18 34 28.833    &       $-0$7 54 31.76  &       4.15$\,\times\,$2.85;   21.03&        $\sim$13        &\multicolumn{2}{c}{---}&       4.03$\,\times\,$3.34;   -5.64 &       $\sim$3.2  \\
\object{G23.98} &       18 34 27.823    &       $-0$7 53 28.76  &       5.12$\,\times\,$4.06;   10.53&        $\sim$14        &\multicolumn{2}{c}{---}&       4.03$\,\times\,$3.33;   -7.06 &       $\sim$3.3  \\
\object{G23.44} &       18 34 39.253    &       $-0$8 31 36.23  &\multicolumn{2}{c}{---}&       2.65$\,\times\,$1.60;   17.16&        $\sim$11        &       4.18$\,\times\,$3.42;   -7.07   &       $\sim$3.6  \\
\object{G23.97S}        &       18 35 22.160    &       $-0$8 01 26.53  &\multicolumn{2}{c}{---}&       2.68$\,\times\,$1.59;   18.89&        $\sim$10        &       4.04$\,\times\,$3.34;   -3.95   &       $\sim$2.8  \\
\object{G25.38} &       18 38 08.108    &       $-0$6 46 54.93  &\multicolumn{2}{c}{---}&       2.68$\,\times\,$1.61;   22.63&        $\sim$10        &       4.10$\,\times\,$3.41;   -7.48   &       $\sim$2.6  \\
\object{G25.71} &       18 38 03.184    &       $-0$6 24 14.30  &       5.01$\,\times\,$3.44;   16.79&        $\sim$13        &\multicolumn{2}{c}{---}&       4.06$\,\times\,$3.43;   -6.29 &       $\sim$2.6  \\
\hline
\end{tabular}
\end{table*}

\section{Observations and data reduction}
\label{sect_data}

The IRAM\footnote{IRAM is supported by INSU/CNRS (France), MPG (Germany) and IGN (Spain).} PdBI and NRAO\footnote{The National Radio Astronomy Observatory is a facility of the National Science Foundation operated under cooperative agreement by Associated Universities, Inc.} VLA observations at 1.3\,mm, 3.5\,mm, and 1.3\,cm continuum have been described and presented in \citet{Paper3}. Here we expand on the spectral observations and data.

\subsection{PdBI observations}

The spectra in the PdBI observations were observed simultaneously along with the continuum in Mar.\,25 $-$ Apr.\,12, 2005 and Feb.\,27 $-$ Mar.\,17, 2006 \citep[see details in][]{Paper3}, but in separated correlator windows. Receiver 1 for covering the 1$_{11}$-1$_{01}$ lines of o-NH$_2$D at 85.926\,GHz was tuned to 40\,MHz bandwidth with 460 channels, leading to a velocity resolution of around 0.27\,$\kms$ per channel. The rms noise of spectra was around 23\,$\mjyb$ for NH$_2$D in the C+D configuration observations in sources G18.17, G18.21, G23.97N, G23.98, and G25.71, and around 12\,$\mjyb$ for NH$_2$D in the B+C+D configuration observations in sources G23.44, G23.97S, and G25.38 (see also Table\,\ref{tab_pdbi}).

The IRAM software package GILDAS\footnote{\url{http://www.iram.fr/IRAMFR/GILDAS/}} was used for the data reduction. The continuum contributions were subtracted from the spectral $uv$-table data sets, which were then cleaned with natural weighting, and imaged to obtain spectral images and maps. The region of a twice primary beam size was searched for cleaning components. No polygon was introduced to avoid any biased cleaning. The primary beam was around 58.5$''$ at 86.086\,GHz. The data have been corrected for primary beam attenuation. The detailed observations, data calibration and reduction have been presented in \citet{Paper3}.

\subsection{VLA observations}

The spectra ($J, K$) = (1,\,1) and (2,\,2) transitions of NH$_3$ were simultaneously covered in the eight clumps, using the 2-IF spectral line mode of the correlator with NRAO\footnote{The National Radio Astronomy Observatory is a facility of the National Science Foundation operated under cooperative agreement by Associated Universities, Inc.} VLA D-configuration on November 2005 (project ID AW0669). The bandwidth was 6.25\,MHz, with 127 channels of around 49\,kHz ($\sim$0.617\,$\kms$) each. The NH$_3$ (1,\,1) and (2,\,2) transitions have five and three hyperfine structure (HfS) lines, respectively, and the frequencies of the strongest HfS lines are at 23.6945 and 23.7263\,GHz, respectively. Due to a narrow bandwidth of 6.25\,MHz at around 23.7\,GHz, it can just cover, at most, four HfS lines in five NH$_3$ (1,\,1). The primary beam was about 2$'$, and the typical synthesized beam size was about $3.0''\times2.5''$ at 1.3\,cm. The raw data from the observations was exported to MIRIAD\footnote{\url{http://www.cfa.harvard.edu/sma/miriad/}} and GILDAS by AIPS\footnote{\url{http://www.aips.nrao.edu/index.shtml}} for calibration and imaging. The spectra and continuum were calibrated with primary beam correction. The rms noise of the spectra was between 2.5 and 3.5\,$\mjyb$ for NH$_3$ (1,\,1) and (2,\,2) data (see also Table\,\ref{tab_pdbi}). The detailed observations, data calibration, and reduction can be found in \citet{Paper3}.

\section{Observational results}
\label{sect_results}

\begin{figure}
\centering
\includegraphics[width=0.45\textwidth, angle=0]{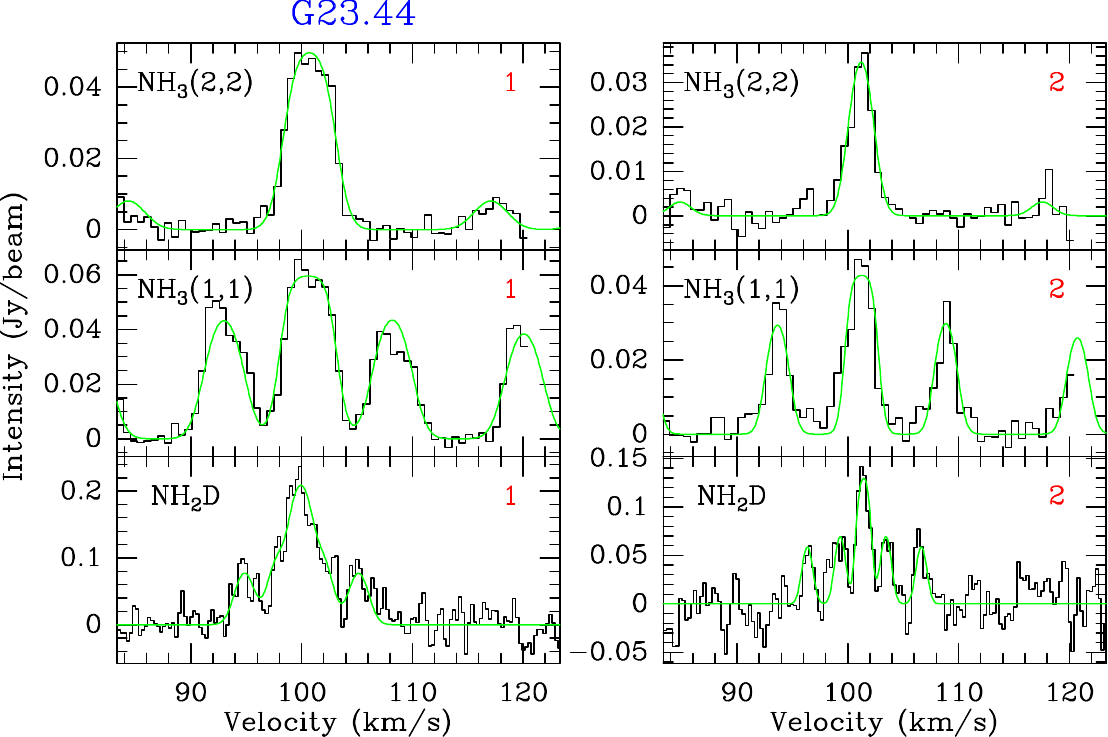}
\caption{Spectra NH$_2$D, NH$_3$ (1,\,1) and (2,\,2) overlaid with their HfS fits for the first two NH$_2$D cores (see Table\,\ref{tab_nh2d_para1}). The spectra are derived by averaging the lines within each NH$_2$D core scale. Other sources and spectra are presented in Appendix Figure\,\ref{Fig_spectra_app}.}
\label{Fig_spectra}
\end{figure}

\subsection{NH$_2$D}
\label{sect_nh2d}

\begin{figure}
\centering
\includegraphics[width=0.45\textwidth, angle=0]{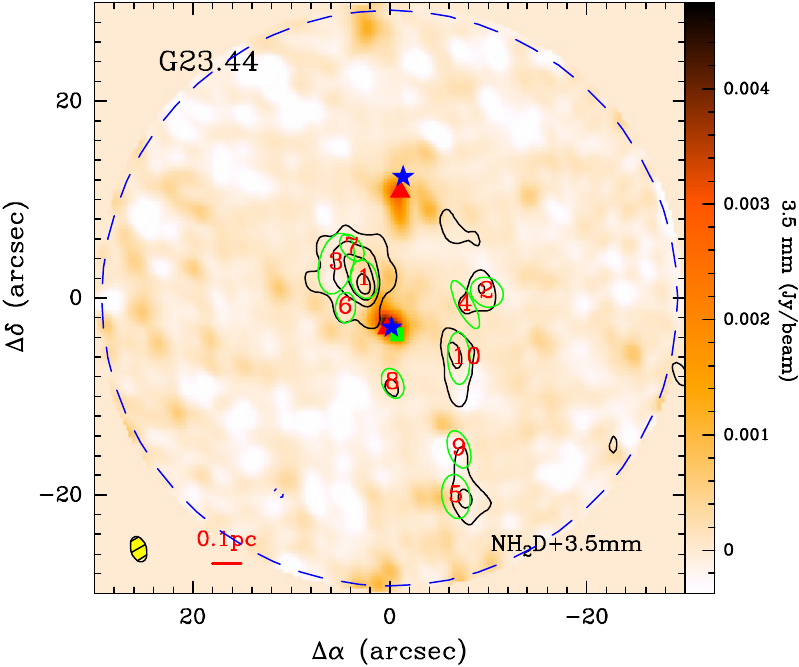}
\caption{NH$_2$D integrated-intensity contours overlaid on a 3.5\,mm continuum with velocity range covering all the six HfS lines. The contour levels start at $-3\sigma$ in steps of $3\sigma$ for NH$_2$D with $\sigma = 33.6\,\mjybkms$. The green ellipses with red numbers indicate the positions of extracted NH$_2$D cores. The symbols ``$\blacktriangle$'', ``$\blacksquare$'', and ``$\bigstar$'' indicate the positions of masers, \HII regions, and infrared sources, respectively. The synthesized beam sizes are indicated at the bottom-left corner. The dashed circle indicates the primary beam of the PdBI observations at 3.5\,mm. Other sources are presented in Appendix Figure\,\ref{Fig_nh2d_app}.}
\label{Fig_nh2d}
\end{figure}

\begin{figure}
\centering
\includegraphics[width=0.45\textwidth, angle=0]{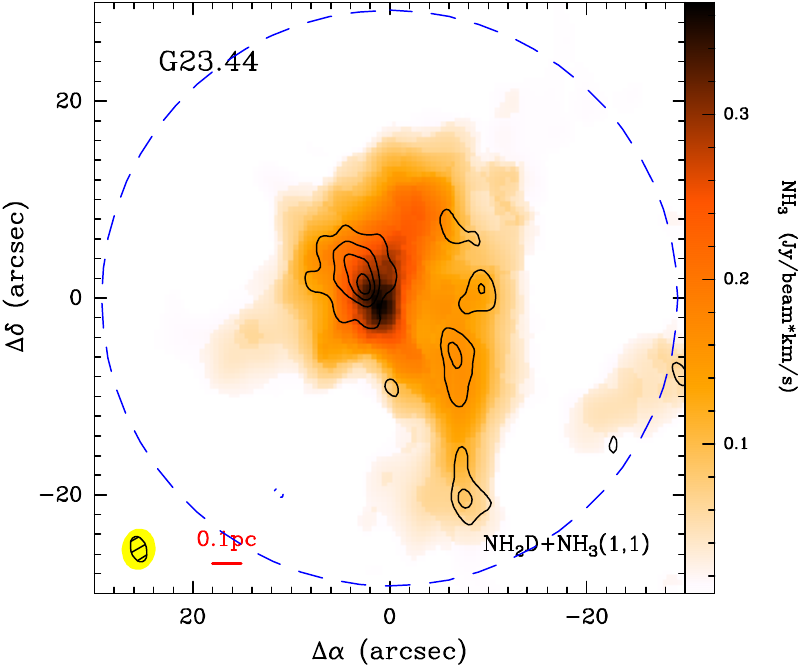}
\caption{NH$_2$D integrated-intensity contours overlaid on an NH$_3$ (1,\,1) integrated-intensity image with velocity range covering all the six HfS lines. The contour levels start at $-3\sigma$ in steps of $3\sigma$ for NH$_2$D with $\sigma = 33.6\,\mjybkms$. The synthesized beam sizes are indicated at the bottom-left corner. The dashed circle indicates the primary beam of the PdBI observations at 3.5\,mm. Other sources are presented in Appendix Figure\,\ref{Fig_nh2d_nh3_app}.}
\label{Fig_nh2d_nh3}
\end{figure}

\begin{table}
\caption{Parameters of o-NH$_2$D HfS lines.}
\label{tab_hfs} \centering \footnotesize
\begin{tabular}{cccccc}
\hline \hline
o-NH$_2$D  & Frequency   & Relative velocity & Relative intensity \\
$1_{1,1}-1_{0,1}$   & MHz         & $\kms$            & \\
\hline
$\rm F=0-1$ & 85924.7829 &  5.189 & 0.111 \\
$\rm F=2-1$ & 85925.7031 &  1.979 & 0.139 \\
$\rm F=2-2$ & 85926.2703 &  0.000 & 0.417 \\
$\rm F=1-1$ & 85926.3165 & -0.162 & 0.083 \\
$\rm F=1-2$ & 85926.8837 & -2.140 & 0.139 \\
$\rm F=1-0$ & 85927.7345 & -5.108 & 0.111 \\
\hline
\end{tabular}
\end{table}

The 1$_{11}$-1$_{01}$ transitions of o-NH$_2$D at around 85.926\,GHz have six HfS lines \citep{Tine2000,Mueller2005,Daniel2016b}. However, two of them blend into one at current spectral resolution (see Figure\,\ref{Fig_spectra}). Therefore we can only distinguish its five emission lines. Table\,\ref{tab_hfs} lists their frequencies, relative velocities, and relative intensities in theoretical calculations \citep{Tine2000}.

The NH$_2$D cores are extracted with 3D \texttt{GAUSSCLUMPS} algorithm (see Section\,\ref{sect_extraction}). The core positions and sizes are indicated using green ellipses with core numbers in Figure\,\ref{Fig_nh2d}. The NH$_2$D spectra of only the first two cores are shown in Figure\,\ref{Fig_spectra} for source G23.44 (other sources are shown in Appendix Figure\,\ref{Fig_spectra_app}), but the whole corresponding parameters are listed in Table\,\ref{tab_nh2d_para1} including velocity, line width, brightness temperature, and opacity\footnote{In Figure\,\ref{Fig_spectra_app}, some main HfS lines in NH$_2$D show multi-velocity components (e.g., G18.17 No.\,2, G18.17 No.\,5, and G23.44 Nos,\,1, 3, 7, 8). For comparison, we only consider the strongest velocity components associated with the corresponding NH$_3$ line. This could bring some error into the line width.}. We also present the integrated-intensity contours of NH$_2$D superimposed on the 3.5\,mm continuum emission image in Figure\,\ref{Fig_nh2d}, and on the NH$_3$ (1,\,1) integrated-intensity maps in Figure\,\ref{Fig_nh2d_nh3}. The integrated velocity range covers all the six HfS lines of the NH$_2$D \citep{Tine2000}. We find that the NH$_2$D peak positions are often not consistent with the 3.5\,mm emission, and there exist obvious offsets between them. In Figures\,\ref{Fig_mom1} and \ref{Fig_mom2}, the integrated-intensity contours of the NH$_2$D are also overlaid on line-center velocity and line-width maps of the NH$_3$ (1,\,1) for further investigating their dynamics.

\subsection{NH$_3$ (1,\,1) and (2,\,2)}
\label{sect_nh3}

In Figure\,\ref{Fig_spectra}, we also present the spectra of NH$_3$ (1,\,1) and (2,\,2) from each corresponding NH${_2}$D core (see Section\,\ref{sect_extraction}). These spectra were derived from the average within a corresponding core size. In Table\,\ref{tab_nh2d_para1}, we list their line width and brightness temperature by spectral Gaussian fitting derived in assumption of local thermodynamic equilibrium (LTE) conditions. In our previous work, \citet{Paper3}, we present the integrated-intensity contours of NH$_3$ (1,\,1) and (2,\,2) overlaying on a 3.5\,mm emission image, respectively. The NH$_3$ peak positions are almost consistent with the 3.5\,mm emission, but the NH$_3$ (1,\,1) and (2,\,2) have much more extended structure than the 3.5\,mm emission distribution. This is mainly because, independent of observations\footnote{The VLA and PdBI observations have different sensitivities to the same spatial structures, but it is not the main reason.}, the 3.5\,mm continuum might be more compact because of the internal heating sources, and NH$_3$ is also optically thick, as we show in Table\,\ref{tab_pdbi}, which also means it is possible to   see more extended NH$_3$ emission.

\begin{figure}
\centering
\includegraphics[width=0.45\textwidth, angle=0]{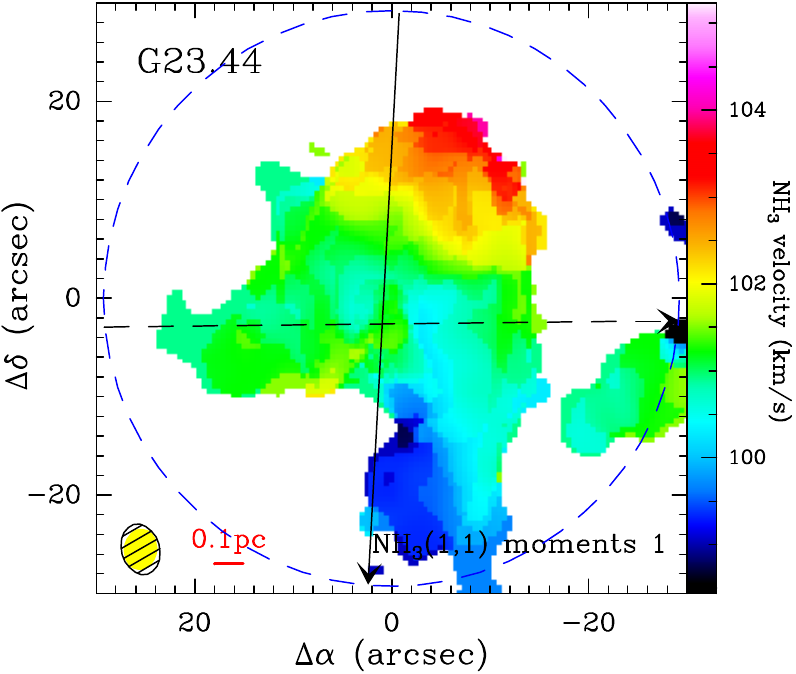}
\caption{Velocity distribution (moment\,1) of NH$_3$ (1,\,1) line overlaid with integrated-intensity contours (moment\,0) of NH$_2$D line with velocity range covering all the six HfS lines. The contour levels start at $-3\sigma$ in steps of $3\sigma$ for NH$_2$D with $\sigma = 84.1\,\mjybkms$. The synthesized beam sizes are indicated at the bottom-left corner. The dashed circle indicates the primary beam of the PdBI observations at 3.5\,mm. The lines with arrows show the position-velocity cutting direction in Figure\,\ref{Fig_pv}. Other sources are presented in Appendix Figure\,\ref{Fig_mom1_app}.}
\label{Fig_mom1}
\end{figure}

\begin{figure}
\centering
\includegraphics[width=0.45\textwidth, angle=0]{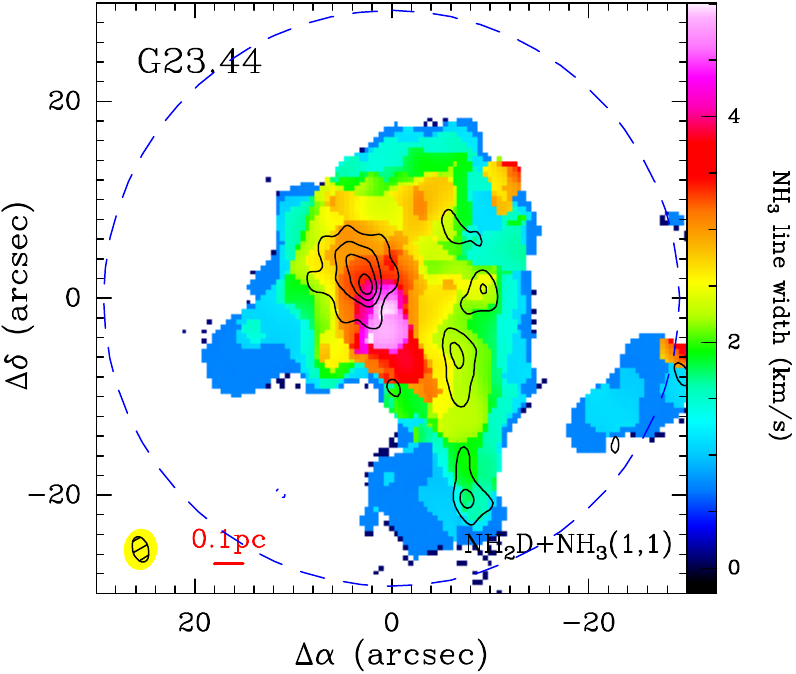}
\caption{Line width distribution (moment\,2) of NH$_3$ (1,\,1) line overlaid with integrated-intensity contours (moment\,0) of NH$_2$D line with velocity range covering all the six HfS lines. The contour levels start at $-3\sigma$ in steps of $3\sigma$ for NH$_2$D with $\sigma = 84.1\,\mjybkms$. The synthesized beam sizes are indicated at the bottom-left corner. The dashed circle indicates the primary beam of the PdBI observations at 3.5\,mm. Other sources are presented in Appendix Figure\,\ref{Fig_mom2_app}.}
\label{Fig_mom2}
\end{figure}

In Figure\,\ref{Fig_mom1}, we present velocity distributions (moment\,1) of NH$_3$ (1,\,1) superimposed on NH$_2$D emission. It is very obvious that there is a steep velocity gradient in the eight molecular clumps, such as G18.17 and 23.97N in east-west direction, G23.44, G25.38, and G25.71 in north-south direction, and the other sources (G23.98 and G23.97S) having two velocity components crossing into together. In Figure\,\ref{Fig_mom2}, we present line width distribution (moment\,2) of NH$_3$ (1,\,1) line overlaid with integrated-intensity contours (moment\,0) of NH$_2$D line. We can see that three high-mass star forming regions (G23.44, G23.97S, and G25.38) have clearly NH$_2$D emission offset from the largest line broadening, indicating that the NH$_2$D emission is often devoid of the dynamically dominated and active regions. Figures\,\ref{Fig_spectra_cont} and \ref{Fig_spectra_cont_app} show the spectra NH$_3$ (1,\,1) and (2,\,2) at the peak position of 3.5\,mm continuum distribution for each source. The NH$_3$ (1,\,1) and (2,\,2) lines present high velocity (wings) emission. The integrated-intensity maps of the blue- and red-shifted spectral wings of NH$_3$ (1,\,1) are shown in Figure\,\ref{Fig_outflow} and \ref{Fig_outflow_app}, where the background is 3.5\,mm emission. To judge the possibility of outflow or rotation movements, Figures\,\ref{Fig_pv} and \ref{Fig_pv_app} show their position-velocity diagrams using the main HfS line of NH$_3$ (1,\,1) along the position-velocity slice indicated with the solid and dashed lines in Figures\,\ref{Fig_outflow} and \ref{Fig_outflow_app}, respectively.

\section{Analysis}
\label{sect_analysis}

\subsection{NH$_2$D core extraction}
\label{sect_extraction}

Assuming that the flux density of each NH$_2$D core can be approximated by a Gaussian distribution, the three dimensional (3D) \texttt{GAUSSCLUMPS} procedure \citep{Stutzki1990,Kramer1996,Kramer1998} in the GILDAS software package was used to characterize them. This methodology has been described in detail and successfully applied in \citet{Kramer1996,Kramer1998}. We consider the sources with line intensity above $5\sigma$ (see $\sigma$ in Table\,\ref{tab_pdbi}) before primary beam correction and line width more than three channels, and fit a Gaussian shaped 3D core with Gaussian size larger than beam size to the surrounding region. For sources G18.17, G18.21, G23.97N, G23.98, and G25.71, we identify NH$_2$D cores using CD configuration observations, and for sources G23.44, G23.97S, and G25.38, we identify them using BCD configuration observations. The identified cores are overlaid onto 2D integrated intensity maps (see Figures\,\ref{Fig_nh2d} and \ref{Fig_nh2d_app}). The velocity, line width, brightness temperature, and opacity are derived from the spectral average within the measured Gaussian size of each core by fitting the HfS lines of NH$_2$D and NH$_3$. The detailed extraction steps are also presented in \citet{Paper3}. The core parameters are listed in Tables\,\ref{tab_nh2d_para1} and \ref{tab_nh2d_para2}.

\subsection{HfS fitting}

The transitions of NH$_3$ (1,\,1) and (2,\,2) at around 23.6945 and 23.7263\,GHz have five and three groups of HfS lines \citep{Kukolich1967,Ho1977}, respectively. This allows for the investigation of spectral profiles and the estimation of line parameters. The outer pair of HfS lines of the NH$_3$ (2,\,2) are too weak to be identified (see Figure\,\ref{Fig_spectra}) with the current sensitivity. We thus do not fit the HfS of NH$_3$ (2,\,2), but fit the HfS of NH$_3$ (1,\,1) to obtain the optical depth and a single component Gaussian profile for the NH$_3$ (2,\,2) main line. We use command ``\texttt{METHOD NH$_3$ (1,\,1)}'' in the CLASS module of the GILDAS package to do Gaussian fitting for the HfS lines assuming in LTE condition. We also estimate the line parameters of the NH$_2$D, by fitting its six HfS lines assuming in LTE condition (see their relative intensities in Table\,\ref{tab_hfs}).

Some spectra of NH$_3$ (1,\,1) shown in Figures\,\ref{Fig_spectra} and \ref{Fig_spectra_app} display anomalies in the inner satellite lines on the blue side (e.g., G18.17 Nos.\,1 and 2, G23.97N No.\,1, G23.98 No.\,1) and in the outer satellite lines on the red side (e.g., G18.21 Nos.\,1--8), which may indicate a non-LTE condition. While the ``\texttt{METHOD NH$_3$ (1,\,1)}'' in CLASS assumes an LTE condition, these anomalies cannot be fitted with the current method. The anomalies of one of the inner satellites could be explained due to systematic motions \citep{Park2001}. On the other hand, the anomaly with the outer satellites being brighter is indicative of non-LTE condition due to HfS selective photon trapping \citep[see, e.g.,][]{Stutzki1985}. Some spectra of NH$_2$D also display such anomalies (e.g., G18.17 No.\,1 and 9, G23.97N No.\,1). This also could be explained by systematic motions or HfS selective photon trapping.

\subsection{Optical depth}

\begin{figure}
\centering
\includegraphics[width=0.33\textwidth, angle=0]{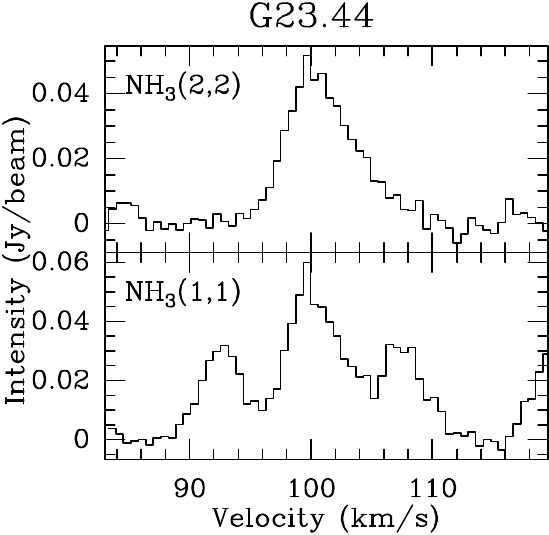}
\caption{Spectra NH$_3$ (1,\,1) and (2,\,2) obtained toward a single and the brightest pixel at 3.5\,mm continuum. The coordinates of the spectra are indicated at the lower panel of Figure\,\ref{Fig_pv}. Other sources are presented in Appendix Figure\,\ref{Fig_spectra_cont_app}.}
\label{Fig_spectra_cont}
\end{figure}

\begin{figure}
\centering
\includegraphics[width=0.45\textwidth, angle=0]{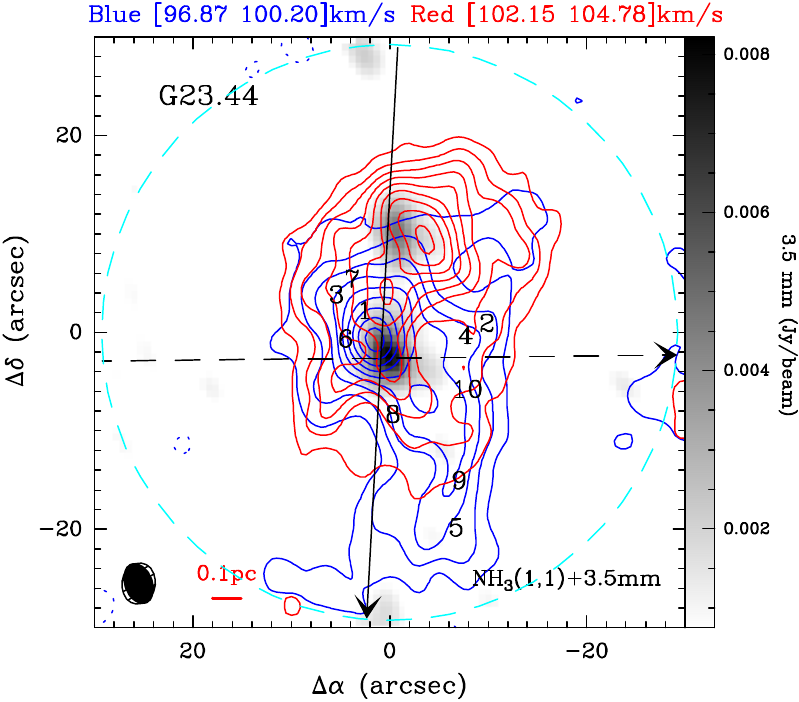}
\caption{Blueshifted and redshifted NH$_3$ (1,\,1) integrated-intensity contours overlaid on a 3.5\,mm continuum. The blue and red contours are the blueshifted and redshifted velocity components, respectively. The blue contour levels start at $-3\sigma$ in steps of $3\sigma$ for NH$_3$ (1,\,1) with $\sigma = 7.2\,\mjybkms$, and the red ones with $\sigma =4.8\,\mjybkms$. The black numbers indicate the positions of extracted NH$_2$D cores. The synthesized beam sizes are indicated at the bottom-left corner. The dashed circle indicates the primary beam of the PdBI observations at 3.5\,mm. The lines with arrows show the position-velocity cutting direction in Figure\,\ref{Fig_pv}. Other sources are presented in Appendix Figure\,\ref{Fig_outflow_app}.}
\label{Fig_outflow}
\end{figure}

\begin{figure}
\centering
\includegraphics[width=0.40\textwidth, angle=0]{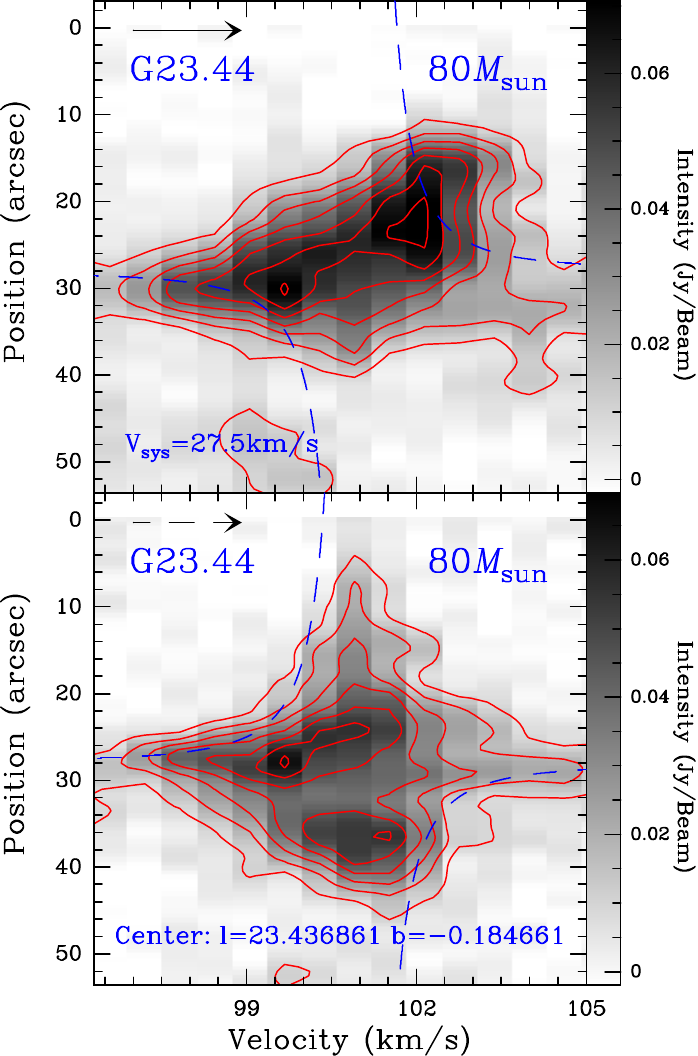}
\caption{Position-velocity diagrams of the main line of NH$_3$ (1,\,1) HfS along the position-velocity slice indicated with solid and dashed lines in Figure\,\ref{Fig_mom1} (see also Figure\,\ref{Fig_outflow}). The arrows show the position-velocity cutting direction. Contour levels start at $3\sigma$ level and increase in steps of $3\sigma$ with $\sigma = 3.3\,\mjyb$ for source G23.44. Blue dashed lines show a possible rotating toroids curve. The central mass, the central position, and the systemic velocity are indicated in the panel. Other sources are presented in Appendix Figure\,\ref{Fig_pv_app}.}
\label{Fig_pv}
\end{figure}

\begin{figure}
\centering
\includegraphics[width=0.45\textwidth, angle=0]{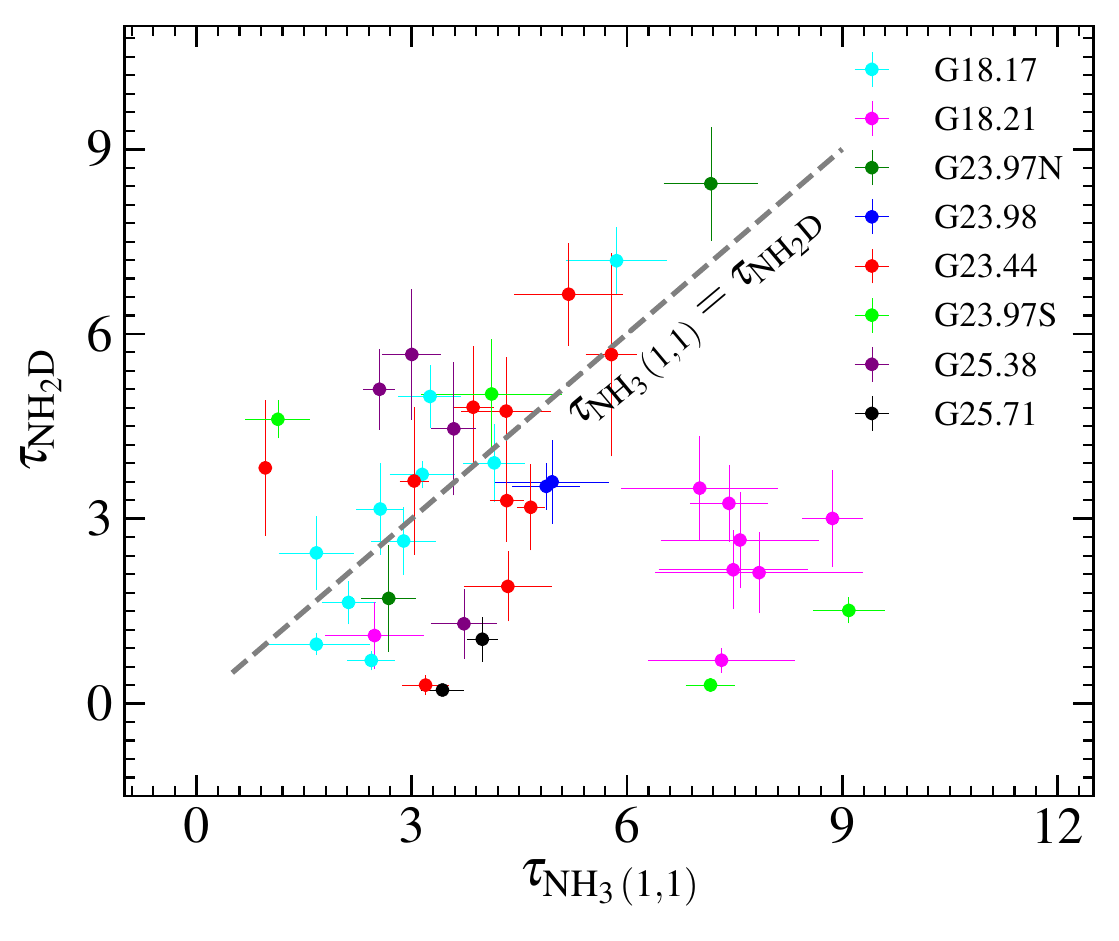}
\caption{Relation between optical depths $\tau$ of NH$_2$D and NH$_3$ (1,\,1) main groups for all NH$_2$D cores. The dashed line corresponds to $\tau_{\rm NH_3 (1,\,1)} = \tau_{\rm NH_2D}$.}
\label{Fig_tau-tau}
\end{figure}

The optical depths of NH$_2$D and NH$_3$ (1,\,1) are derived by HfS fitting and listed in Table\,\ref{tab_nh2d_para1}. Figure\,\ref{Fig_tau-tau} displays the optical depth $\tau$ distribution between NH$_2$D and NH$_3$ (1,\,1). The distribution does not follow any linear relation. The optical depths of the NH$_3$ (1,\,1) range from 1.0 to 9.1 with a median width of $4.05\pm0.04$, indicating that the NH$_3$ (1,\,1) is often optically thick in the cores. The optical depths in the NH$_2$D cores range from 0.2 to 8.4 with a median width of $3.22\pm0.10$, most of which have $\tau_{\rm NH_2D} \gtrsim 1$. Therefore, both NH$_3$ and NH$_2$D are usually optically thick in the dense sources.

\subsection{Excitation temperature}
\label{sect_excitation}

Figure\,\ref{Fig_Tex_Tex} presents the relation between excitation temperatures $T_{\rm ex}$ of NH$_2$D and NH$_3$ (1,\,1) main groups for all NH$_2$D cores. The excitation temperatures of the NH$_3$ (1,\,1) range from 7.0 to 13.0\,K with a median width of $10.21\pm0.28$\,K, while the excitation temperatures in the NH$_2$D cores range from 3.9 to 10.0\,K with a median width of $4.92\pm0.09$\,K. Therefore, the NH$_2$D have lower excitation temperature than the NH$_3$ in the cores.

\subsection{Kinetic temperature}
\label{sect_rot}

Temperature in the dense core is vital in determining the chemical reaction rate of deuteration \citep{Millar1989,Roberts2000}. Ammonia rotational lines NH$_3$ (1,\,1) and (2,\,2) belong to the most useful tracers of the dense cores of molecular clouds, owing to the excitation and chemical properties \citep{Ho1983,Benson1989,Tafalla2002,Friesen2009}. They can remain gaseous in and nearby the cold, dense interior parts of starless and prestellar cores. They  can thus be used as a precise tracer in probing the dust temperature ($\lesssim$\,30\,K) of dense and cold clumps, and in detecting the dynamical motions including outflow and rotation.

\begin{figure*}
\centering
\includegraphics[width=0.45\textwidth, angle=0]{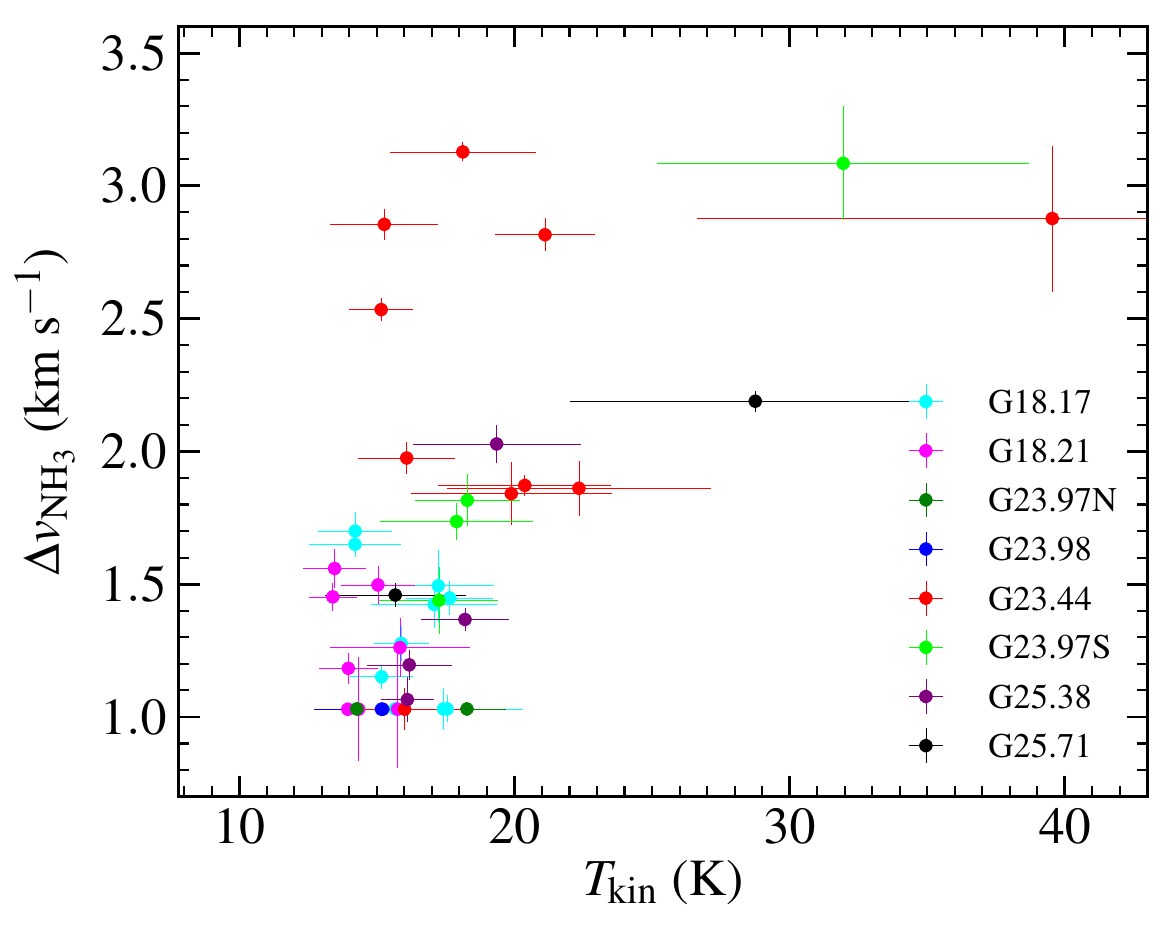}
\includegraphics[width=0.45\textwidth, angle=0]{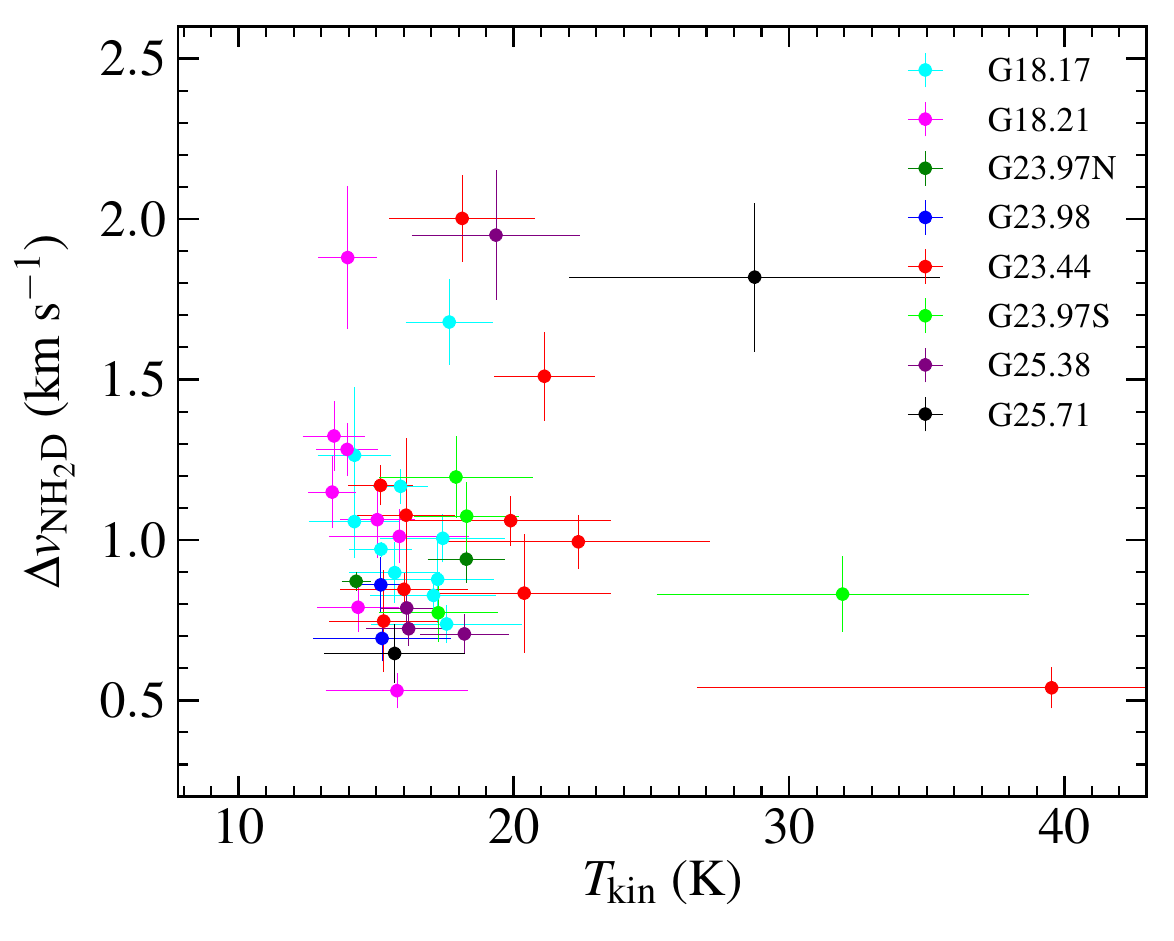}
\includegraphics[width=0.45\textwidth, angle=0]{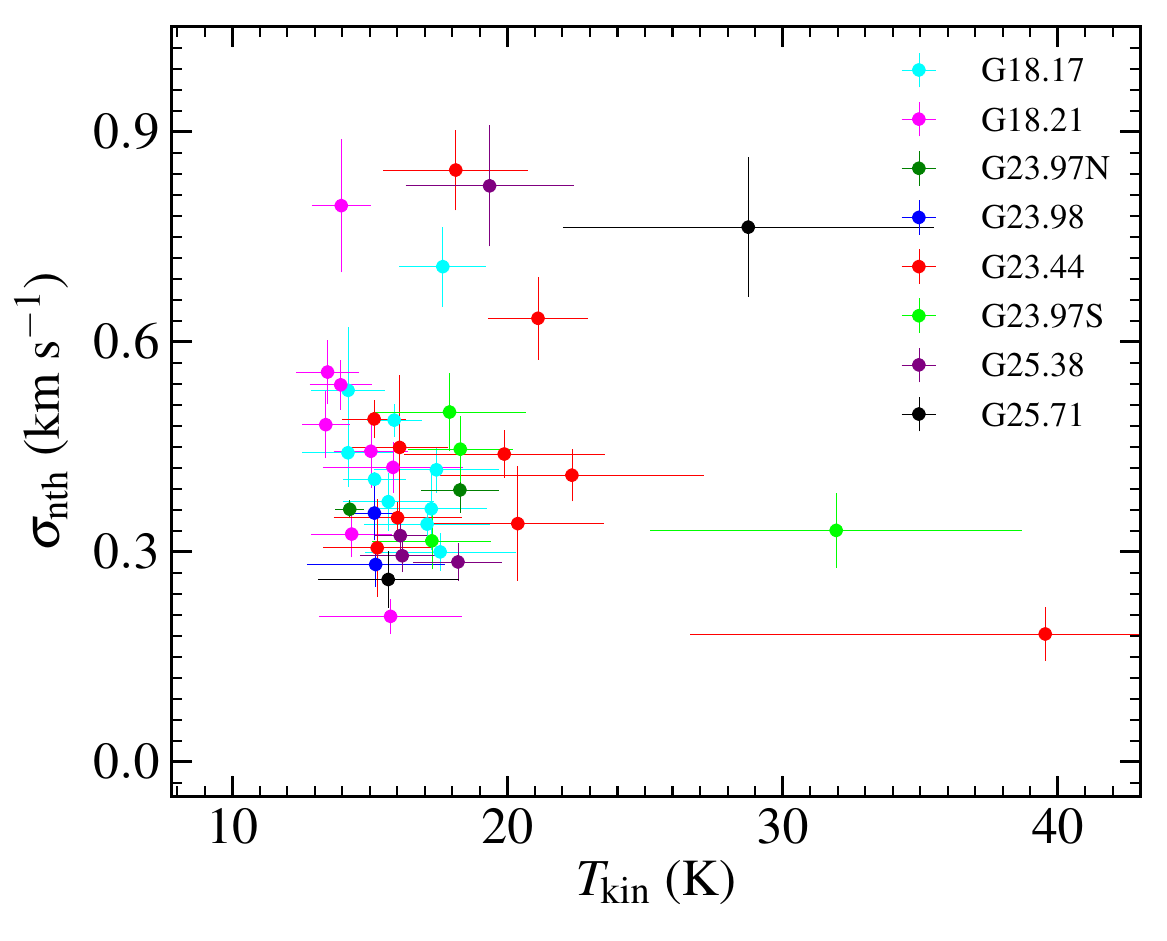}
\includegraphics[width=0.45\textwidth, angle=0]{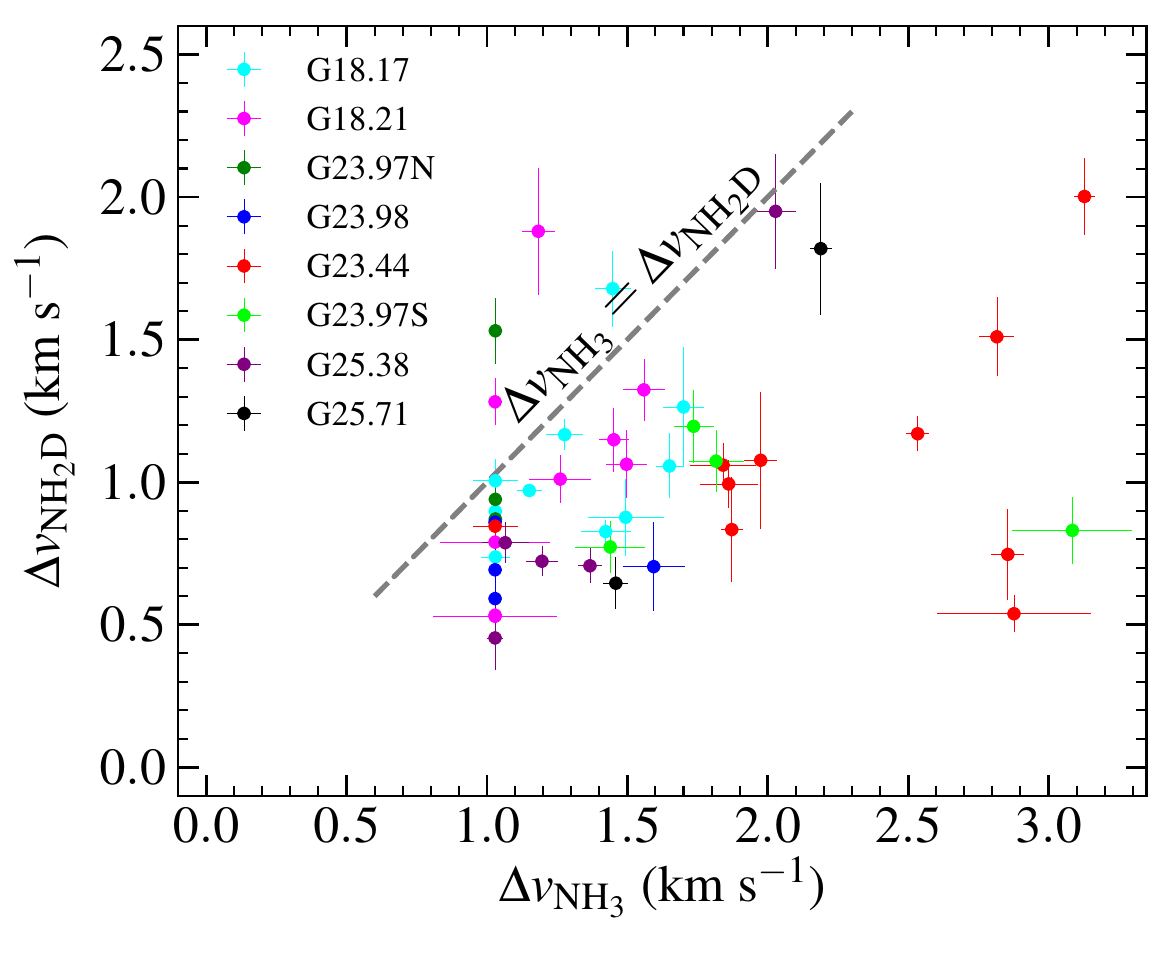}
\caption{Relation between velocity line widths $\Delta v$ and kinetic temperatures $T_{\rm kin}$ for all NH$_2$D cores. The dashed line in lower right panel corresponds to $\Delta v_{\rm NH_3 (1,\,1)} = \Delta v_{\rm NH_2D}$.}
\label{Fig_width-temperature}
\end{figure*}

The kinetic temperature $T_{\rm kin}$ can be estimated from the rotational temperature $T_{\rm rot}$ by using NH$_3$ (1,\,1) and (2,\,2) transitions \citep{Ott2011}, assuming that the NH$_2$D cores have the same temperature condition as the NH$_3$ location. The calculation procedure of the kinetic temperature for each NH$_2$D core is the same as the estimation for the continuum cores in \citet{Paper3}. The derived kinetic temperatures are listed in Table\,\ref{tab_nh2d_para2} and presented in Figures\,\ref{Fig_width-temperature} and \ref{Fig_hist-tempera}. It is quite evident that the NH$_2$D cores have a colder condition than the continuum cores. Most NH$_2$D cores have a temperature ranging from 13.5 to 18.5\,K with a median width of $16.1\pm0.5$\,K. Few NH$_2$D cores have kinetic temperatures above 20\,K. These cores with temperature above 20\,K are always close to the central protostellar cores (see the right panel in Figure\,\ref{Fig_offset}), traced by strong 3.5\,mm and 1.3\,cm continuum. The statistics shows that the number of NH$_2$D cores becomes small in a condition of relatively high temperature, for example these close to the central hot protostellar cores. It is possible that the NH$_2$D excitation have been inhibited to some extent in such condition. Furthermore, the protostellar cores have a kinetic temperature ranging from around 10 to 40\,K, and the median width is $22.1\pm4.3$\,K, which is obviously higher than the kinetic temperature in the NH$_2$D cores.

\subsection{Density and mass}
\label{sect_density}

\begin{figure}
\centering
\includegraphics[width=0.45\textwidth, angle=0]{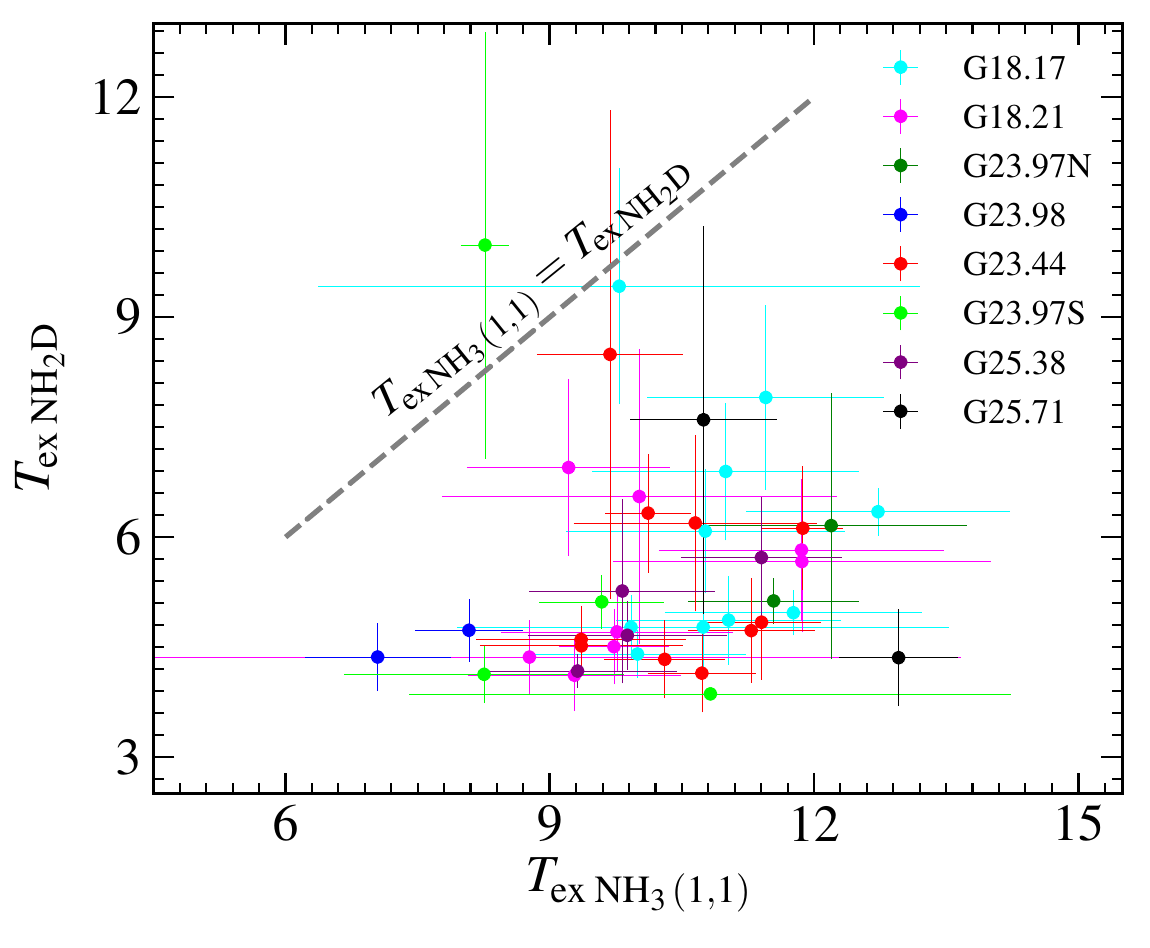}
\caption{Relation between excitation temperatures $T_{\rm ex}$ of NH$_2$D and NH$_3$ (1,\,1) main groups for all NH$_2$D cores. The dashed line corresponds to $T_{\rm ex\,NH_3 (1,\,1)} = T_{\rm ex\,NH_2D}$.}
\label{Fig_Tex_Tex}
\end{figure}

We adopted the analysis routine described in Appendix of \citet{Pillai2007} to estimate the NH$_2$D column density (see Table\,\ref{tab_nh2d_para2}). The derivation of NH$_3$ column density (see Table\,\ref{tab_nh2d_para2}) follows the standard formulation in \citet{Bachiller1987}. The H$_2$ densities of continuum cores are discussed in \citet{Paper3}.

Due to the fact that there is no reliable abundance ratio available between NH$_2$D and molecular hydrogen H$_2$, we use the measured continuum flux within the Gaussian size of each NH$_2$D core to derive a corresponding H$_2$ column density and core mass (see Table\,\ref{tab_nh2d_para2}). If the continuum flux is lower than $3\sigma$, we use the $3\sigma$ as an upper limit. The H$_2$ volume density is estimated by assuming the NH$_2$D cores are in a spherical structure. The derived densities are listed in Table\,\ref{tab_nh2d_para2}. The derived H$_2$ volume density for the NH$_2$D cores ranges from $1.8\times10^{5}$ to $2.4\times10^{6}$\,cm$^{-3}$ with a median width of $(5.3\pm1.4)\times10^{5}$\,cm$^{-3}$, while that for the continuum cores ranges from $1.5\times10^{5}$ to $4.6\times10^{6}$\,cm$^{-3}$ with a median width of $(1.4\pm0.1)\times10^{6}$\,cm$^{-3}$. Therefore, the NH$_2$D emission distributions are in a relatively less dense condition than the continuum cores.

The masses of NH$_2$D cores are also estimated with a corresponding 3.5\,mm continuum emission within the Gaussian size of each NH$_2$D core. We calculate the mass of NH$_2$D cores, using 3.5\,mm dust opacity 0.002\,$\rm cm^2g^{-1}$ \citep{Ossenkopf1994}, dust emissivity 1.7, and gas-to-dust mass ratio 100, and the derived kinetic temperature. The calculation processes have been shown in Section\,4.4 of \citet{Paper3}. The derived parameters are listed in Table\,\ref{tab_nh2d_para2}. Figure\,\ref{Fig_radius-mass} shows $M_{\rm NH_2D}$-$R_{\rm eff}$ distributions of all continuum and NH$_2$D cores for comparisons. According to \citet{Kauffmann2010}, we also plot a threshold between high-mass and low-mass star candidates in Figure\,\ref{Fig_radius-mass}, as it can be used to determine whether the NH$_2$D cores are high-mass star formation candidates. The statistics shows that masses of the NH$_2$D cores at a scale of $R_{\rm eff}\approx0.05$\,pc range from 5.9 to 54.0\,$\Msun$ with a median width of $13.8\pm0.6\,\Msun$. This indicates that some of the NH$_2$D cores may be intermediate- or high-mass candidates unless they further fragment.

\subsection{Deuterium fractionation}
\label{sect_D-frac}

Deuterium fractionation is defined as $D_{\rm frac} = N_{\rm NH_2D}/N_{\rm NH_3}={\rm [NH_2D]/[NH_3]}$ (see calculations for NH$_3$ and NH$_2$D column densities in Section\,\ref{sect_density}). We follow the analytical method of spectra NH$_3$ and NH$_2$D in \citet{Pillai2007} to estimate the deuterium fractionation for our detected NH$_2$D cores. The derived results are listed in Table\,\ref{tab_nh2d_para2}. The deuterium fractionation range in $0.03 \leqslant D_{\rm frac} \leqslant 1.41$ with a median width of $0.48\pm0.01$. Seven of these cores have $D_{\rm frac} > 1.0$.

\subsection{Thermal and non-thermal velocities}
\label{sect_ther-tur}

In a gas at kinetic temperature $T_{\rm kin}$, individual atoms will have random motions away from or towards the observer, leading to red- or blue-wards frequency shifts. The thermal $\sigma_{\rm ther}$ and non-thermal $\sigma_{\rm nth}$ one-dimensional velocity dispersion in each source arising from a Maxwellian velocity distribution is:
        \begin{eqnarray}
        \label{equa_ther}
    \sigma_{\rm ther} = \sqrt{\frac{k T_{\rm kin}}{m_{\rm NH_2D}}},
        \end{eqnarray}
        \begin{eqnarray}
        \label{equa_tur}
    \sigma_{\rm nth} = \sqrt{\frac{\Delta v^{2}_{\rm NH_2D}}{\rm 8ln(2)}-\sigma^{2}_{\rm ther}},
        \end{eqnarray}
where $k$ is the Boltzmann constant, $m_{\rm NH_2D}$ is the molecular mass of the deuterated ammonia, and $\Delta v_{\rm NH_2D}$ is the Gaussian line width FWHM of the NH$_2$D.

The NH$_2$D cores have quite narrow line widths with a median width of $0.98\pm0.02\,\kms$. Based on Equations\,\ref{equa_ther} and \ref{equa_tur}, the thermal and non-thermal velocity dispersion have a median width of $0.09\pm0.01$ and $0.41\pm0.01\,\kms$, respectively (see also Figure\,\ref{Fig_width-temperature}), indicating that only a very small part of thermal velocity contribute into the NH$_2$D line width. Therefore, the non-thermal line broadening takes much higher weighting than the thermal velocity contribution. Comparing the line widths of the NH$_2$D cores with the extracted 3.5\,mm continuum cores in \citet{Paper3}, it is obvious that the 3.5\,mm continuum cores have larger velocity dispersion than the NH$_2$D cores within a similar source size. Therefore the NH$_2$D cores are less turbulent than the 3.5\,mm continuum cores.

\subsection{Position-velocity diagrams}
\label{sect_pv}

\begin{figure}
\centering
\includegraphics[width=0.45\textwidth, angle=0]{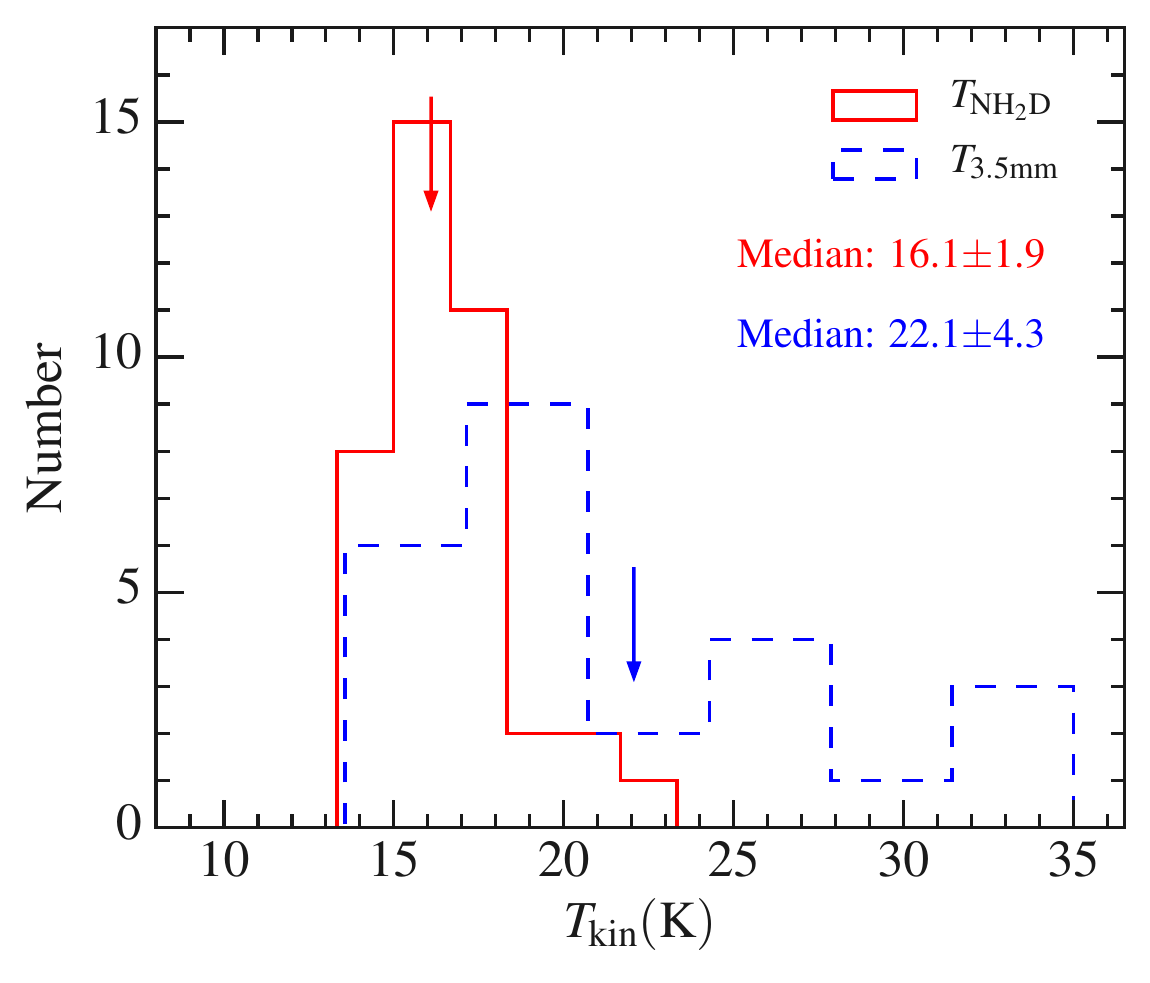}
\caption{Histogram of the kinetic temperatures $T_{\rm kin}$ estimated with lines NH$_3$ (1,\,1) and (2,\,2) for NH$_2$D and 3.5\,mm cores, respectively. The two downwards arrows show the corresponding median width.}
\label{Fig_hist-tempera}
\end{figure}

In Figure\,\ref{Fig_pv}, we present position-velocity diagrams in two different directions along the position-velocity slice indicated with the solid and dashed lines in Figure\,\ref{Fig_outflow}, respectively. We also present possible Keplerian rotation curves in Figure\,\ref{Fig_pv} via
        \begin{eqnarray}
        \label{equa_keplerian}
  v_{\rm kep}(r) = \sqrt{\frac{GM_{\rm core}}{r}},
        \end{eqnarray}
where $v_{\rm kep}$ is the Keplerian velocity, $G$ is the gravitational constant, $M_{\rm core}$ is the continuum core mass, and $r$ is the radius from the central continuum peak position. In this work, the possible central mass within Keplerian orbit is up to $80\,\Msun$. The existence of circumstellar disks ($>30\,\Msun$) has remained elusive up to now. This observational result is likely to prove   unsettling in the areas of theory and simulations. Therefore, the rotating structures are referred to as toroids \citep{Beltran2016}, so as to distinguish them clearly from accretion disks in Keplerian rotation.

The diagrams in Figures\,\ref{Fig_pv} and \ref{Fig_pv_app} show what are obviously dynamical features of rotating toroids \citep{Beltran2016}, such as G23.44, G23.97S, G25.38, and G25.71. We also present NH$_{3}$ (1,\,1) and (2,\,2) lines at the peak position of 3.5\,mm continuum distributions in Figures\,\ref{Fig_spectra_cont} and \ref{Fig_spectra_cont_app}. The spectra show broadening line widths with somewhat blueshifted profiles, such as G23.44, G23.97S, G25.38, and G25.71. The characteristics shown in Figures\,\ref{Fig_spectra_cont}, \ref{Fig_spectra_cont_app}, \ref{Fig_pv}, and \ref{Fig_pv_app} indicate that such sources are dynamically active and that their envelopes (traced by NH$_{3}$) may be rotating and infalling into the central dense cores (traced by 3.5\,mm). It is likely that the central dense cores are boosting their masses by accretion to form a high-mass star in future. For the other four sources (G18.17, G18.21, G23.97N, and G23.98), however, their dynamical motions are relatively quiescent with a little narrower line width than the other four sources (see Figure\,\ref{Fig_spectra_cont_app}). We also present their possible rotating toroids structure in Figure\,\ref{Fig_pv_app}. It seems that the G18.21 shows some dynamical features of rotating toroids. Although the massive gas clumps (e.g., G23.97N and G23.98) do not have any embedded protostellar source down to \textit{Herschel} far-infrared detection limits, the fragmentation and dynamical properties of the gas and dust are consistent with early collapse motion and clustered star formation, which was also argued by \citet{Beuther2013}. Additionally, the central continuum cores in G18.17, G18.21, G23.97N, and G23.98 have relatively quiescent dynamical movements, but their large-scale gas distributions beyond the core size show a large velocity gradient (see Figures\,\ref{Fig_mom1_app} and \ref{Fig_pv_app}).

The possible central mass within Keplerian orbit velocity for each source is roughly estimated and indicated in Figure\,\ref{Fig_pv_app}. The sources G23.44, G23.97S, G25.38, and G25.71 have relatively large central mass with around 80\,$\Msun$, while the masses in sources G18.17, G18.21, G23.97N, and G23.98 range from 10 to 30\,$\Msun$. The evidence in Figure\,\ref{Fig_pv_app} may suggest that the accretion has started in prestellar core stage (e.g., G23.97N and G23.98), and the accretion rate continues to increase in protostellar stages (e.g., G23.44 and G23.97S).

\subsection{Virial parameter}

The virial theorem can be used to test whether one NH$_2$D core is in a stable state. We assume a simple spherical fragment with a density distribution of $\rho\propto r^{-2}$, where $r$ is the radius of spherical fragment. If ignoring magnetic fields, bulk motions, and external pressure of the gas, the virial mass of a fragment can be estimated with the formula \citep{MacLaren1988,Evans1999}:
        \begin{eqnarray}
        \label{equa_virial-mass}
  M_{\rm vir} \simeq 126\, R_{\rm eff}\, \Delta v^{2}_{\rm nth}\, (\Msun),
        \end{eqnarray}
where $R_{\rm eff}$ is the source effective radius in pc and $\Delta v_{\rm nth}$ is the non-thermal line width for NH$_2$D main line (see also Equation\,\ref{equa_tur}). A similar derivation for virial mass can be found in \citet{Paper3}. However, one exception is that the velocity dispersion in this work was estimated with NH$_2$D non-thermal velocity rather than NH$_3$. The corresponding parameters are listed in Table\,\ref{tab_nh2d_para2}.

\begin{figure}
\centering
\includegraphics[width=0.45\textwidth, angle=0]{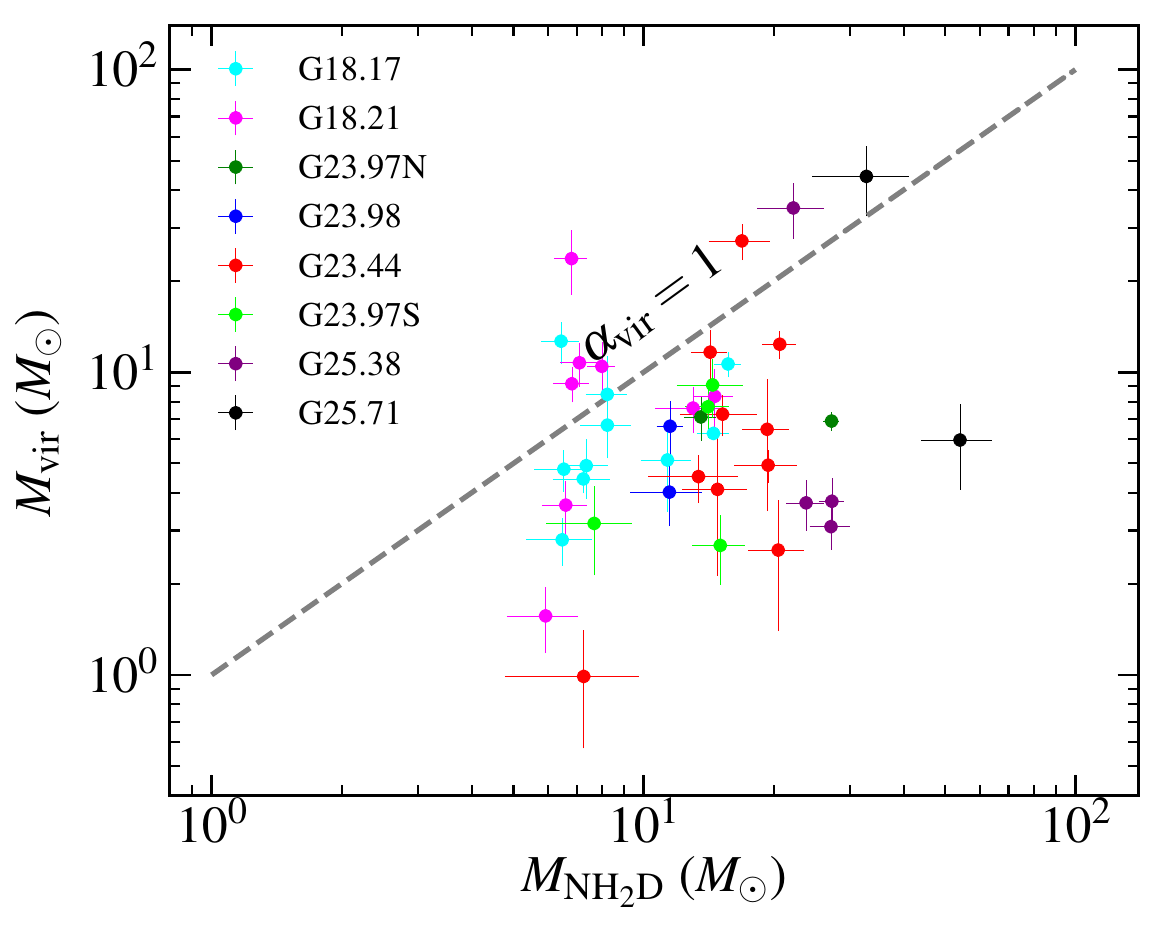}
\caption{$M_{\rm vir}$-$M_{\rm NH_2D}$ distributions of all NH$_2$D cores. The corresponding data are listed in Table\,\ref{tab_nh2d_para2}. The dashed gray line shows a threshold of virial parameter $\alpha=M_{\rm vir}/M_{\rm NH_2D}=1$.}
\label{Fig_mass-virial}
\end{figure}

Comparing the virial mass $M_{\rm vir}$ with the NH$_2$D core $M_{\rm NH_2D}$, if the virial parameter $\alpha_{\rm vir} = M_{\rm vir}/M_{\rm NH_2D} < 1$, the dense source is gravitationally bound, potentially unstable, and to collapse; if $\alpha_{\rm vir} > 1$, the source is not gravitationally bound, in a stable or expanding state. In Figure\,\ref{Fig_mass-virial}, we show $M_{\rm vir}$-$M_{\rm NH_2D}$ distributions of all NH$_2$D cores. The statistics shows that the virial parameters range from 0.11 to 3.48 with a median width of $0.55\pm0.02$. This indicates that the NH$_2$D cores are mostly gravitationally bound.

\section{Discussion}
\label{sect_discuss}

\begin{figure}
\centering
\includegraphics[width=0.45\textwidth, angle=0]{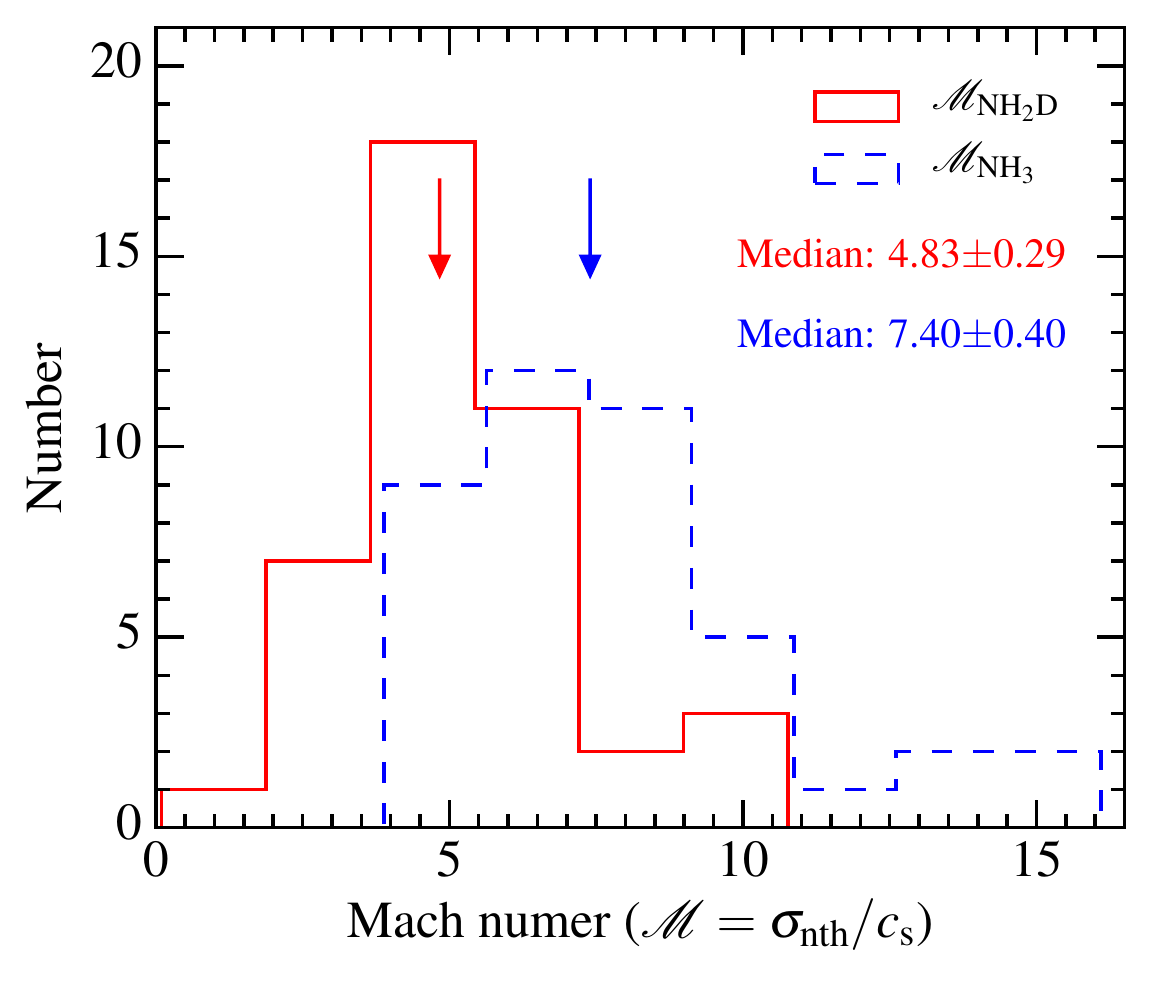}
\caption{Histogram for the ratio (Mach number $\mathcal{M}=\sigma_{\rm nth}/c_{\rm s}$) of the non-thermal velocity dispersion $\sigma_{\rm nth}$ to the local sound speed $c_{\rm s}$ estimated with lines NH$_2$D and NH$_3$ (1,\,1) for NH$_2$D cores, respectively. The two downwards arrows show the corresponding median width.}
\label{Fig_hist-width}
\end{figure}

\subsection{Complex gas dynamics}
\label{sect_dynamics}

NH$_3$ (1,\,1) is a good tracer of relatively dense gas and extended molecular clouds, and often used as dynamical tracer \citep[e.g.,][]{Galvan2009,Zhang2014}. To investigate the dynamical structure of the molecular clumps, we present the velocity distributions (moment\,1) of NH$_3$ (1,\,1) in Figure\,\ref{Fig_mom1}, which shows large velocity gradient and complicated distribution. Steep velocity gradient clearly exists in the eight molecular clumps, such as G18.17 and 23.97N in east-west direction, G23.44, G25.38, and G25.71 in north-south direction, and the other sources (G23.98 and G23.97S) having two velocity components crossing into together. The integrated-intensity maps in Figure\,\ref{Fig_outflow} show multiple emission peaks, even the blue- and red-shifted components present crossed distributions. It seems to be common to observe this dynamical phenomenon not only in the prestellar stage (e.g., G23.98) but  in the protostellar stage as well (e.g., G23.97S). \citet{Csengeri2011a,Csengeri2011b} explained this movement as a convergent flow. It is also likely that the molecular clumps are colliding with each other or simply overlapping in the plane of the sky but still physically separated in the third spatial dimension. In clumps G18.21, G23.98, G23.44, and G23.97S, the interaction regions of the possible convergent or colliding flows show relatively broad line width with no evidence of elevated gas temperatures (see Figures\,\ref{Fig_mom2} and \ref{Fig_mom2_app}), while in clump G23.97S, we find a relatively high temperature as evidence of convergent or colliding flows \citep[see the temperature distribution in Figures\,4 and D.10 of][]{Paper3}. Furthermore, Figures\,\ref{Fig_pv} and \ref{Fig_pv_app} show possible rotating toroids signatures in all the eight sample. Thus, the convergent flow, the colliding flow, and the rotating toroids are  coexistent in the clumps in a complicated way.

In Figures\,\ref{Fig_spectra} and \ref{Fig_spectra_app}, the hyperfine lines of NH$_3$ (1,\,1) in some cores exhibit anomalous intensity ratios. For example, G18.17 Nos.\,1, 2, and 6, G23.97N No.\,1, G23.98 Nos.\,1 and 4, G23.44 Nos.\,1, 3, 5, 6, G23.97S No.\,1, G25.38 No.\,3, and G25.71 No.\,1 show obviously stronger inner satellites on the blue side than on the red side, while G18.21 Nos.\,1--9 have stronger outer satellites on the red side. The anomaly is only partially attributable to a non-LTE effect on the hyperfine transitions \citep{Park2001,Stutzki1985}. It was suggested by \citet{Park2001} that the hyperfine line intensity ratios could be tracing a systematic motion inside the dense cores. \citet{Park2001} also found that expansion (outward motion) can strengthen the inner as well as outer satellite lines on the red (blue) side, while suppressing those (inward motion) on the other side, and that the line anomaly becomes prominent as the gas density increases. The anomalies further suggest complicated dynamical motions in the dense cores.

Since the detected NH$_2$D lines are very narrow ($<$\,1.0\,$\kms$), and just cover several channels, it is really difficult to discuss the dynamics of the NH$_2$D lines. We tried to use the blue- and red-shifted wings of the NH$_2$D line to check whether one can find evidence of flow motions, but nothing was found. For the NH$_2$D cores, the Mach number (the ratio of the non-thermal velocity dispersion $\sigma_{\rm nth}$ to the sound speed $c_{\rm s}$ in Figure\,\ref{Fig_hist-width}) traced by NH$_2$D has a median width of $4.83\pm0.29$, which is 1.5 times smaller than the Mach number traced by NH$_3$. Therefore, the NH$_2$D cores have a much more quiescent dynamics than the NH$_3$ cores (see also the Mach number in Figure\,\ref{Fig_hist-width}). This suggests that the NH$_2$D distributions have a very small velocity gradient. Figures\,\ref{Fig_mom2} and \ref{Fig_mom2_app} also shows that the NH$_2$D cores are often located at dynamically quiescent regions, relatively, for example in sources G23.44, G23.97S, and G25.38. In Figure\,\ref{Fig_outflow}, the blue- and red-shifted spectral wing emission seems to be correlated with each 3.5\,mm source separately, such as G18.17, G18.21, G23.44, and G25.71. It is likely that such individual sources have multiple velocity components. For clumps G23.97N, G23.97S, and G25.38, we find that the central continuum cores are located at the shearing positions of blue- and red-shifted components. It is likely that the central continuum cores are the power sources, which produce a large velocity dispersion of about 5\,$\kms$ probably derived by outflow motions.

\subsection{Offset between NH$_{2}$D cores and continuum peak}
\label{sect_offset}

\begin{figure}
\centering
\includegraphics[width=0.45\textwidth, angle=0]{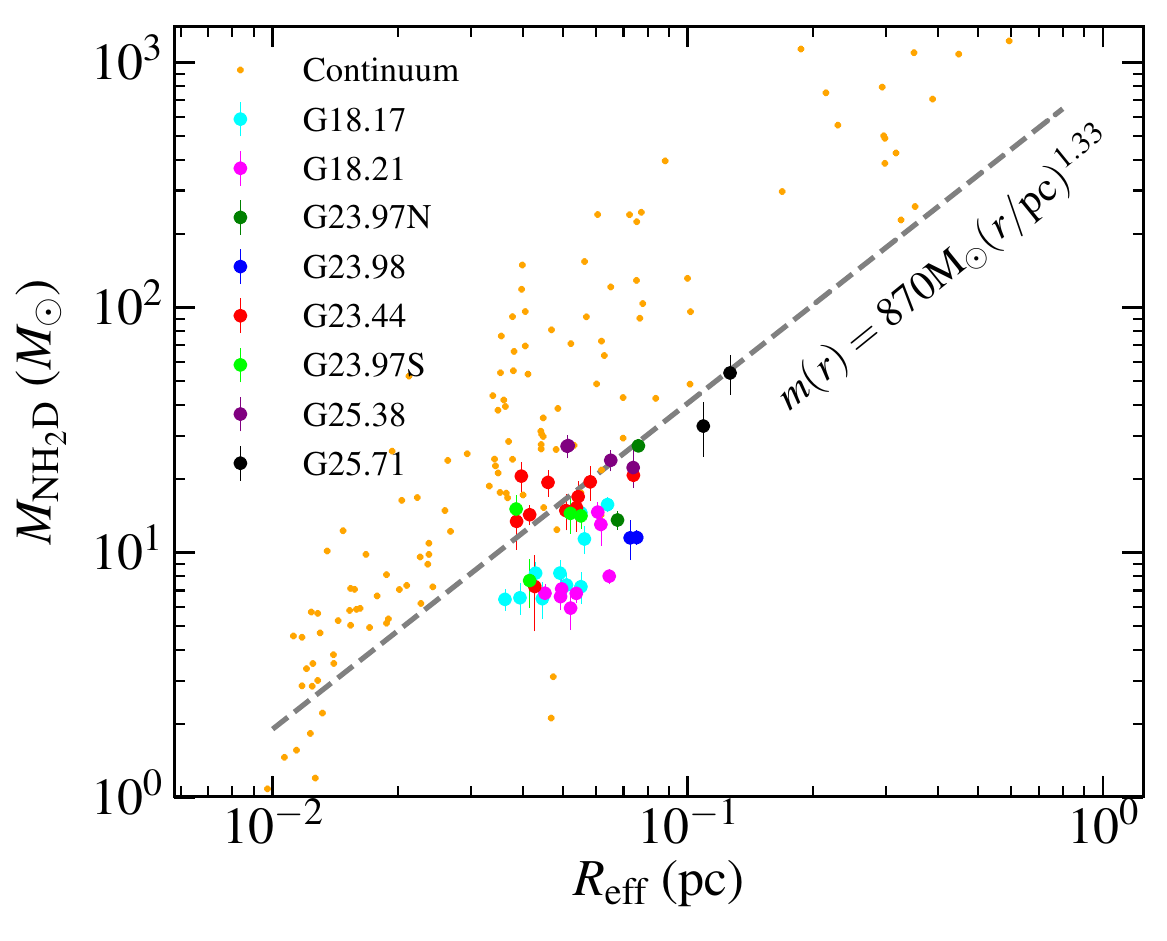}
\caption{$M_{\rm NH_2D}$-$R_{\rm eff}$ distributions of all continuum and NH$_2$D cores. The masses are derived from the integrated flux within a measured Gaussian FWHM using \texttt{GAUSSCLUMPS}, and the effective radius is $R_{\rm eff} = {\rm FWHM}/(2\sqrt{\rm ln2})$. The dashed gray line shows a threshold between high-mass and low-mass star candidates \citep{Kauffmann2010}. The yellow and other color points indicate the continuum and NH$_2$D cores, respectively.}
\label{Fig_radius-mass}
\end{figure}

\begin{figure*}
\centering
\includegraphics[height=0.36\textwidth, angle=0]{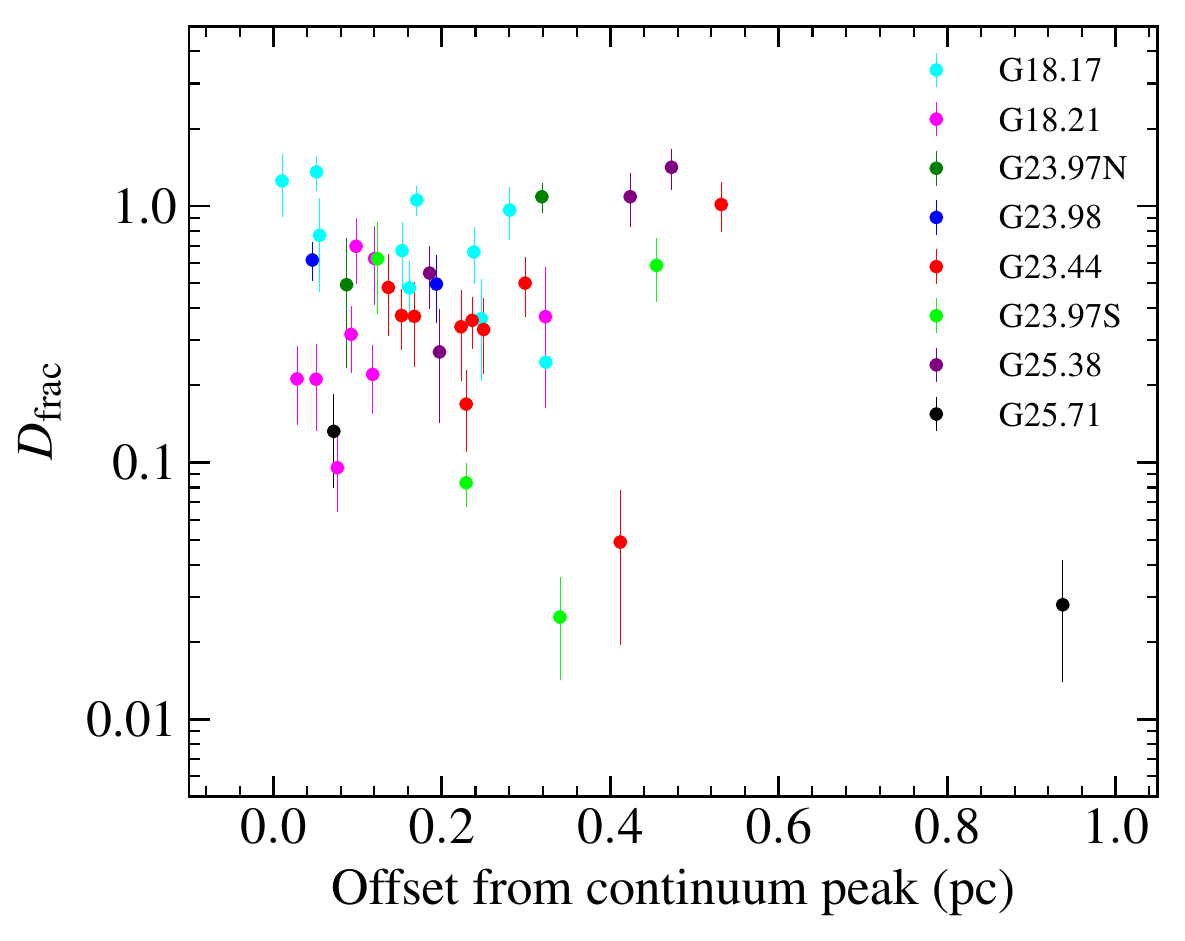}
\includegraphics[height=0.36\textwidth, angle=0]{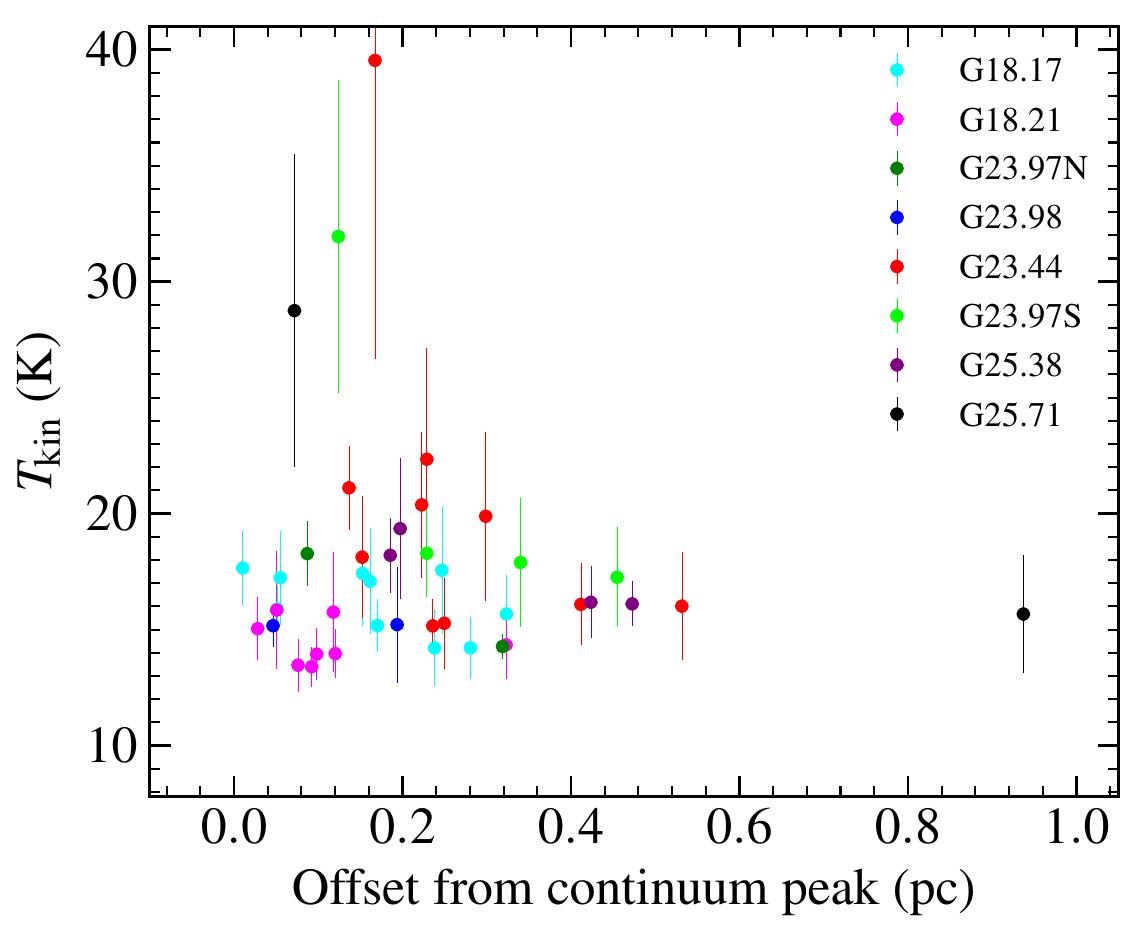}
\caption{Deuterium fractionation $D_{\rm frac}$ and kinetic temperature $T_{\rm kin}$ versus the projected offset distance to 3.5\,mm continuum peak position for each NH$_2$D core.}
\label{Fig_offset}
\end{figure*}

Seen from low-angular resolution of single-dish observations, deuterated species often have a good positional association with the cold cores in early stage. However, \citet{Roueff2005} found that deuterated species do not peak in protostars themselves, but at offset positions, and suggested that protostellar activity decreases deuteration built in the prestellar phase. \citet{Pillai2012} reported that the H$_2$D$^+$ peak is not associated with either a dust continuum or N$_2$D$^+$ peak. \citet{Friesen2014} revealed that there exists offset between H$_2$D$^+$ core and continuum peak positions, probably due to heating from undetected, young, low luminosity protostellar source or first hydrostatic core, or HD depletion in the cold center of the condensation in their opinion. As has also been argued by \citet{Pillai2011}, the cold ($\rm < 20\,K$) and dense ($\rm > 10^{6}\, cm^{-3}$) situations are two necessary conditions for producing a high NH$_{2}$D abundance.

In Figures\,\ref{Fig_nh2d} and \ref{Fig_nh2d_nh3}, we overlaid NH$_{2}$D integrated intensity contours on NH$_{3}$ integrated intensity and 3.5\,mm continuum images, respectively. Figure\,\ref{Fig_offset} displays the relation of the deuterium fractionation and kinetic temperature versus the projected offset distance to 3.5\,mm continuum peak position for each NH$_2$D core. We find that the NH$_2$D peak positions are often not associated with either dust continuum or NH$_3$ emission peak positions. Clumps G18.17, G18.21, G23.97N, and G23.98 have very weak infrared and millimeter emission, but strong and extended NH$_2$D emission distributions. For clumps G23.44, G23.97S, and G25.38, the NH$_{2}$D distributions are extended and surrounding the 3.5\,mm continuum peak positions, and their minimum projected offset distances to the continuum peak nearby are 0.13, 0.12, and 0.17\,pc, respectively (see also Figure\,\ref{Fig_offset}). In Figure\,\ref{Fig_nh2d}, for source G25.71, we only detected two weak NH$_{2}$D emission. One core is located at the continuum peak position, another is far away from the continuum peak. Considering that G25.71 is an evolved source with an embedded protostellar core, it is really strange that the NH$_{2}$D core (No.\,1) has almost no projected offset from the peaked bright continuum core, which has a relatively high dust temperature of $28.8\pm6.7$\,K. It may be possible that this NH$_{2}$D core is just located in line of sight toward the continuum core. It is also likely that the 3.5\,mm core in G25.71 has a very cold and thick NH$_{2}$D envelope covering the hot dust insides. Generally, large projected offsets exist between the NH$_{2}$D core and continuum peak positions, and the projected offsets are larger in evolved objects (e.g., G23.44, G23.97S, and G25.38) than those in the earlier evolutionary stages, (e.g., G23.97N and G23.98).

By measuring the kinetic temperatures of NH$_{2}$D cores (see Figure\,\ref{Fig_hist-tempera}), we find a suitable condition for producing a high-level abundance of NH$_{2}$D: dust temperature between 13.0\,K and 22.0\,K (see also Section\,\ref{sect_chemistry}), and the corresponding column density derived from 3.5\,mm continuum ranges from $4.0\rm \times10^{22}$ to $36.0\rm \times10^{22}\, cm^{-2}$. The NH$_{2}$D distributions are also devoid of a bright infrared emission \citep[see their infrared distributions in][]{Paper3}, masers, and \HII regions (see Figures\,\ref{Fig_nh2d} and \ref{Fig_nh2d_app}). Based on the analysis above, we suggest that the NH$_{2}$D emission close to the central bright continuum core (protostellar core) has been destroyed by an embedded young stellar object (YSO) due to its heating. The detected NH$_{2}$D cores may be just the fragments of the cold and dense envelope associated with high-mass star-forming core insides. It is also very likely that the NH$_{2}$D cores are massive starless seeds, some of which are possible to form future high-mass stars (see Figure\,\ref{Fig_radius-mass}).

\subsection{Very high deuterium fractionation}
\label{sect_chemistry}

Deuterium fractionation is believed to be a fossil of cold chemistry in the early cold evolutionary phase \citep{Parise2009,Pillai2012}. Using single-dish observations, \citet{Pillai2007} found that 65\% of the observed sample have strong NH$_2$D emission with a high deuterium fractionation of $0.1 \leqslant D_{\rm frac} \leqslant 0.7$. Toward G29.96e and G35.20w with interferometer, \citet{Pillai2011} obtained another deuterium fractionation of $0.06 \leqslant D_{\rm frac} \leqslant 0.37$. Recently, \citet{Busquet2010} reported a high value of $D_{\rm frac} \sim 0.8$ in a pre-protostellar core close to high-mass star-forming region IRAS 20293+3952. By modeling the observed spectra, \citet{Harju2017} derived the fractionation ratios with $D_{\rm frac} \sim 0.4$.

High deuteration is mainly produced by two pathways: gas-phase ion-molecule chemistry and ice-grain surface chemistry \citep[e.g.,][]{Rodgers2001,Millar2002,Millar2003,Hatchell2003,Roueff2005,Pillai2007}. The root ion-neutral fractionation reaction is:
 \begin{eqnarray}
{\rm H_3^+ + HD \rightarrow H_2D^+ + H_2 + 230\,K},
 \end{eqnarray}
which dominates at a temperature of $<$\,20\,K, generally \citep[e.g.,][]{Millar1989,Ceccarelli2014,Harju2017,Simone2018}. Neutral molecules like CO can destroy $\rm H_2D^+$, thereby lowering the deuterium enhancement. \citet{Roberts2000} suggested that at around 10\,K, accretion of neutrals onto the dust grains, especially CO, leads to the formation of doubly deuterated molecules. Based on the above, we expect to see a correlation between the deuteration fractionation $D_{\rm frac}$ and temperature $T_{\rm kin}$. Figure\,\ref{Fig_fractionation} displays deuterium fractionation versus kinetic temperature for all the extracted NH$_2$D cores. We find that the distribution between deuterium fractionation and kinetic temperature\footnote{However, we have to note that the NH$_2$D excitation temperature is smaller than the NH$_3$ excitation temperature (see also Figure\,\ref{Fig_Tex_Tex}). Then by using the NH$_3$ temperatures the NH$_2$D column density will be overestimated and in turn the deuterium fractionation.} shows a number density peak at around $T_{\rm kin}=16.1$\,K and $D_{\rm frac}\sim0.4$, and the NH$_2$D cores are mainly located at a temperature range of 13.0 -- 22.0\,K (see also the histogram of the kinetic temperatures in Figure\,\ref{Fig_hist-tempera}).

Figure\,\ref{Fig_fractionation} also displays a gas phase model predictions from \citet{Roueff2005} for comparison. Most of our sample have much higher deuterium fractionation than the model. Therefore, current models of gas phase reaction even under conditions of high depletion are not capable of explaining the high fractionation observed in this work. The gas-grain chemical reactions are expected to explain the production of the high deuterium fractionation. However, few corresponding gas-grain chemical models are currently available. Additionally, deuterium fractionation $D_{\rm frac}>1.0$  warrant closer scrutiny. We attribute such anomalous values to missing short spacing information in our data. However, comparing with single-dish observations in \citet{Pillai2007}, the deuterium fractionation has been up to 0.7 in clump G18.17. Therefore, it is reasonable for the three higher deuterium fractionations (e.g., $D_{\rm frac}\sim1.06$ for NH$_{2}$D core No.\,1, $D_{\rm frac}\sim1.37$ for No.\,4, and $D_{\rm frac}\sim1.26$ for No.\,6 in clump G18.17) with higher spatial resolution from interferometer observations.

As seen from the diagram between the deuterium fractionation and kinetic temperature in Figure \ref{Fig_fractionation}, these data points are just scattering from 13.0 to 22.0\,K, and the median value is around 16.1\,K. Therefore, the suitable condition for NH$_2$D production mainly ranges from 13.0 to 22.0\,K, and the deuterium fractionations will reach up to the maximum at 16.1\,K. For these higher than 22.0\,K, the activity of NH$_2$D production is likely inhibited, and maybe they have been dissociated \citep[e.g.,][]{Rodgers2001,Millar2002,Millar2003,Roueff2005}. When it is less than 13.0\,K, it is possible that NH$_2$D also tends to be frozen out onto dust grain \citep{Brown1989a,Brown1989b,Fedoseev2015b}. However, we also should not ignore that the tracer of kinetic temperature, NH$_3$, may have been seriously frozen onto the dust grain before NH$_2$D has \citep{Fedoseev2015a}. Therefore, it may be not suitable to use NH$_3$ and NH$_2$D as tracers to study dense gas at a temperature condition of $T_{\rm kin}<$\,13.0\,K.

In Figure\,\ref{Fig_offset}, we plot the relation in the deuterium fractionation and kinetic temperature versus the projected offset distance to 3.5\,mm continuum peak position for each NH$_2$D core. We find that the NH$_2$D cores are located within a projected radius region between 0.02 and 0.5\,pc. It seems that the deuterium fractionation or kinetic temperature distribution does not vary significantly with the changes of the projected offset distances between between 0.02 and 0.5\,pc. In other projected offset areas, we detected few NH$_2$D cores.  This diverges from what has been suggested by \citet{Friesen2018}. It is possible that the cold dust envelopes are extremely thick enough, leading to that most of heating from the central hot source has been cooled down by the envelopes.

\subsection{Deuteration is a poor evolutionary indicator of star formation in evolved stage}

\begin{figure}
\centering
\includegraphics[width=0.45\textwidth, angle=0]{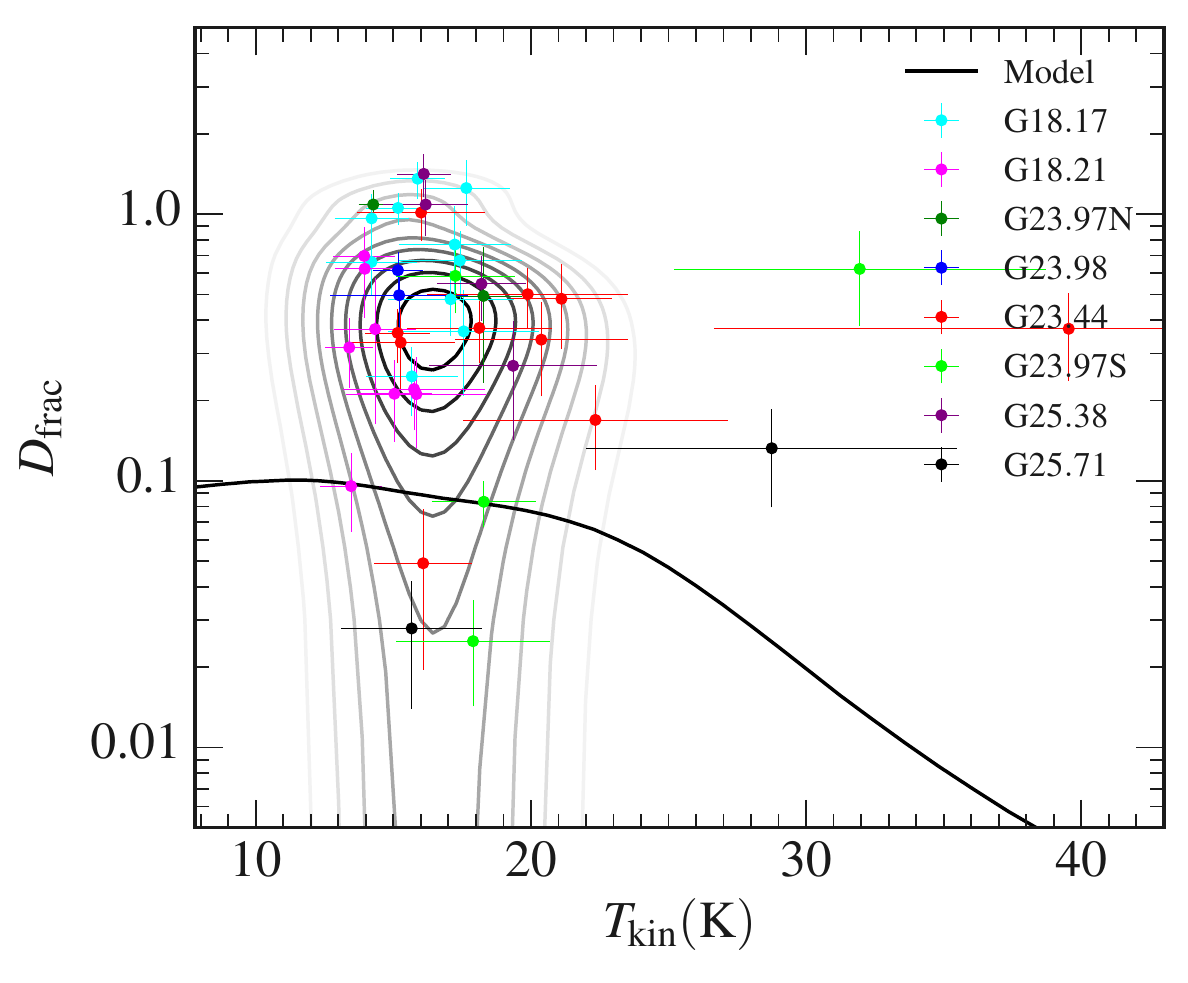}
\caption{Deuterium fractionation $D_{\rm frac}$ versus kinetic temperature $T_{\rm kin}$ for all NH$_2$D cores. The solid line is the latest gas phase model predictions from \citet{Roueff2005}. The contours show the number density distribution of the cores. The data points with error bars are derived from this work.}
\label{Fig_fractionation}
\end{figure}

Many previous works \citep[e.g.,][]{Fontani2011,Brunken2014,Ceccarelli2014} argued that the deuteration can be used as an evolutionary indicator of star formation in a wide range of evolutionary stages, for example, from high-mass starless core candidates (HMSCs) to high-mass protostellar objects (HMPOs) and ultracompact (UC) \HII regions (for a definition of these stages see e.g. \citealt{Beuther2007}). \citet{Busquet2010} found that in a high-mass star-forming region (harbouring an UC \HII region), the deuterium fractionation increases until the onset of star formation and decreases afterwards. The method is to check the changes of deuterium abundance in different evolutionary stages. Since the evolved sources often have high dust temperature ($>$\,30\,K), the growth of deuterium fractionation will be inhibited easily (see Setion\,\ref{sect_chemistry}). So, what is the nature of the deuterium emission that still can be detected even in evolved objects (e.g., HMPOs and UC \HII regions)? We think that previous observations mainly focused on using single-dish telescopes or interferometers with relatively low spatial resolution. They were not able to resolve the distributions of deuterium species from central bright sources, for example HMSCs, HMPOs, and UC\,\HII regions.

In our PdBI NH$_2$D observations, we find that the positions detected with NH$_2$D emission are often offset far away from the protostellar cores, traced by bright 3.5\,mm and 1.3\,cm continuum (see Section\,\ref{sect_offset}). \citet{Fontani2006} proposed two scenarios: in the first one, the cold gas is distributed in an external shell not yet heated up by the high-mass protostellar object, a remnant of the parental massive starless core, due to heavily thick envelope of high-mass star formation; in the second one, the cold gas is located in cold and dense cores close to the high-mass protostellar cores but not associated with them. We also argue that the NH$_2$D that we detected does not emit from the evolved objects indeed. Considering that the high-mass stars often form in clusters \citep{Tutukov1978,Kurtz2000}, the detected NH$_2$D cores may be just some cold and dense fragments of the neighbouring evolved objects. Basically, the detected NH$_2$D cores belong to prestellar or starless objects (see Section\,\ref{sect_offset}). Therefore, these deuterium production should have few correlation with the HMPOs and UC \HII regions.

In this work, we do not see much discrepancy (see Figure\,\ref{Fig_fractionation}) in deuteration fractionation between the early evolutionary stage of sources (e.g., G18.17, G18.21) and the evolved objects (e.g., G23.44, G23.97S). This further suggests that the habitat conditions where NH$_2$D remains gaseous have no direct correlation with different evolutionary stages but, rather, they depend mainly on the temperature conditions between 13.0 and 22.0\,K and the density condition of around $5.3\times10^{5}$\,cm$^{-3}$. The HMPOs and UC \HII regions often have high dust temperature of $>$\,30\,K, which is too high for significant deuterium fractionation. It is for this reason that, in principle, the observed cold and dense gas responsible for the NH$_2$D emission may be not associated with the high-mass evolved objects. Therefore, NH$_2$D is a poor evolutionary indicator of high-mass star formation in evolved stages, but a useful tracer in the starless and prestellar cores.

\section{Summary}
\label{sect summary}

At the early stages (e.g., prestellar or starless core stages) of star formation, most species tend to be frozen out onto dust grains, except deuterium molecules and ions. Using the PdBI and the VLA, we presented o-NH$_2$D 1$_{11}$-1$_{01}$ and NH$_3$ (1,\,1), (2,\,2) observations in eight massive precluster protocluster clumps including G18.17, G18.21, G23.97N, G23.98, G23.44, G23.97S, G25.38, and G25.71. We used 3D \texttt{GAUSSCLUMPS} to extract NH$_2$D cores and provided a statistical view of their deuterium chemistry.

We detected seven instances of extremely high deuterium fractionation of $1.0 \leqslant D_{\rm frac} \leqslant 1.41$ in the NH$_2$D cores. Current gas phase models have difficulty for explaining the high fractionation observed in this work. The gas-grain chemical reactions are needed to explain the production of the high deuterium fractionation. In addition, we found that the distribution between deuterium fractionation and kinetic temperature shows a number density peak at around $T_{\rm kin}=16.1$\,K, and the NH$_2$D cores are mainly located at a temperature range of 13.0 to 22.0\,K. The 3.5\,mm continuum cores have a kinetic temperature with the median width of $22.1\pm4.3$\,K, which is obviously higher than the temperature in NH$_2$D cores. This suggests that the high deuterium fractionation strongly depends on the temperature condition.

We found that the NH$_2$D emission is often not associated with either a dust continuum or NH$_3$ emission peak positions. For the protocluster clumps G23.44, G23.97S, and G25.38, the NH$_{2}$D distributions are extended and surrounding the 3.5\,mm continuum peak positions, and their minimum projected offset distances to the continuum peak nearby are 0.13, 0.12, and 0.17\,pc, respectively. We also found that large projected offsets exist between the NH$_{2}$D core and continuum peak positions, and the projected offsets are larger in the more evolved objects, for example G23.44, G23.97S, and G25.38 in protostellar core stage than those in the earlier evolutionary stages, for example G23.97N and G23.98 in prestellar core stage.

We found that the NH$_3$ and NH$_2$D are often optically thick in these clumps with a median width of $4.05\pm0.04$ and $3.22\pm0.10$, respectively. The masses of the NH$_2$D cores at a scale of $R_{\rm eff}\approx0.05$\,pc range from 5.9 to 54.0\,$\Msun$ with a median width of $13.8\pm0.6\,\Msun$. The NH$_2$D cores are mostly gravitationally bound ($\alpha_{\rm vir}<1$), are likely prestellar or starless, and can potentially form intermediate-mass or high-mass stars in future.

The derived volume density of the NH$_2$D cores is between $1.8\times10^{5}$ and $2.4\times10^{6}$\,cm$^{-3}$ with a median width of $(5.3\pm1.4)\times10^{5}$\,cm$^{-3}$, while that of continuum cores ranges from $1.5\times10^{5}$ to $4.6\times10^{6}$\,cm$^{-3}$ with a median width of $(1.4\pm0.1)\times10^{6}$\,cm$^{-3}$. Therefore, the NH$_2$D distributions are in a relatively less dense condition than the continuum cores.

The detected NH$_2$D line widths are very narrow with a median width of $0.98\pm0.02\,\kms$, where the thermal and non-thermal velocity dispersion have a median width of $0.09\pm0.01$ and $0.41\pm0.01\,\kms$ in the NH$_2$D cores, respectively. Therefore, non-thermal motions still contribute significantly to the line width of NH$_2$D.

We found that the detected NH$_2$D cores belong to prestellar or starless object stages. The association between the NH$_2$D cores and the evolved objects is not significant. The remaining of NH$_2$D mainly depends on the suitable temperature of around 13.0 to 22.0\,K and the density of $\sim$5.3$\times10^{5}$\,cm$^{-3}$. Therefore, we suggest that the NH$_2$D is a useful tracer in prestellar or starless cores, but cannot be used as a precise indicator in other evolved stages.

Using NH$_3$ (1,\,1) as a dynamical tracer, we found evidence of very complicated dynamical movement in all the eight clumps, either outflow and rotation or convergent flow and colliding each other, not only in earlier stage of clumps but also in evolved objects. The velocity signatures that indicate to rotating toroids are also identified. The sample partly present obvious dynamical characteristic of rotating toroids, suggesting that accretion has started and continues to increase gradually from the prestellar core stage (e.g., G23.97N and G23.98) to the protostellar stage (e.g., G23.44 and G23.97S). Additionally, the central continuum cores in G18.17, G18.21, G23.97N, and G23.98 have relatively quiescent dynamical movements, but their large-scale gas distributions beyond the core size show a large velocity gradient.

\begin{acknowledgements}
We thank the anonymous referees for constructive comments that improved the manuscript. This work is supported by the National Natural Science Foundation of China No.\,11703040, and the National Key Basic Research Program of China (973 Program) No.\,2015CB857101. C.-P. Zhang acknowledges support by the NAOC Nebula Talents Program and the China Scholarship Council in Germany as a postdoctoral researcher (No.\,201704910137).
\end{acknowledgements}
\bibliographystyle{aa}
\bibliography{references}

\appendix
\section{Tables and Figures}

\begin{table*}
\caption{Coordinates of infrared sources, \HII regions, and masers.}
\label{tab_coordinates} \centering \footnotesize
\begin{tabular}{lcccc}
\hline \hline
Source &        $\alpha$ (J2000) & $\delta$ (J2000)     & $l$  & $b$ \\
&       h~~m~~s~~       &       $^\circ~~'~~''~~$       & $^\circ$  & $^\circ$ \\
\hline
\multicolumn{2}{l}{Infrared sources}  \\
G18.17   &      18 25 07.60     & $-$13 14 31.73    &   18.175196   &  $-$0.298597  \\
G18.21   &      18 25 21.56     & $-$13 13 37.80    &   18.214882   &  $-$0.341650  \\
G23.97N  &      18 34 28.81     & $-$07 54 30.77    &   23.966176   &  $+$0.139036  \\
G23.98   &      18 34 27.90     & $-$07 53 29.19    &   23.979640   &  $+$0.150223  \\
G23.44-l &      18 34 39.24     & $-$08 31 39.19    &   23.436575   &  $-$0.184295   \\
G23.44-u &      18 34 39.16     & $-$08 31 23.88    &   23.440278   &  $-$0.182195   \\
G23.97S  &      18 35 22.11     & $-$08 01 24.47    &   23.965439   &  $-$0.109116   \\
G25.38-l &      18 38 08.09     & $-$06 46 53.73    &   25.382845   &  $-$0.147931  \\
G25.38-u &      18 38 08.04     & $-$06 46 30.62    &   25.387294   &  $-$0.145579  \\
G25.71-l &      18 38 03.15     & $-$06 24 15.22    &   25.709478   &  $+$0.043806  \\
G25.71-u &      18 38 02.77     & $-$06 23 46.82    &   25.715834   &  $+$0.048689  \\
\hline
\multicolumn{2}{l}{\HII regions}  \\
G23.44-l &      18 34 39.20     & $-$08 31 39.91    &   23.436396   &  $-$0.184384   \\
G23.97S  &      18 35 22.28     & $-$08 01 22.76    &   23.966253   &  $-$0.109648   \\
G25.71-l &      18 38 03.12     & $-$06 24 15.27    &   25.709481   &  $+$0.043770  \\
G25.71-u &      18 38 02.78     & $-$06 23 47.17    &   25.715764   &  $+$0.048615  \\
\hline
\multicolumn{2}{l}{Methanol masers}  \\
G23.44-l &      18 34 39.27     & $-$08 31 39.30    &   23.436606   &  $-$0.184422   \\
G23.44-u &      18 34 39.18     & $-$08 31 25.40    &   23.439862   &  $-$0.182314   \\
G23.97S  &      18 35 22.21     & $-$08 01 22.50    &   23.966121   &  $-$0.109239   \\
G25.71-l &      18 38 03.15     & $-$06 24 14.90    &   25.709557   &  $+$0.043843  \\
\hline
\end{tabular}
\end{table*}

\begin{sidewaystable*}
\caption{Parameters of the identified NH$_2$D cores: position, velocity, line width, brightness temperature, optical depth and excitation temperature.}
\label{tab_nh2d_para1} \centering \scriptsize
\begin{tabular}{ccccccccccccc}
\hline \hline
Sources &  Offsets &    $v_{\rm NH_2D}$         &  $\Delta v_{\rm NH_2D}$       &       $\Delta
v_{\rm NH_3(1,\,1)}$    & $\Delta v_{\rm NH_3 (2,2)}$   &  $T_{\rm MB_{\rm NH_2D}}$  & $T_{\rm MB_{\rm NH_3(1,\,1)}}$
& $T_{\rm MB_{\rm NH_3(2,2)}}$ & $\tau_{\rm NH_2D}$     & $\tau_{\rm NH_3 (1,\,1)}$ & $T_{\rm ex_{\rm NH_2D}}$    & $T_{\rm ex_{\rm NH_3 (1,\,1)}}$  \\
No. &   arcsec  &       km\,s$^{-1}$    &       km\,s$^{-1}$    &       km\,s$^{-1}$    & km\,s$^{-1}$ & K & K & K & & & K & K \\
\hline
G18.17 & \multicolumn{3}{l}{Center: R.A.=18 25 07.534, DEC.=$-$13 14 32.75}\\
1 &$ ( 4.40 , 3.09 ) $&$ 49.98 \pm 0.01 $&$ 0.97 \pm 0.02 $&$ 1.15 \pm 0.04 $&$ 1.20 \pm 0.14 $&$ 3.49 \pm 0.08 $&$ 11.20 \pm 0.76 $&$ 8.05 \pm 0.26 $&$ 7.19 \pm 0.55 $&$ 5.86 \pm 0.70 $&$ 6.35 \pm 0.33 $&$ 12.73 \pm 1.50 $ \\
2 &$ ( 1.10 , 4.44 ) $&$ 49.03 \pm 0.02 $&$ 0.83 \pm 0.04 $&$ 1.42 \pm 0.09 $&$ 1.07 \pm 0.09 $&$ 2.94 \pm 0.21 $&$ 8.38 \pm 1.05 $&$ 6.14 \pm 0.46 $&$ 2.64 \pm 0.55 $&$ 2.89 \pm 0.45 $&$ 6.90 \pm 0.94 $&$ 11.00 \pm 1.52 $ \\
3 &$ ( -4.31 , 9.40 ) $&$ 49.61 \pm 0.02 $&$ 0.74 \pm 0.06 $&$ 1.03 \pm 0.05 $&$ 1.11 \pm 0.21 $&$ 2.47 \pm 0.10 $&$ 6.74 \pm 0.94 $&$ 3.30 \pm 0.01 $&$ 0.96 \pm 0.18 $&$ 1.67 \pm 0.75 $&$ 9.42 \pm 1.61 $&$ 9.79 \pm 3.42 $ \\
4 &$ ( -3.64 , -1.86 ) $&$ 49.41 \pm 0.03 $&$ 1.17 \pm 0.06 $&$ 1.28 \pm 0.07 $&$ 1.03 \pm 0.11 $&$ 2.10 \pm 0.10 $&$ 9.23 \pm 0.26 $&$ 6.18 \pm 0.18 $&$ 4.99 \pm 0.51 $&$ 3.25 \pm 0.44 $&$ 4.97 \pm 0.31 $&$ 11.77 \pm 1.46 $ \\
5 &$ ( 1.24 , 8.80 ) $&$ 50.37 \pm 0.05 $&$ 1.06 \pm 0.11 $&$ 1.65 \pm 0.05 $&$ 1.51 \pm 0.32 $&$ 1.97 \pm 0.24 $&$ 8.06 \pm 1.32 $&$ 4.58 \pm 0.65 $&$ 3.91 \pm 0.64 $&$ 4.15 \pm 0.43 $&$ 4.77 \pm 0.44 $&$ 9.92 \pm 0.89 $ \\
6 &$ ( -1.27 , -3.77 ) $&$ 49.36 \pm 0.07 $&$ 1.68 \pm 0.13 $&$ 1.45 \pm 0.06 $&$ 1.29 \pm 0.14 $&$ 1.89 \pm 0.13 $&$ 8.03 \pm 0.53 $&$ 5.34 \pm 0.25 $&$ 3.16 \pm 0.74 $&$ 2.56 \pm 0.33 $&$ 4.87 \pm 0.61 $&$ 11.03 \pm 1.28 $ \\
7 &$ ( 6.28 , 11.95 ) $&$ 49.48 \pm 0.04 $&$ 0.90 \pm 0.10 $&$ 1.03 \pm 0.02 $&$ 1.03 \pm 0.20 $&$ 1.61 \pm 0.19 $&$ 8.41 \pm 0.74 $&$ 4.16 \pm 0.50 $&$ 0.70 \pm 0.16 $&$ 2.44 \pm 0.34 $&$ 7.90 \pm 1.26 $&$ 11.45 \pm 1.35 $ \\
8 &$ ( 1.24 , -3.80 ) $&$ 49.30 \pm 0.04 $&$ 0.88 \pm 0.14 $&$ 1.49 \pm 0.14 $&$ 1.15 \pm 0.15 $&$ 1.64 \pm 0.16 $&$ 6.61 \pm 0.19 $&$ 4.03 \pm 0.47 $&$ 2.45 \pm 0.59 $&$ 1.67 \pm 0.53 $&$ 4.77 \pm 0.67 $&$ 10.74 \pm 2.80 $ \\
9 &$ ( -2.47 , 4.41 ) $&$ 48.99 \pm 0.03 $&$ 1.00 \pm 0.07 $&$ 1.03 \pm 0.08 $&$ 1.03 \pm 0.26 $&$ 1.83 \pm 0.26 $&$ 7.03 \pm 0.40 $&$ 3.95 \pm 0.11 $&$ 1.64 \pm 0.35 $&$ 2.12 \pm 0.38 $&$ 6.08 \pm 0.85 $&$ 10.77 \pm 1.58 $ \\
10 &$ ( 0.64 , 11.32 ) $&$ 49.24 \pm 0.11 $&$ 1.26 \pm 0.21 $&$ 1.70 \pm 0.07 $&$ 1.61 \pm 0.19 $&$ 1.83 \pm 0.20 $&$ 7.59 \pm 0.93 $&$ 3.59 \pm 0.59 $&$ 3.72 \pm 0.23 $&$ 3.15 \pm 0.45 $&$ 4.41 \pm 0.33 $&$ 10.00 \pm 1.23 $ \\
G18.21 & \multicolumn{3}{l}{Center: R.A.=18 25 21.558, DEC.=$-$13 13 39.56}\\
1 &$ ( -1.23 , 3.70 ) $&$ 46.63 \pm 0.04 $&$ 1.01 \pm 0.08 $&$ 1.26 \pm 0.11 $&$ 1.69 \pm 0.15 $&$ 2.01 \pm 0.07 $&$ 9.86 \pm 1.15 $&$ 7.62 \pm 0.90 $&$ 2.12 \pm 0.66 $&$ 7.84 \pm 1.44 $&$ 5.67 \pm 0.96 $&$ 11.86 \pm 2.14 $ \\
2 &$ ( 0.01 , 1.90 ) $&$ 46.62 \pm 0.05 $&$ 1.06 \pm 0.12 $&$ 1.50 \pm 0.07 $&$ 2.06 \pm 0.15 $&$ 2.02 \pm 0.06 $&$ 10.20 \pm 0.94 $&$ 7.08 \pm 0.51 $&$ 2.17 \pm 0.64 $&$ 7.48 \pm 1.04 $&$ 5.82 \pm 0.97 $&$ 11.86 \pm 1.62 $ \\
3 &$ ( 3.78 , -15.76 ) $&$ 46.68 \pm 0.03 $&$ 0.79 \pm 0.08 $&$ 1.03 \pm 0.20 $&$ 1.03 \pm 0.04 $&$ 1.56 \pm 0.10 $&$ 6.53 \pm 0.80 $&$ 2.65 \pm 0.32 $&$ 1.10 \pm 0.54 $&$ 2.48 \pm 0.69 $&$ 6.55 \pm 2.01 $&$ 10.02 \pm 2.24 $ \\
4 &$ ( 5.08 , 5.02 ) $&$ 47.00 \pm 0.27 $&$ 1.32 \pm 0.11 $&$ 1.56 \pm 0.07 $&$ 1.08 \pm 0.14 $&$ 1.48 \pm 0.11 $&$ 7.43 \pm 0.58 $&$ 5.89 \pm 0.48 $&$ 0.70 \pm 0.20 $&$ 7.32 \pm 1.03 $&$ 6.95 \pm 1.20 $&$ 9.21 \pm 1.16 $ \\
5 &$ ( -3.14 , 7.55 ) $&$ 46.84 \pm 0.03 $&$ 0.53 \pm 0.06 $&$ 1.03 \pm 0.22 $&$ 1.03 \pm 0.06 $&$ 1.60 \pm 0.10 $&$ 6.49 \pm 0.86 $&$ 5.48 \pm 0.14 $&$ 3.25 \pm 0.62 $&$ 7.42 \pm 0.54 $&$ 4.51 \pm 0.51 $&$ 9.73 \pm 0.62 $ \\
6 &$ ( -1.21 , -3.75 ) $&$ 46.14 \pm 0.10 $&$ 1.88 \pm 0.22 $&$ 1.18 \pm 0.06 $&$ 1.62 \pm 0.27 $&$ 1.02 \pm 0.09 $&$ 7.32 \pm 0.34 $&$ 4.46 \pm 0.39 $&$ 2.65 \pm 0.78 $&$ 7.58 \pm 1.10 $&$ 4.12 \pm 0.49 $&$ 9.28 \pm 1.21 $ \\
7 &$ ( 5.04 , -1.29 ) $&$ 46.85 \pm 0.05 $&$ 1.15 \pm 0.11 $&$ 1.45 \pm 0.05 $&$ 1.83 \pm 0.21 $&$ 1.36 \pm 0.07 $&$ 6.94 \pm 0.47 $&$ 4.44 \pm 0.49 $&$ 3.00 \pm 0.78 $&$ 8.86 \pm 0.42 $&$ 4.36 \pm 0.50 $&$ 8.77 \pm 4.90 $ \\
8 &$ ( -3.14 , -0.64 ) $&$ 46.73 \pm 0.04 $&$ 1.28 \pm 0.08 $&$ 1.03 \pm 0.01 $&$ 1.10 \pm 0.18 $&$ 1.63 \pm 0.11 $&$ 7.86 \pm 0.46 $&$ 5.27 \pm 0.46 $&$ 3.50 \pm 0.85 $&$ 7.01 \pm 1.10 $&$ 4.71 \pm 0.54 $&$ 9.76 \pm 1.32 $ \\
G23.97N & \multicolumn{3}{l}{Center: R.A.=18 34 28.833, DEC.=$-$07 54 31.76}\\
1 &$ ( 12.70 , -3.11 ) $&$ 77.32 \pm 0.02 $&$ 0.87 \pm 0.03 $&$ 1.03 \pm 0.00 $&$ 1.04 \pm 0.03 $&$ 2.62 \pm 0.19 $&$ 10.00 \pm 0.38 $&$ 7.31 \pm 0.13 $&$ 8.44 \pm 0.93 $&$ 7.17 \pm 0.66 $&$ 5.13 \pm 0.31 $&$ 11.54 \pm 0.97 $ \\
2 &$ ( 2.63 , -1.29 ) $&$ 77.72 \pm 0.03 $&$ 0.94 \pm 0.07 $&$ 1.03 \pm 0.01 $&$ 1.15 \pm 0.08 $&$ 1.93 \pm 0.18 $&$ 9.56 \pm 0.40 $&$ 5.89 \pm 0.31 $&$ 1.71 \pm 0.87 $&$ 2.68 \pm 0.39 $&$ 6.16 \pm 1.81 $&$ 12.19 \pm 1.54 $ \\
G23.98 & \multicolumn{3}{l}{Center: R.A.=18 34 27.823, DEC.=$-$07 53 28.76}\\
1 &$ ( 4.46 , -0.62 ) $&$ 82.59 \pm 0.04 $&$ 0.86 \pm 0.09 $&$ 1.03 \pm 0.01 $&$ 1.12 \pm 0.15 $&$ 1.09 \pm 0.04 $&$ 5.70 \pm 0.14 $&$ 3.63 \pm 0.17 $&$ 3.53 \pm 0.38 $&$ 4.88 \pm 0.47 $&$ 4.73 \pm 0.43 $&$ 8.08 \pm 0.61 $ \\
2 &$ ( -0.66 , -9.44 ) $&$ 81.65 \pm 0.04 $&$ 0.69 \pm 0.07 $&$ 1.03 \pm 0.01 $&$ 1.03 \pm 0.48 $&$ 0.91 \pm 0.03 $&$ 4.57 \pm 0.12 $&$ 3.10 \pm 0.03 $&$ 3.60 \pm 0.68 $&$ 4.96 \pm 0.79 $&$ 4.36 \pm 0.46 $&$ 7.05 \pm 0.83 $ \\
G23.44 & \multicolumn{3}{l}{Center: R.A.=18 34 39.253, DEC.=$-$08 31 36.23}\\
1 &$ ( 2.58 , 1.92 ) $&$ 99.94 \pm 0.07 $&$ 2.00 \pm 0.14 $&$ 3.13 \pm 0.04 $&$ 3.40 \pm 0.11 $&$ 1.84 \pm 0.13 $&$ 10.20 \pm 0.94 $&$ 8.00 \pm 0.74 $&$ 3.19 \pm 0.70 $&$ 4.66 \pm 0.19 $&$ 6.12 \pm 0.84 $&$ 11.87 \pm 0.46 $ \\
2 &$ ( -9.87 , 0.61 ) $&$ 101.44 \pm 0.05 $&$ 1.06 \pm 0.08 $&$ 1.84 \pm 0.12 $&$ 2.30 \pm 0.29 $&$ 1.18 \pm 0.10 $&$ 6.86 \pm 0.65 $&$ 5.32 \pm 0.39 $&$ 4.75 \pm 0.88 $&$ 4.32 \pm 0.62 $&$ 4.61 \pm 0.46 $&$ 9.36 \pm 1.19 $ \\
3 &$ ( 5.42 , 3.52 ) $&$ 100.02 \pm 0.03 $&$ 1.17 \pm 0.06 $&$ 2.53 \pm 0.04 $&$ 2.71 \pm 0.12 $&$ 1.98 \pm 0.05 $&$ 8.00 \pm 0.95 $&$ 4.82 \pm 0.32 $&$ 3.29 \pm 0.67 $&$ 4.33 \pm 0.23 $&$ 6.33 \pm 0.81 $&$ 10.12 \pm 0.49 $ \\
4 &$ ( -7.69 , -0.64 ) $&$ 101.42 \pm 0.04 $&$ 0.99 \pm 0.09 $&$ 1.86 \pm 0.10 $&$ 2.70 \pm 0.15 $&$ 1.35 \pm 0.08 $&$ 8.36 \pm 0.65 $&$ 6.62 \pm 0.61 $&$ 1.90 \pm 0.57 $&$ 4.34 \pm 0.61 $&$ 6.19 \pm 1.20 $&$ 10.65 \pm 1.38 $ \\
5 &$ ( -6.74 , -20.15 ) $&$ 99.92 \pm 0.03 $&$ 0.85 \pm 0.05 $&$ 1.03 \pm 0.08 $&$ 1.41 \pm 0.41 $&$ 0.87 \pm 0.12 $&$ 7.65 \pm 0.72 $&$ 4.83 \pm 0.33 $&$ 6.64 \pm 0.84 $&$ 5.19 \pm 0.76 $&$ 4.52 \pm 0.34 $&$ 9.36 \pm 1.15 $ \\
6 &$ ( 4.48 , -0.96 ) $&$ 100.23 \pm 0.08 $&$ 1.51 \pm 0.14 $&$ 2.82 \pm 0.06 $&$ 3.30 \pm 0.22 $&$ 1.19 \pm 0.08 $&$ 8.93 \pm 0.32 $&$ 6.57 \pm 0.30 $&$ 3.61 \pm 1.20 $&$ 3.03 \pm 0.20 $&$ 4.84 \pm 0.79 $&$ 11.40 \pm 0.67 $ \\
7 &$ ( 3.83 , 5.12 ) $&$ 100.00 \pm 0.05 $&$ 0.75 \pm 0.16 $&$ 2.85 \pm 0.06 $&$ 2.62 \pm 0.10 $&$ 1.52 \pm 0.07 $&$ 8.27 \pm 0.96 $&$ 5.21 \pm 0.99 $&$ 4.81 \pm 0.99 $&$ 3.86 \pm 0.28 $&$ 4.33 \pm 0.53 $&$ 10.30 \pm 0.68 $ \\
8 &$ ( -0.31 , -8.64 ) $&$ 100.72 \pm 0.03 $&$ 0.54 \pm 0.06 $&$ 2.88 \pm 0.28 $&$ 7.20 \pm 0.50 $&$ 1.10 \pm 0.02 $&$ 4.87 \pm 0.78 $&$ 2.34 \pm 0.12 $&$ 3.83 \pm 1.10 $&$ 0.96 \pm 0.04 $&$ 4.73 \pm 0.71 $&$ 11.29 \pm 0.72 $ \\
9 &$ ( -7.05 , -15.36 ) $&$ 99.49 \pm 0.11 $&$ 1.08 \pm 0.24 $&$ 1.98 \pm 0.06 $&$ 2.22 \pm 0.14 $&$ 0.57 \pm 0.08 $&$ 7.64 \pm 1.02 $&$ 4.21 \pm 0.52 $&$ 0.30 \pm 0.16 $&$ 3.19 \pm 0.33 $&$ 8.49 \pm 3.33 $&$ 9.68 \pm 0.83 $ \\
10 &$ ( -7.03 , -6.09 ) $&$ 101.31 \pm 0.06 $&$ 0.83 \pm 0.18 $&$ 1.87 \pm 0.04 $&$ 2.37 \pm 0.23 $&$ 1.06 \pm 0.04 $&$ 8.84 \pm 0.53 $&$ 7.67 \pm 0.34 $&$ 5.67 \pm 1.65 $&$ 5.78 \pm 0.36 $&$ 4.14 \pm 0.53 $&$ 10.73 \pm 0.62 $ \\
G23.97S & \multicolumn{3}{l}{Center: R.A.=18 35 22.160, DEC.=$-$08 01 26.53}\\
1 &$ ( 4.11 , -10.50 ) $&$ 73.39 \pm 0.07 $&$ 1.20 \pm 0.13 $&$ 1.74 \pm 0.07 $&$ 1.33 \pm 0.11 $&$ 1.07 \pm 0.06 $&$ 4.95 \pm 0.14 $&$ 4.75 \pm 0.16 $&$ 0.30 \pm 0.12 $&$ 7.16 \pm 0.35 $&$ 9.98 \pm 2.91 $&$ 8.26 \pm 0.28 $ \\
2 &$ ( 1.19 , -5.95 ) $&$ 74.02 \pm 0.27 $&$ 1.07 \pm 0.11 $&$ 1.82 \pm 0.10 $&$ 3.60 \pm 0.17 $&$ 0.77 \pm 0.12 $&$ 6.38 \pm 0.34 $&$ 5.14 \pm 0.29 $&$ 1.51 \pm 0.21 $&$ 9.09 \pm 0.50 $&$ 5.12 \pm 0.37 $&$ 9.59 \pm 0.71 $ \\
3 &$ ( 3.23 , -15.74 ) $&$ 72.86 \pm 0.04 $&$ 0.77 \pm 0.09 $&$ 1.44 \pm 0.12 $&$ 2.15 \pm 0.33 $&$ 0.85 \pm 0.07 $&$ 5.87 \pm 0.26 $&$ 3.46 \pm 0.11 $&$ 5.03 \pm 0.90 $&$ 4.11 \pm 0.98 $&$ 4.13 \pm 0.39 $&$ 8.25 \pm 1.59 $ \\
4 &$ ( 1.10 , 9.52 ) $&$ 69.74 \pm 0.27 $&$ 0.83 \pm 0.12 $&$ 3.08 \pm 0.21 $&$ 3.81 \pm 0.25 $&$ 0.71 \pm 0.04 $&$ 4.73 \pm 0.27 $&$ 3.54 \pm 0.29 $&$ 4.62 \pm 0.31 $&$ 1.14 \pm 0.45 $&$ 3.86 \pm 0.09 $&$ 10.82 \pm 3.42 $ \\
G25.38 & \multicolumn{3}{l}{Center: R.A.=18 38 08.108, DEC.=$-$06 46 54.93}\\
1 &$ ( 0.13 , -4.70 ) $&$ 96.39 \pm 0.02 $&$ 0.71 \pm 0.06 $&$ 1.37 \pm 0.04 $&$ 1.55 \pm 0.13 $&$ 2.83 \pm 0.21 $&$ 9.35 \pm 0.56 $&$ 6.48 \pm 0.34 $&$ 4.46 \pm 1.08 $&$ 3.59 \pm 0.31 $&$ 5.72 \pm 0.83 $&$ 11.40 \pm 0.91 $ \\
2 &$ ( -7.02 , -12.09 ) $&$ 96.32 \pm 0.03 $&$ 0.72 \pm 0.05 $&$ 1.20 \pm 0.06 $&$ 1.18 \pm 0.22 $&$ 2.02 \pm 0.07 $&$ 7.08 \pm 0.36 $&$ 4.15 \pm 0.24 $&$ 5.67 \pm 1.06 $&$ 3.00 \pm 0.41 $&$ 4.66 \pm 0.47 $&$ 9.88 \pm 1.13 $ \\
3 &$ ( -7.75 , 2.91 ) $&$ 97.36 \pm 0.11 $&$ 1.95 \pm 0.20 $&$ 2.03 \pm 0.07 $&$ 2.24 \pm 0.14 $&$ 1.21 \pm 0.13 $&$ 8.02 \pm 0.69 $&$ 6.08 \pm 0.57 $&$ 1.29 \pm 0.56 $&$ 3.73 \pm 0.46 $&$ 5.26 \pm 1.26 $&$ 9.82 \pm 1.06 $ \\
4 &$ ( -3.22 , -15.10 ) $&$ 95.94 \pm 0.05 $&$ 0.79 \pm 0.07 $&$ 1.06 \pm 0.08 $&$ 1.03 \pm 0.07 $&$ 1.03 \pm 0.15 $&$ 6.29 \pm 0.28 $&$ 3.44 \pm 0.20 $&$ 5.10 \pm 0.66 $&$ 2.55 \pm 0.22 $&$ 4.17 \pm 0.23 $&$ 9.32 \pm 1.13 $ \\
G25.71 & \multicolumn{3}{l}{Center: R.A.=18 38 03.184, DEC.=$-$06 24 14.30}\\
1 &$ ( 0.50 , 0.02 ) $&$ 97.54 \pm 0.23 $&$ 1.82 \pm 0.23 $&$ 2.19 \pm 0.04 $&$ 3.03 \pm 0.15 $&$ 0.35 \pm 0.06 $&$ 11.00 \pm 0.61 $&$ 9.69 \pm 0.47 $&$ 1.04 \pm 0.36 $&$ 3.98 \pm 0.22 $&$ 4.36 \pm 0.66 $&$ 12.96 \pm 0.68 $ \\
2 &$ ( -6.98 , 18.32 ) $&$ 98.86 \pm 0.04 $&$ 0.65 \pm 0.09 $&$ 1.46 \pm 0.04 $&$ 1.35 \pm 0.16 $&$ 1.33 \pm 0.01 $&$ 8.35 \pm 1.38 $&$ 5.15 \pm 1.08 $&$ 0.22 \pm 0.10 $&$ 3.43 \pm 0.30 $&$ 7.60 \pm 2.65 $&$ 10.75 \pm 0.83 $ \\
\hline
\end{tabular}
\begin{flushleft}
\textbf{Note:} \\
Offsets: Derived from NH$_2$D core extraction using the \texttt{GAUSSCLUMPS}.\\
Other parameters: Derived by HfS fittings of NH$_2$D and NH$_3$ using the GILDAS. \\
$\tau_{\rm NH_2D}$ and $\tau_{\rm NH_3 (1,\,1)}$: Main group opacity of HfS components.\\
\end{flushleft}
\end{sidewaystable*}

\begin{table*}
\caption{Parameters of the identified NH$_2$D cores: size, temperature, mass, density, and deuterium fractionation.}
\label{tab_nh2d_para2} \centering \scriptsize
\begin{tabular}{cccccccccccc}
\hline \hline
Sources & FWHM & $R_{\rm eff}$  & $T_{\rm kin}$ & $M_{\rm H_2}$ & $M_{\rm vir}$ & $\alpha_{\rm vir}$ & $N_{\rm NH_2D}$ & $N_{\rm NH_3}$ & $N_{\rm H_2}$ & $n_{\rm H_2}$ & $D_{\rm frac}$ \\
No. & arcsec & pc & K & $\Msun$ &  $\Msun$ & pc & $\rm 10^{14} cm^{-2}$ & $\rm 10^{14} cm^{-2}$ & $\rm
10^{22} cm^{-2}$  & $\rm 10^{5} cm^{-3}$ & [NH$_2$D]/[NH$_3$] \\
\hline
G18.17 \\
1 &$5.09 $&$ 0.055 $&$ 15.2 \pm 1.1 $&$ 14.5 \pm 1.3 $&$ 6.3 \pm 0.3 $&$ 0.43 \pm 0.04 $&$ 30.0 \pm 2.4 $&$ 28.5 \pm 3.0 $&$ 11.9 \pm 1.0 $&$ 6.3 \pm 0.5 $&$ 1.05 \pm 0.14 $ \\
2 &$5.09 $&$ 0.055 $&$ 17.1 \pm 2.3 $&$ 7.3 \pm 1.1 $&$ 4.4 \pm 0.5 $&$ 0.61 \pm 0.11 $&$ 9.4 \pm 2.0 $&$ 19.6 \pm 3.3 $&$ 5.9 \pm 0.9 $&$ 3.1 \pm 0.5 $&$ 0.48 \pm 0.13 $ \\
3 &$4.11 $&$ 0.045 $&$ 17.6 \pm 2.7 $&$ 6.5 \pm 1.1 $&$ 2.8 \pm 0.5 $&$ 0.43 \pm 0.10 $&$ 3.1 \pm 0.6 $&$ 8.4 \pm 3.1 $&$ 8.1 \pm 1.4 $&$ 5.3 \pm 0.9 $&$ 0.36 \pm 0.15 $ \\
4 &$5.90 $&$ 0.064 $&$ 15.9 \pm 1.0 $&$ 15.7 \pm 1.1 $&$ 10.6 \pm 1.0 $&$ 0.68 \pm 0.08 $&$ 25.0 \pm 2.8 $&$ 18.4 \pm 2.1 $&$ 9.6 \pm 0.7 $&$ 4.3 \pm 0.3 $&$ 1.36 \pm 0.22 $ \\
5 &$4.53 $&$ 0.049 $&$ 14.2 \pm 1.7 $&$ 8.2 \pm 1.1 $&$ 6.7 \pm 1.5 $&$ 0.81 \pm 0.21 $&$ 17.9 \pm 3.5 $&$ 27.1 \pm 3.8 $&$ 8.5 \pm 1.1 $&$ 5.0 \pm 0.7 $&$ 0.66 \pm 0.16 $ \\
6 &$3.34 $&$ 0.036 $&$ 17.7 \pm 1.6 $&$ 6.4 \pm 0.6 $&$ 12.7 \pm 2.0 $&$ 1.97 \pm 0.37 $&$ 22.8 \pm 5.7 $&$ 18.2 \pm 2.3 $&$ 12.2 \pm 1.2 $&$ 9.8 \pm 1.0 $&$ 1.25 \pm 0.35 $ \\
7 &$4.70 $&$ 0.051 $&$ 15.7 \pm 1.7 $&$ 7.4 \pm 0.9 $&$ 4.9 \pm 1.1 $&$ 0.67 \pm 0.17 $&$ 2.7 \pm 0.7 $&$ 11.0 \pm 1.6 $&$ 7.1 \pm 0.9 $&$ 4.0 \pm 0.5 $&$ 0.25 \pm 0.07 $ \\
8 &$5.19 $&$ 0.056 $&$ 17.2 \pm 2.0 $&$ 11.4 \pm 1.5 $&$ 5.1 \pm 1.7 $&$ 0.45 \pm 0.16 $&$ 9.2 \pm 2.7 $&$ 12.0 \pm 3.3 $&$ 8.9 \pm 1.2 $&$ 4.6 \pm 0.6 $&$ 0.77 \pm 0.31 $ \\
9 &$3.63 $&$ 0.039 $&$ 17.4 \pm 2.3 $&$ 6.5 \pm 1.0 $&$ 4.8 \pm 0.7 $&$ 0.73 \pm 0.15 $&$ 7.1 \pm 1.6 $&$ 10.6 \pm 1.9 $&$ 10.5 \pm 1.5 $&$ 7.8 \pm 1.1 $&$ 0.67 \pm 0.19 $ \\
10 &$3.96 $&$ 0.043 $&$ 14.2 \pm 1.3 $&$ 8.2 \pm 0.9 $&$ 8.4 \pm 2.9 $&$ 1.03 \pm 0.37 $&$ 20.4 \pm 3.6 $&$ 21.2 \pm 3.1 $&$ 11.1 \pm 1.2 $&$ 7.5 \pm 0.8 $&$ 0.96 \pm 0.22 $ \\
G18.21 \\
1 &$5.89 $&$ 0.062 $&$ 15.8 \pm 2.5 $&$ 13.0 \pm 2.4 $&$ 7.6 \pm 1.3 $&$ 0.59 \pm 0.14 $&$ 9.2 \pm 2.9 $&$ 43.6 \pm 8.4 $&$ 8.5 \pm 1.6 $&$ 4.0 \pm 0.7 $&$ 0.21 \pm 0.08 $ \\
2 &$5.79 $&$ 0.061 $&$ 15.0 \pm 1.4 $&$ 14.6 \pm 1.5 $&$ 8.3 \pm 2.0 $&$ 0.57 \pm 0.14 $&$ 9.9 \pm 3.1 $&$ 46.9 \pm 5.9 $&$ 9.9 \pm 1.0 $&$ 4.8 \pm 0.5 $&$ 0.21 \pm 0.07 $ \\
3 &$4.71 $&$ 0.049 $&$ 14.3 \pm 1.5 $&$ 6.6 \pm 0.8 $&$ 3.6 \pm 0.7 $&$ 0.55 \pm 0.13 $&$ 3.8 \pm 1.9 $&$ 10.2 \pm 2.7 $&$ 6.8 \pm 0.8 $&$ 4.0 \pm 0.5 $&$ 0.37 \pm 0.21 $ \\
4 &$4.74 $&$ 0.050 $&$ 13.5 \pm 1.1 $&$ 7.1 \pm 0.7 $&$ 10.7 \pm 1.8 $&$ 1.51 \pm 0.29 $&$ 4.1 \pm 1.2 $&$ 42.8 \pm 5.8 $&$ 7.2 \pm 0.7 $&$ 4.2 \pm 0.4 $&$ 0.10 \pm 0.03 $ \\
5 &$4.98 $&$ 0.052 $&$ 15.8 \pm 2.6 $&$ 5.9 \pm 1.1 $&$ 1.6 \pm 0.4 $&$ 0.26 \pm 0.08 $&$ 7.4 \pm 1.6 $&$ 33.5 \pm 6.8 $&$ 5.4 \pm 1.0 $&$ 3.0 \pm 0.6 $&$ 0.22 \pm 0.07 $ \\
6 &$5.14 $&$ 0.054 $&$ 14.0 \pm 1.1 $&$ 6.8 \pm 0.6 $&$ 23.7 \pm 5.7 $&$ 3.48 \pm 0.89 $&$ 21.7 \pm 6.9 $&$ 34.9 \pm 4.5 $&$ 5.9 \pm 0.5 $&$ 3.2 \pm 0.3 $&$ 0.62 \pm 0.21 $ \\
7 &$6.17 $&$ 0.065 $&$ 13.4 \pm 0.9 $&$ 8.0 \pm 0.6 $&$ 10.5 \pm 2.1 $&$ 1.31 \pm 0.28 $&$ 15.2 \pm 4.2 $&$ 48.0 \pm 3.7 $&$ 4.8 \pm 0.4 $&$ 2.2 \pm 0.2 $&$ 0.32 \pm 0.09 $ \\
8 &$4.32 $&$ 0.045 $&$ 13.9 \pm 1.1 $&$ 6.8 \pm 0.6 $&$ 9.2 \pm 1.2 $&$ 1.34 \pm 0.22 $&$ 19.5 \pm 4.9 $&$ 28.0 \pm 3.7 $&$ 8.3 \pm 0.8 $&$ 5.4 \pm 0.5 $&$ 0.70 \pm 0.20 $ \\
G23.97N \\
1 &$5.58 $&$ 0.076 $&$ 14.3 \pm 0.5 $&$ 27.2 \pm 1.2 $&$ 6.9 \pm 0.5 $&$ 0.25 \pm 0.02 $&$ 31.9 \pm 3.7 $&$ 29.4 \pm 2.1 $&$ 11.8 \pm 0.5 $&$ 4.5 \pm 0.2 $&$ 1.09 \pm 0.15 $ \\
2 &$4.97 $&$ 0.068 $&$ 18.3 \pm 1.4 $&$ 13.6 \pm 1.2 $&$ 7.1 \pm 1.2 $&$ 0.52 \pm 0.10 $&$ 6.9 \pm 3.6 $&$ 14.1 \pm 1.6 $&$ 7.4 \pm 0.6 $&$ 3.2 \pm 0.3 $&$ 0.49 \pm 0.26 $ \\
G23.98 \\
1 &$5.53 $&$ 0.075 $&$ 15.2 \pm 0.9 $&$ 11.5 \pm 0.8 $&$ 6.6 \pm 1.4 $&$ 0.58 \pm 0.13 $&$ 13.0 \pm 1.9 $&$ 21.2 \pm 1.8 $&$ 5.1 \pm 0.4 $&$ 2.0 \pm 0.1 $&$ 0.62 \pm 0.11 $ \\
2 &$5.33 $&$ 0.073 $&$ 15.2 \pm 2.5 $&$ 11.5 \pm 2.2 $&$ 4.0 \pm 0.9 $&$ 0.35 \pm 0.10 $&$ 10.7 \pm 2.3 $&$ 21.6 \pm 4.1 $&$ 5.4 \pm 1.0 $&$ 2.2 \pm 0.4 $&$ 0.50 \pm 0.15 $ \\
G23.44 \\
1 &$3.18 $&$ 0.054 $&$ 18.1 \pm 2.6 $&$ 16.9 \pm 2.8 $&$ 27.2 \pm 3.7 $&$ 1.61 \pm 0.34 $&$ 27.5 \pm 6.3 $&$ 73.6 \pm 10.8 $&$ 14.3 \pm 2.3 $&$ 7.6 \pm 1.2 $&$ 0.37 \pm 0.10 $ \\
2 &$3.15 $&$ 0.054 $&$ 19.9 \pm 3.6 $&$ 15.2 \pm 3.1 $&$ 7.3 \pm 1.1 $&$ 0.48 \pm 0.12 $&$ 22.0 \pm 4.5 $&$ 44.0 \pm 8.1 $&$ 13.1 \pm 2.7 $&$ 7.1 \pm 1.4 $&$ 0.50 \pm 0.13 $ \\
3 &$4.32 $&$ 0.074 $&$ 15.2 \pm 1.2 $&$ 20.7 \pm 1.8 $&$ 12.4 \pm 1.4 $&$ 0.60 \pm 0.08 $&$ 16.6 \pm 3.5 $&$ 46.3 \pm 3.9 $&$ 9.5 \pm 0.8 $&$ 3.7 \pm 0.3 $&$ 0.36 \pm 0.08 $ \\
4 &$2.26 $&$ 0.039 $&$ 22.3 \pm 4.8 $&$ 13.4 \pm 3.2 $&$ 4.5 \pm 0.8 $&$ 0.34 \pm 0.10 $&$ 8.5 \pm 2.7 $&$ 50.3 \pm 10.3 $&$ 22.4 \pm 5.3 $&$ 16.8 \pm 4.0 $&$ 0.17 \pm 0.06 $ \\
5 &$3.40 $&$ 0.058 $&$ 16.0 \pm 2.3 $&$ 19.4 \pm 3.2 $&$ 4.9 \pm 0.6 $&$ 0.25 \pm 0.05 $&$ 24.1 \pm 3.4 $&$ 23.8 \pm 4.0 $&$ 14.3 \pm 2.4 $&$ 7.2 \pm 1.2 $&$ 1.01 \pm 0.22 $ \\
6 &$2.43 $&$ 0.042 $&$ 21.1 \pm 1.8 $&$ 14.3 \pm 1.4 $&$ 11.7 \pm 2.2 $&$ 0.82 \pm 0.17 $&$ 24.2 \pm 8.3 $&$ 50.2 \pm 4.5 $&$ 20.6 \pm 2.0 $&$ 14.4 \pm 1.4 $&$ 0.48 \pm 0.17 $ \\
7 &$2.32 $&$ 0.040 $&$ 15.3 \pm 2.0 $&$ 20.5 \pm 3.0 $&$ 2.6 \pm 1.2 $&$ 0.13 \pm 0.06 $&$ 15.5 \pm 4.6 $&$ 46.8 \pm 6.5 $&$ 32.5 \pm 4.8 $&$ 23.8 \pm 3.5 $&$ 0.33 \pm 0.11 $ \\
8 &$2.50 $&$ 0.043 $&$ 39.5 \pm 12.9 $&$ 7.3 \pm 2.5 $&$ 1.0 \pm 0.4 $&$ 0.14 \pm 0.06 $&$ 11.3 \pm 3.8 $&$ 30.4 \pm 9.1 $&$ 9.9 \pm 3.4 $&$ 6.8 \pm 2.3 $&$ 0.37 \pm 0.14 $ \\
9 &$2.69 $&$ 0.046 $&$ 16.1 \pm 1.8 $&$ 19.3 \pm 2.4 $&$ 6.5 \pm 3.0 $&$ 0.34 \pm 0.16 $&$ 1.4 \pm 0.8 $&$ 28.3 \pm 3.6 $&$ 22.8 \pm 2.8 $&$ 14.4 \pm 1.8 $&$ 0.05 \pm 0.03 $ \\
10 &$2.97 $&$ 0.051 $&$ 20.4 \pm 3.1 $&$ 14.8 \pm 2.5 $&$ 4.1 \pm 2.0 $&$ 0.28 \pm 0.14 $&$ 20.8 \pm 7.6 $&$ 61.4 \pm 9.4 $&$ 14.3 \pm 2.4 $&$ 8.2 \pm 1.4 $&$ 0.34 \pm 0.13 $ \\
G23.97S \\
1 &$3.81 $&$ 0.052 $&$ 17.9 \pm 2.8 $&$ 14.5 \pm 2.5 $&$ 9.1 \pm 2.0 $&$ 0.63 \pm 0.18 $&$ 1.6 \pm 0.6 $&$ 62.0 \pm 9.8 $&$ 13.3 \pm 2.3 $&$ 7.4 \pm 1.3 $&$ 0.03 \pm 0.01 $ \\
2 &$4.04 $&$ 0.055 $&$ 18.3 \pm 1.9 $&$ 14.1 \pm 1.6 $&$ 7.7 \pm 1.6 $&$ 0.55 \pm 0.13 $&$ 7.0 \pm 1.2 $&$ 84.1 \pm 8.8 $&$ 11.5 \pm 1.3 $&$ 6.1 \pm 0.7 $&$ 0.08 \pm 0.02 $ \\
3 &$2.82 $&$ 0.039 $&$ 17.3 \pm 2.2 $&$ 15.1 \pm 2.1 $&$ 2.7 \pm 0.7 $&$ 0.18 \pm 0.05 $&$ 16.7 \pm 3.6 $&$ 28.5 \pm 4.9 $&$ 25.3 \pm 3.6 $&$ 19.1 \pm 2.7 $&$ 0.59 \pm 0.16 $ \\
4 &$3.04 $&$ 0.042 $&$ 31.9 \pm 6.8 $&$ 7.7 \pm 1.7 $&$ 3.2 \pm 1.0 $&$ 0.41 \pm 0.16 $&$ 19.4 \pm 3.4 $&$ 31.2 \pm 10.9 $&$ 11.1 \pm 2.5 $&$ 7.8 \pm 1.8 $&$ 0.62 \pm 0.24 $ \\
G25.38 \\
1 &$4.00 $&$ 0.065 $&$ 18.2 \pm 1.6 $&$ 23.8 \pm 2.4 $&$ 3.7 \pm 0.7 $&$ 0.16 \pm 0.03 $&$ 13.6 \pm 3.5 $&$ 24.9 \pm 2.4 $&$ 14.0 \pm 1.4 $&$ 6.2 \pm 0.6 $&$ 0.55 \pm 0.15 $ \\
2 &$3.14 $&$ 0.051 $&$ 16.2 \pm 1.5 $&$ 27.2 \pm 2.9 $&$ 3.1 \pm 0.5 $&$ 0.11 \pm 0.02 $&$ 17.6 \pm 3.5 $&$ 16.2 \pm 2.1 $&$ 25.9 \pm 2.8 $&$ 14.7 \pm 1.6 $&$ 1.09 \pm 0.26 $ \\
3 &$4.53 $&$ 0.074 $&$ 19.4 \pm 3.1 $&$ 22.2 \pm 3.9 $&$ 34.9 \pm 7.3 $&$ 1.57 \pm 0.43 $&$ 11.0 \pm 4.9 $&$ 40.8 \pm 6.7 $&$ 10.2 \pm 1.8 $&$ 4.0 \pm 0.7 $&$ 0.27 \pm 0.13 $ \\
4 &$3.16 $&$ 0.052 $&$ 16.1 \pm 1.0 $&$ 27.3 \pm 1.9 $&$ 3.7 \pm 0.7 $&$ 0.14 \pm 0.03 $&$ 17.2 \pm 2.7 $&$ 12.2 \pm 1.1 $&$ 25.7 \pm 1.8 $&$ 14.5 \pm 1.0 $&$ 1.41 \pm 0.26 $ \\
G25.71 \\
1 &$3.94 $&$ 0.109 $&$ 28.8 \pm 6.7 $&$ 32.8 \pm 8.3 $&$ 44.4 \pm 11.6 $&$ 1.35 \pm 0.49 $&$ 9.2 \pm 3.5 $&$ 69.8 \pm 16.1 $&$ 6.9 \pm 1.7 $&$ 1.8 \pm 0.5 $&$ 0.13 \pm 0.05 $ \\
2 &$4.57 $&$ 0.126 $&$ 15.7 \pm 2.6 $&$ 54.0 \pm 10.0 $&$ 6.0 \pm 1.9 $&$ 0.11 \pm 0.04 $&$ 0.6 \pm 0.3 $&$ 21.8 \pm 3.8 $&$ 8.5 \pm 1.6 $&$ 1.9 \pm 0.4 $&$ 0.03 \pm 0.01 $ \\
\hline
\end{tabular}
\begin{flushleft}
\textbf{Note:} \\
FWHM and $R_{\rm eff}$: Derived from NH$_2$D core extraction by using the \texttt{GAUSSCLUMPS}. \\
$T_{\rm kin}$: Estimated from the rotational temperature $T_{\rm rot}$ by using NH$_3$ (1,\,1) and (2,\,2) transitions \citep{Ott2011}. \\
$M_{\rm H_2}$, $N_{\rm H_2}$, and $n_{\rm H_2}$: Estimated from the corresponding 3.5\,mm continuum flux to each NH$_2$D core size. \\
$N_{\rm NH_2D}$: Adopting the analyze routine described in Appendix of \citet{Pillai2007} to estimate the NH$_2$D column density. \\
$N_{\rm NH_3}$: Following the standard formulation in \citet{Bachiller1987} to estimate the NH$_3$ column density. \\
$M_{\rm vir} \simeq 126\, R_{\rm eff}\, \Delta v^{2}_{\rm nth}\, (\Msun)$  and $\alpha_{\rm vir} = M_{\rm vir}/M_{\rm NH_2D}$ \citep{MacLaren1988,Evans1999}: The $\Delta v_{\rm nth}$ is the non-thermal line width for NH$_2$D main line. \\
$D_{\rm frac}$: Deuterium fractionation is defined as $D_{\rm frac} = N_{\rm NH_2D}/N_{\rm NH_3}={\rm [NH_2D]/[NH_3]}$.
\end{flushleft}
\end{table*}

\clearpage

%
%
%

\begin{figure*}
\centering
\includegraphics[height=0.32\textwidth,angle=0]{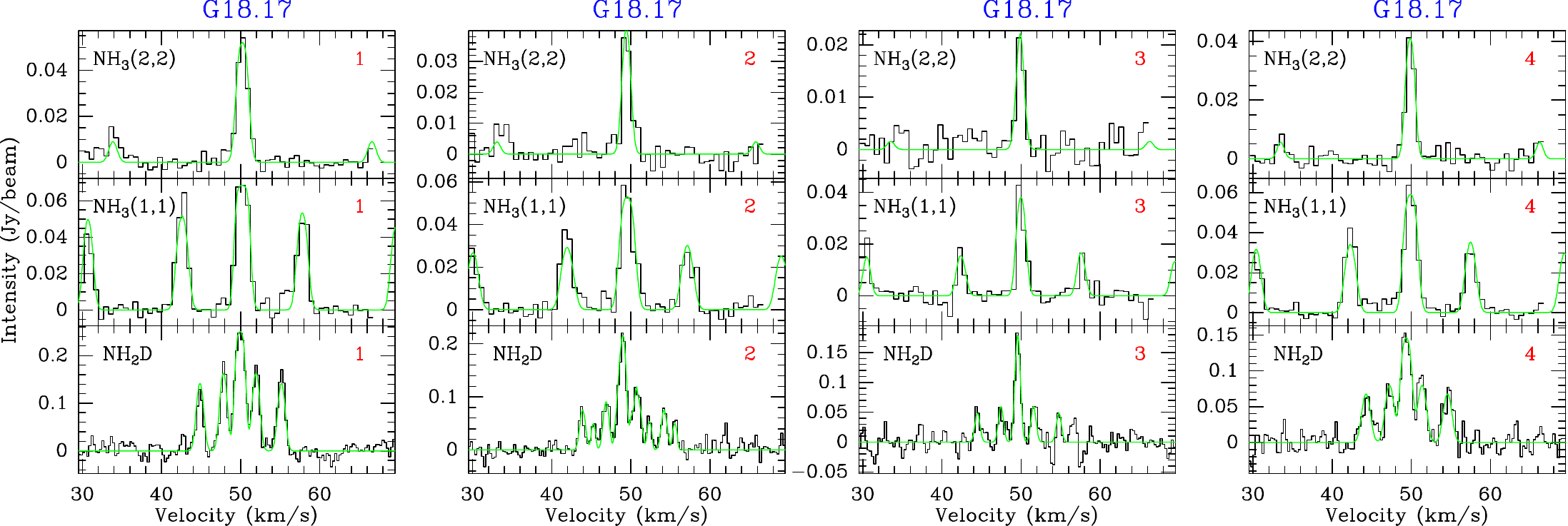}
\includegraphics[height=0.32\textwidth,angle=0]{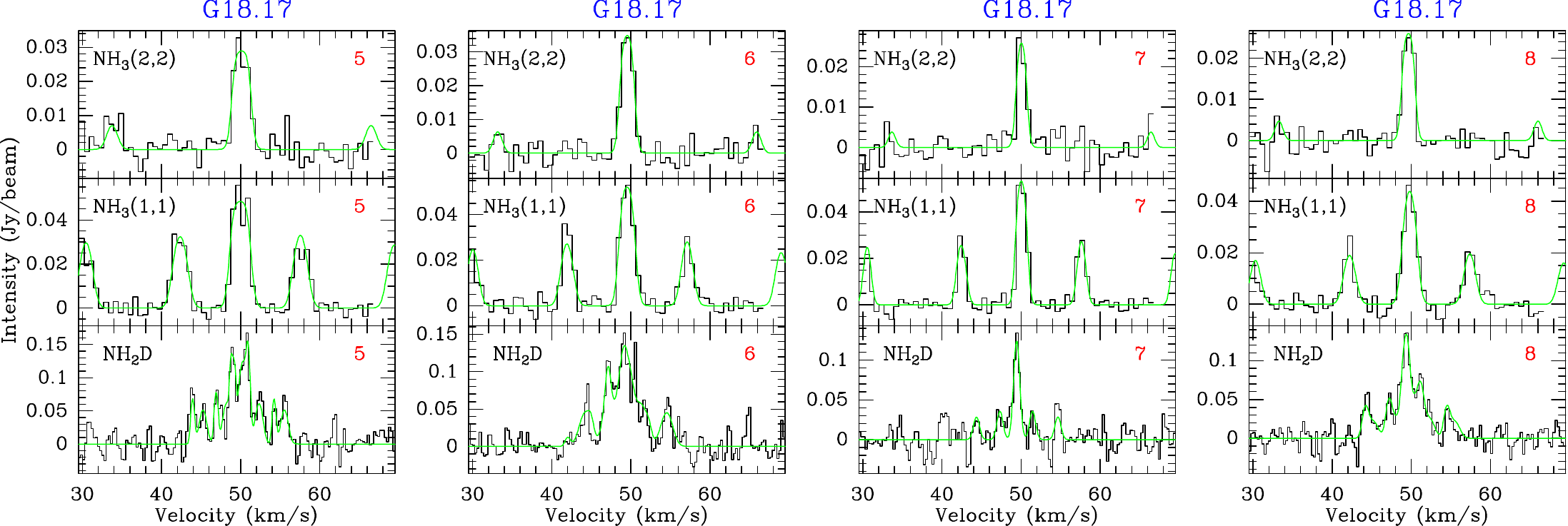}
\includegraphics[height=0.32\textwidth,angle=0]{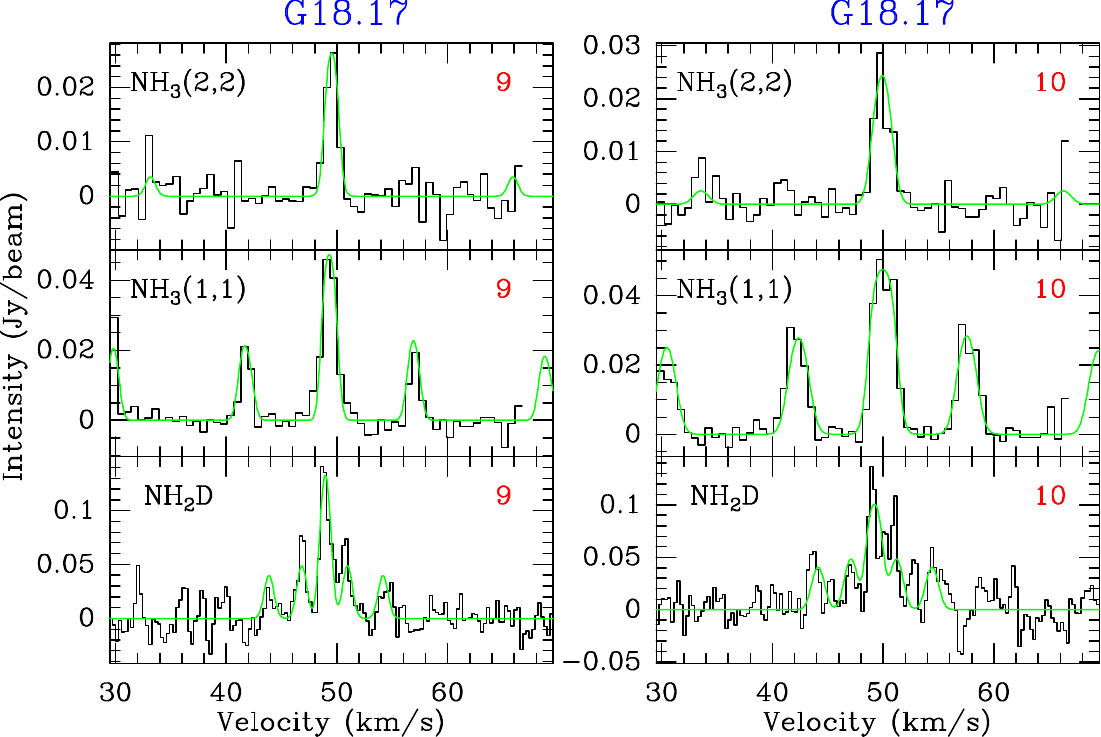}
\includegraphics[height=0.32\textwidth,angle=0]{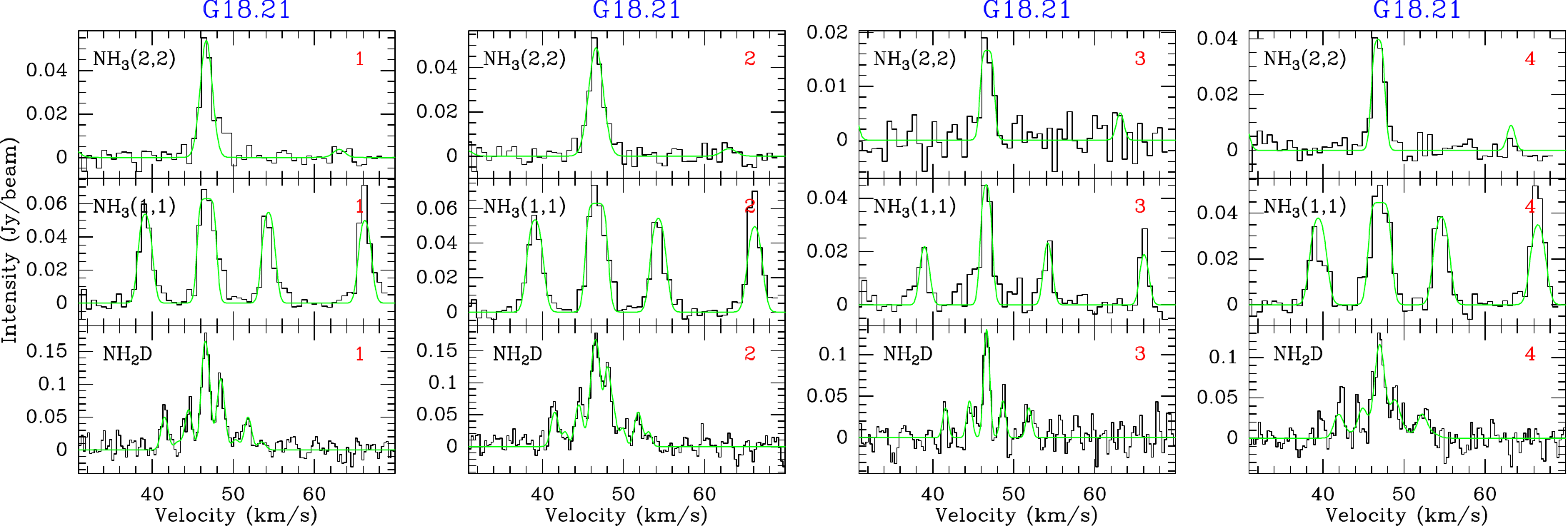}
\caption{Spectra NH$_2$D, NH$_3$ (1,\,1) and (2,\,2) overlaid with their HfS fits for all NH$_2$D cores (see Table\,\ref{tab_nh2d_para1}). The spectra are derived by averaging the lines within each NH$_2$D core scale. }
\label{Fig_spectra_app}
\end{figure*}

\begin{figure*}
\ContinuedFloat
\captionsetup{list=off}
\centering
\includegraphics[height=0.32\textwidth,angle=0]{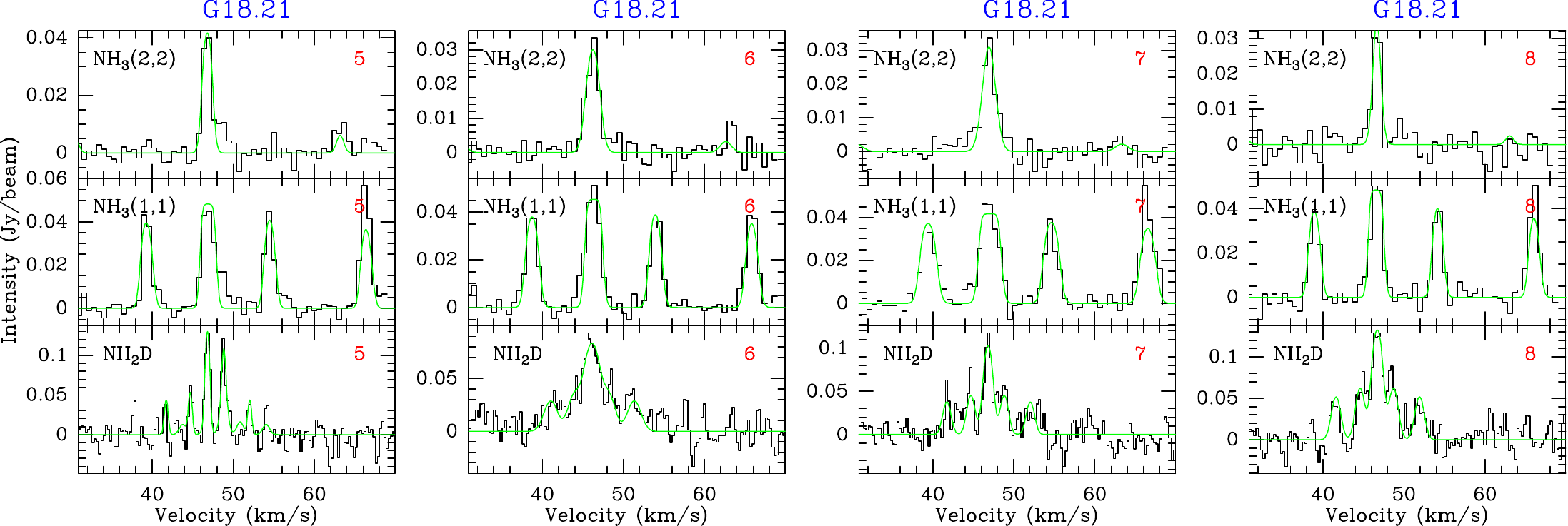}
\includegraphics[height=0.32\textwidth,angle=0]{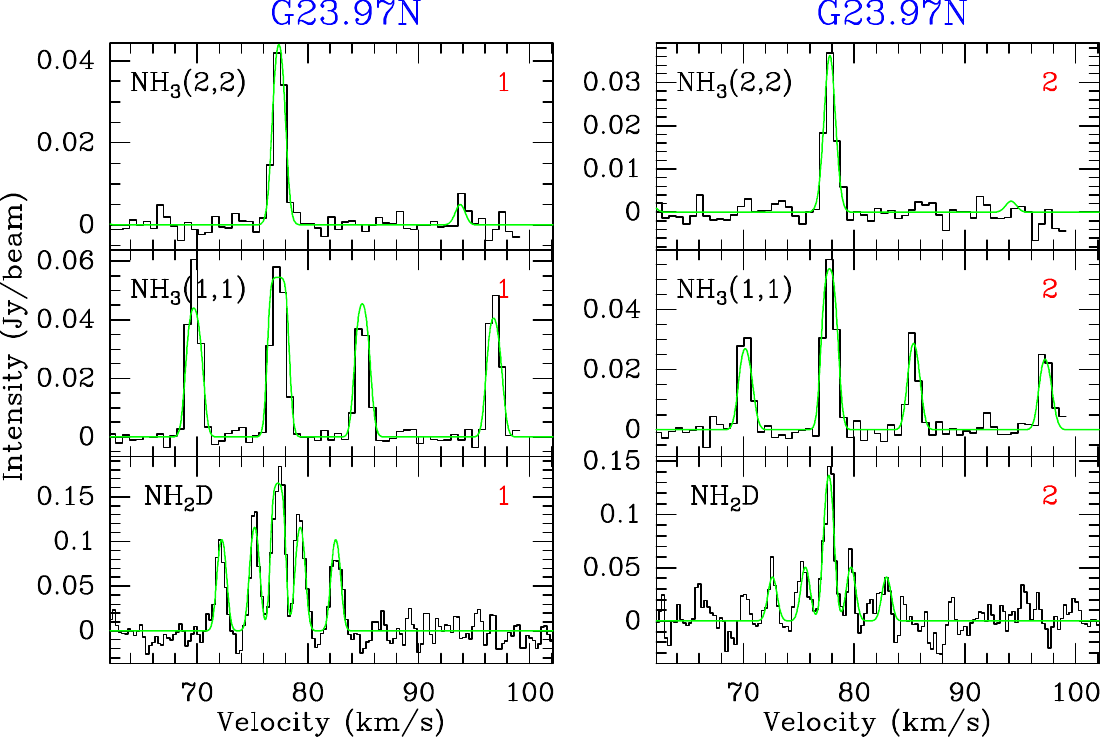}
\includegraphics[height=0.32\textwidth,angle=0]{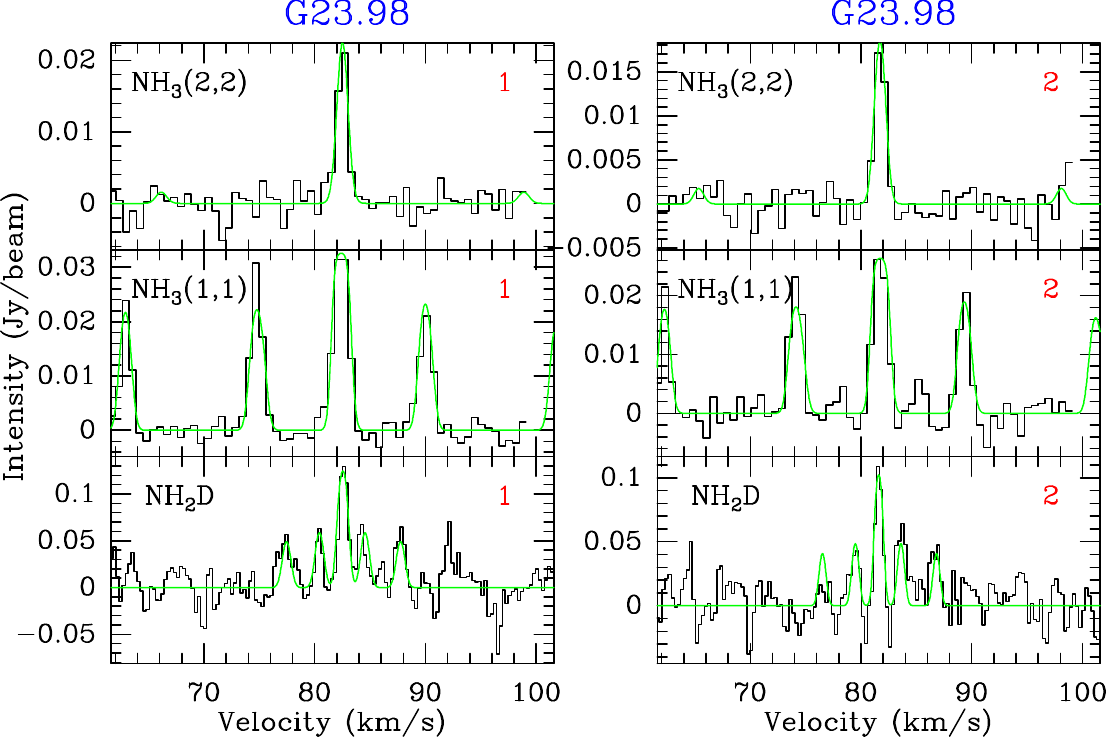}
\includegraphics[height=0.32\textwidth,angle=0]{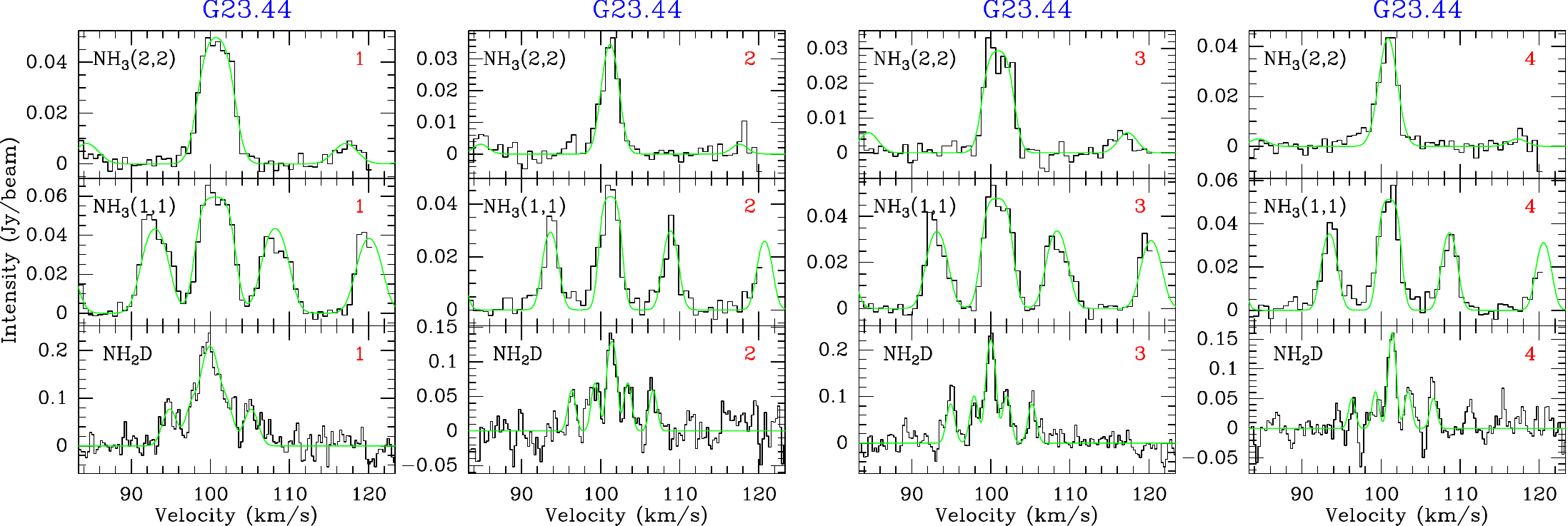}
\includegraphics[height=0.32\textwidth,angle=0]{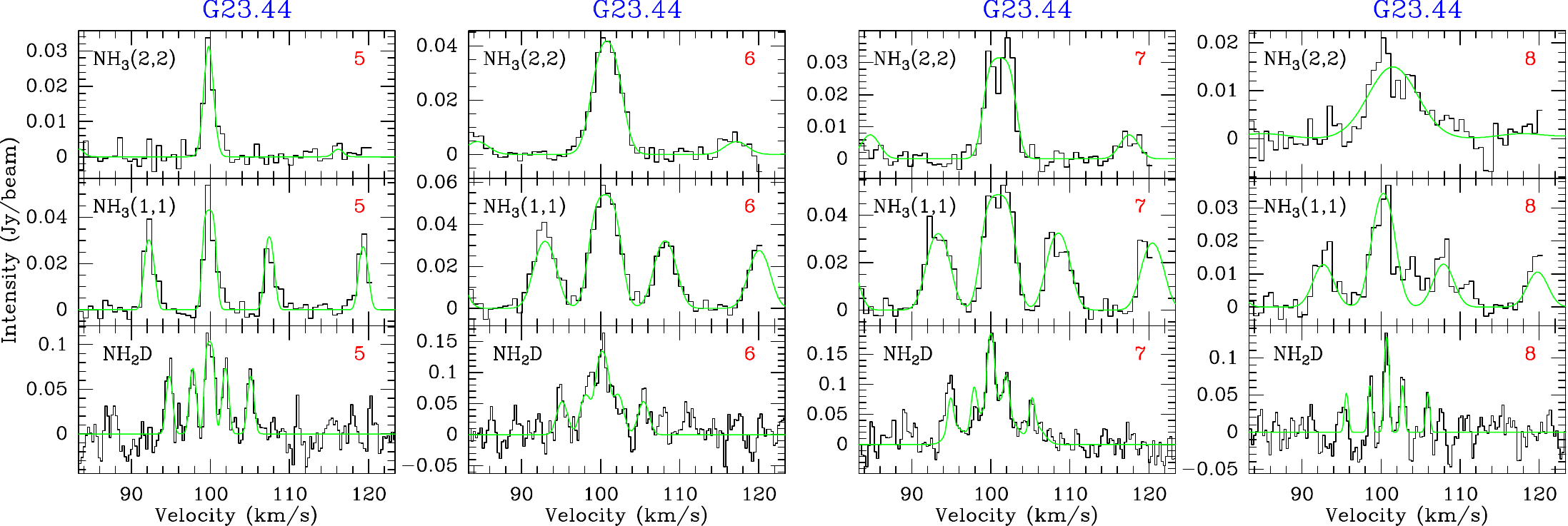}
\caption{Continued. }
\label{Fig_spectra_app}
\end{figure*}

\begin{figure*}
\ContinuedFloat
\captionsetup{list=off}
\centering
\includegraphics[height=0.32\textwidth,angle=0]{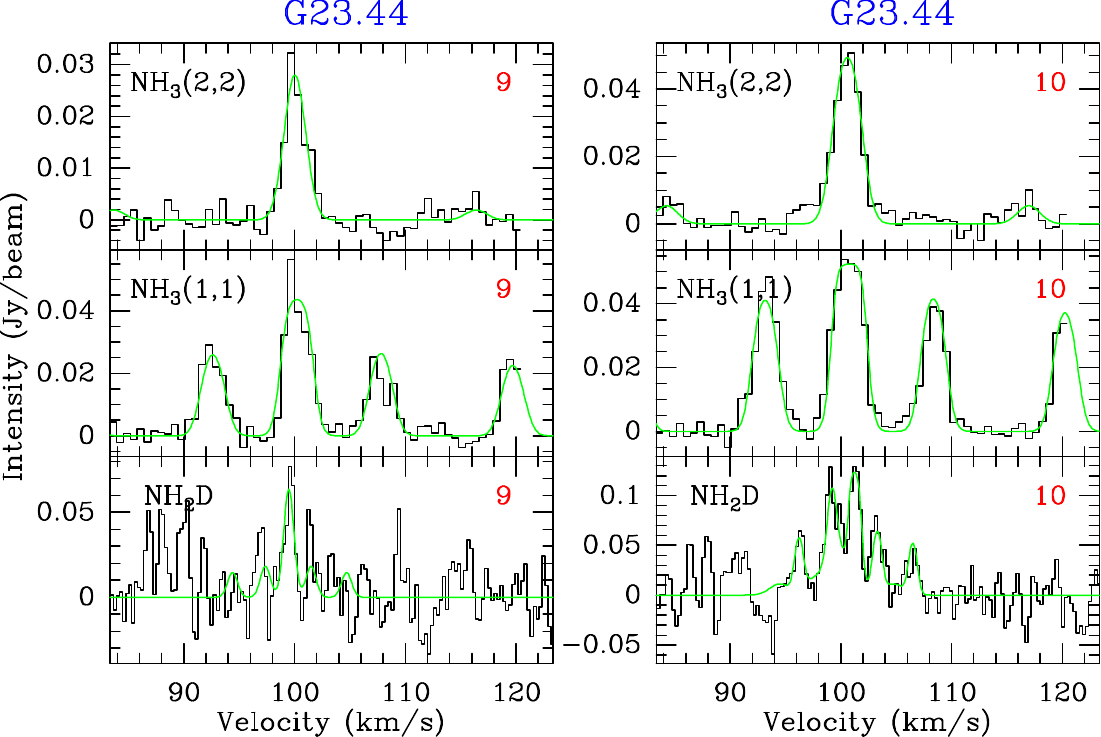}
\includegraphics[height=0.32\textwidth,angle=0]{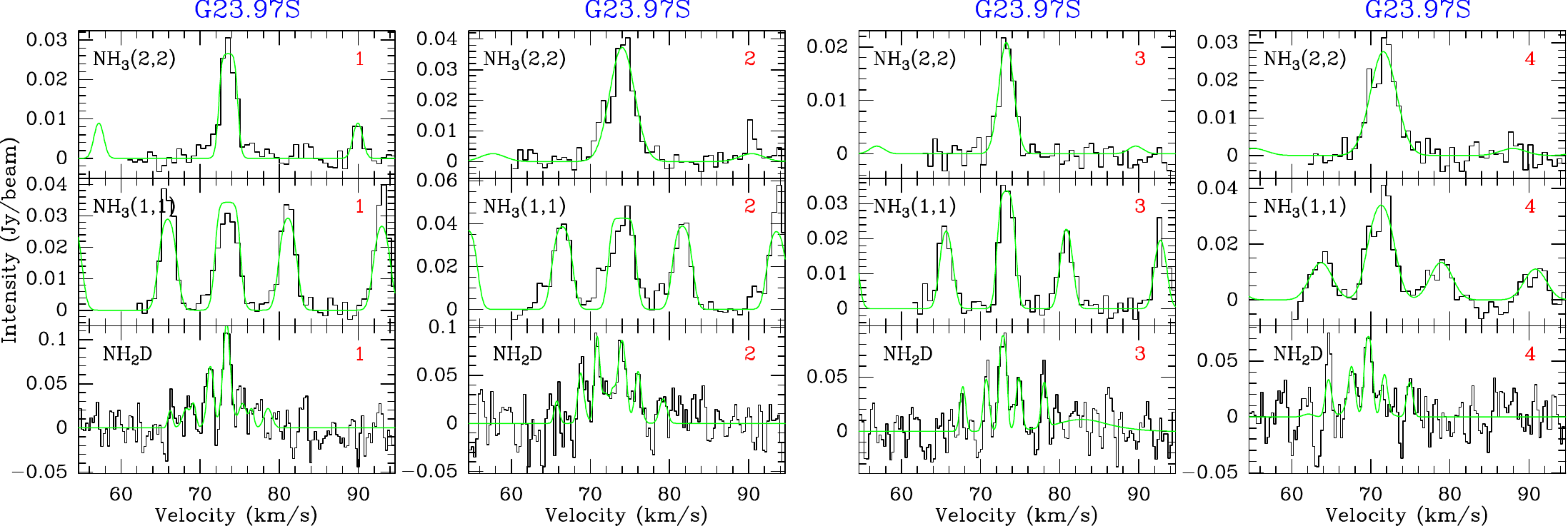}
\includegraphics[height=0.32\textwidth,angle=0]{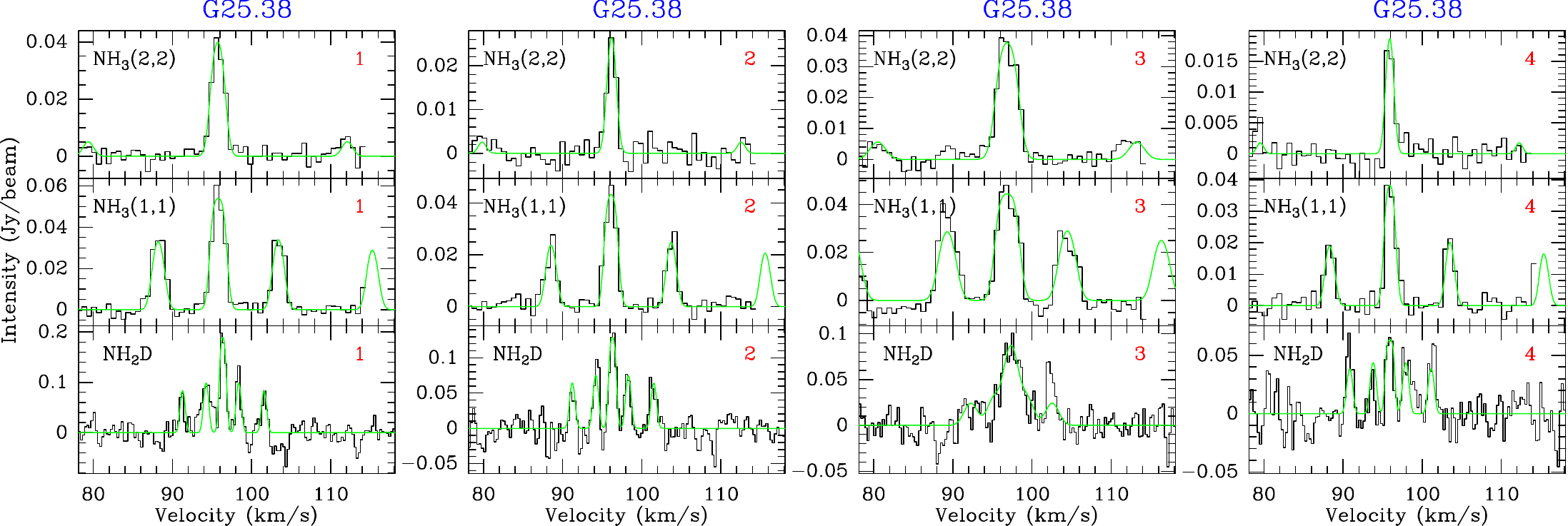}
\includegraphics[height=0.32\textwidth,angle=0]{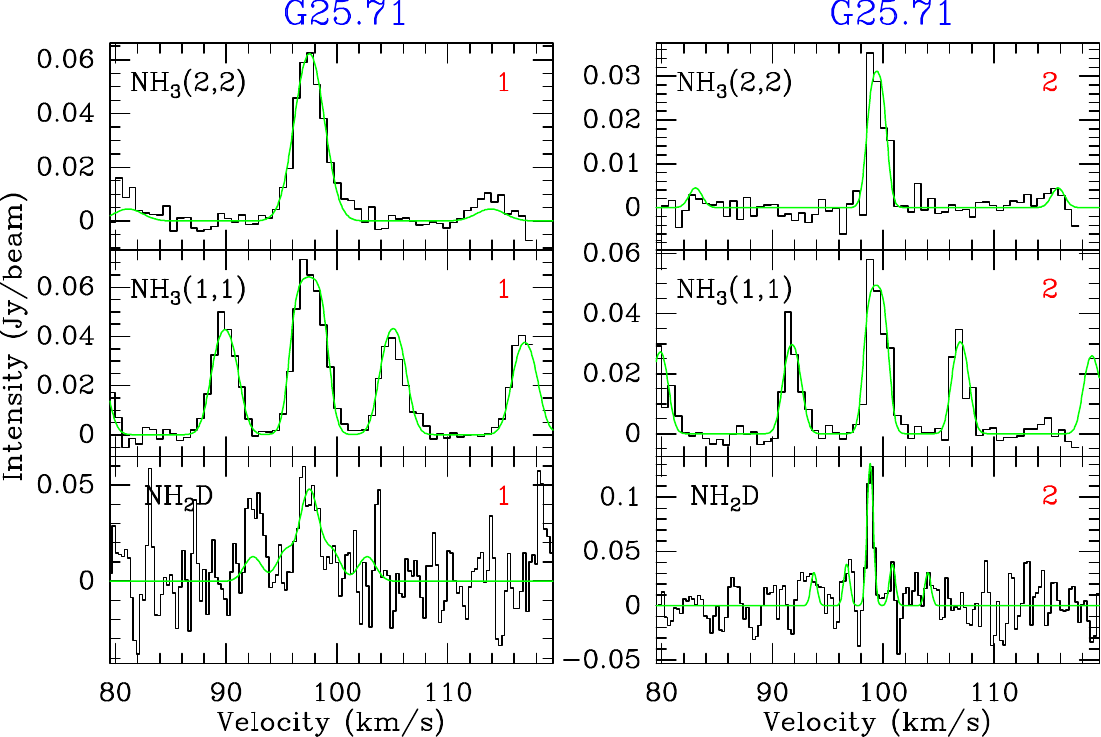}
\caption{Continued. }
\label{Fig_spectra_app}
\end{figure*}

\begin{figure*}
\centering
\subfigure[]{\includegraphics[width=0.33\textwidth,
angle=0]{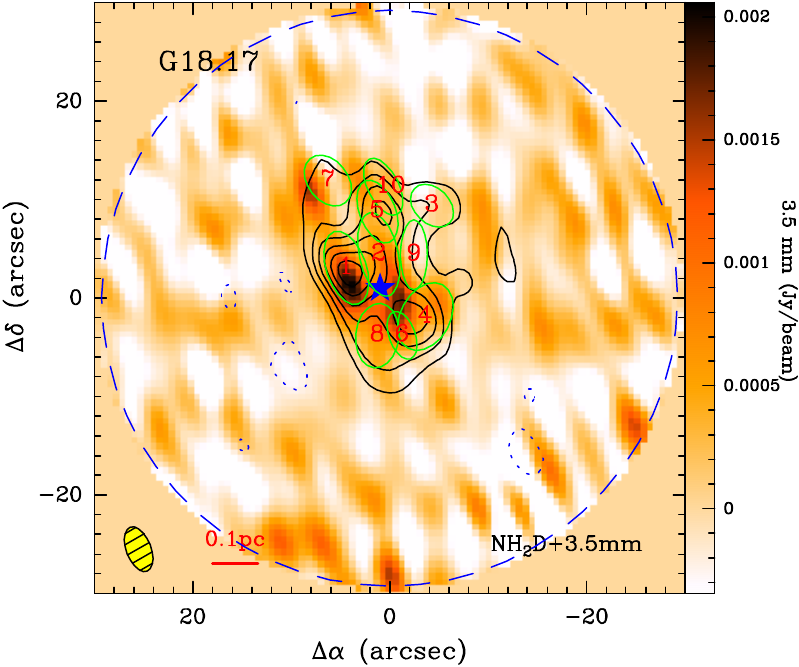}}
\subfigure[]{\includegraphics[width=0.33\textwidth,
angle=0]{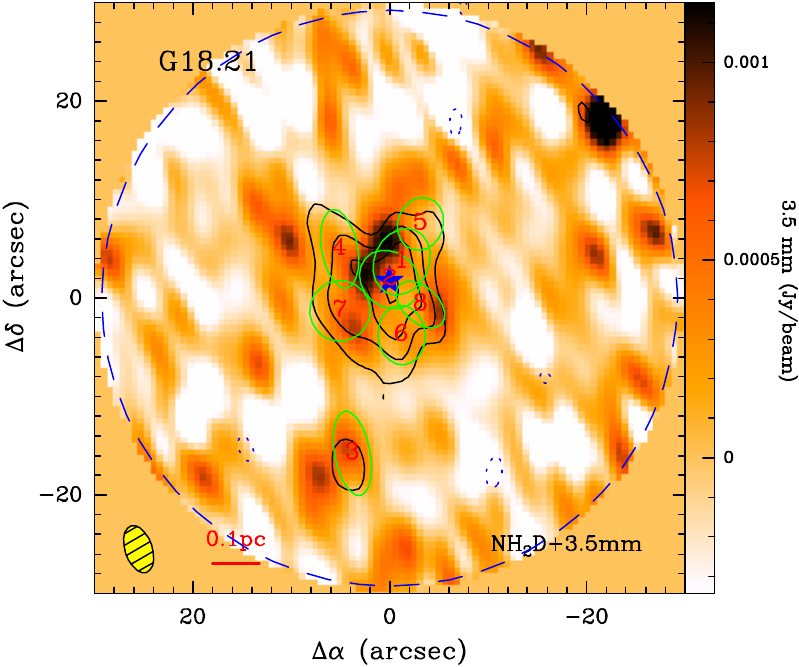}}
\subfigure[]{\includegraphics[width=0.33\textwidth,
angle=0]{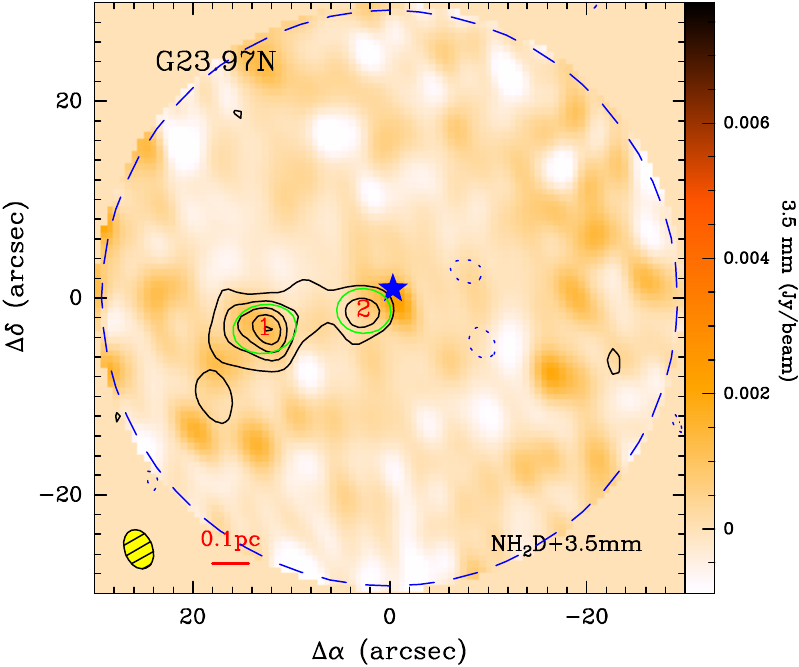}}
\subfigure[]{\includegraphics[width=0.33\textwidth,
angle=0]{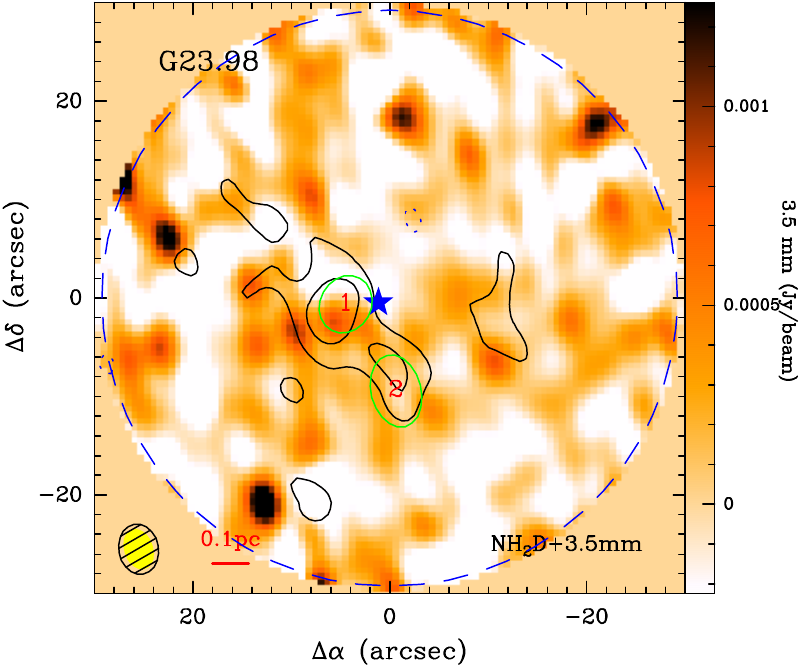}}
\subfigure[]{\includegraphics[width=0.33\textwidth,
angle=0]{figures/nh2d-3mm/g2344_bcd-eps-converted-to.pdf}}
\subfigure[]{\includegraphics[width=0.33\textwidth,
angle=0]{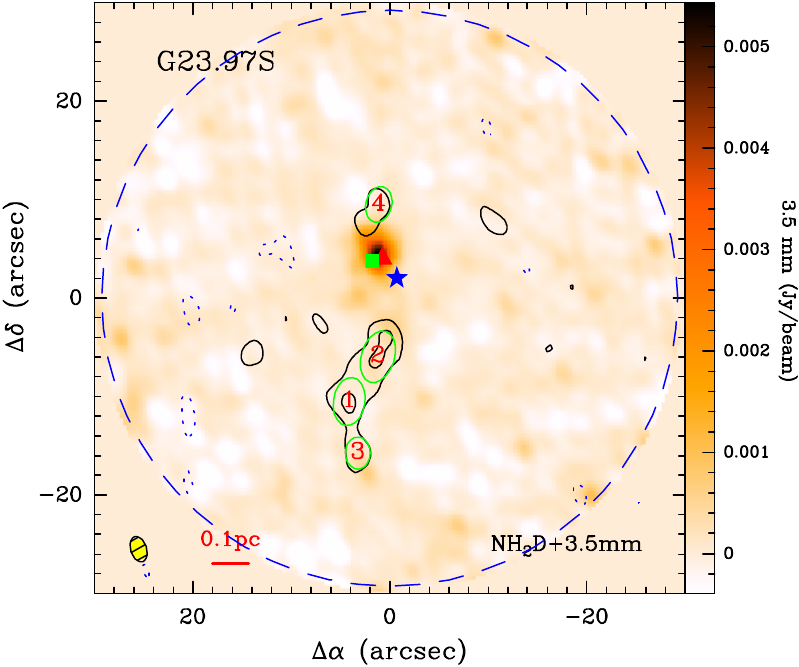}}
\subfigure[]{\includegraphics[width=0.33\textwidth,
angle=0]{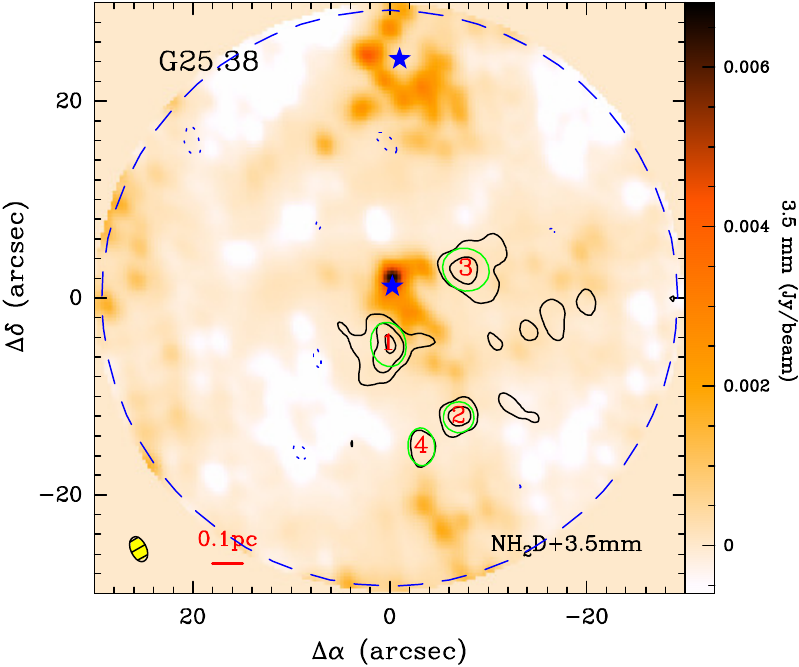}}
\subfigure[]{\includegraphics[width=0.33\textwidth,
angle=0]{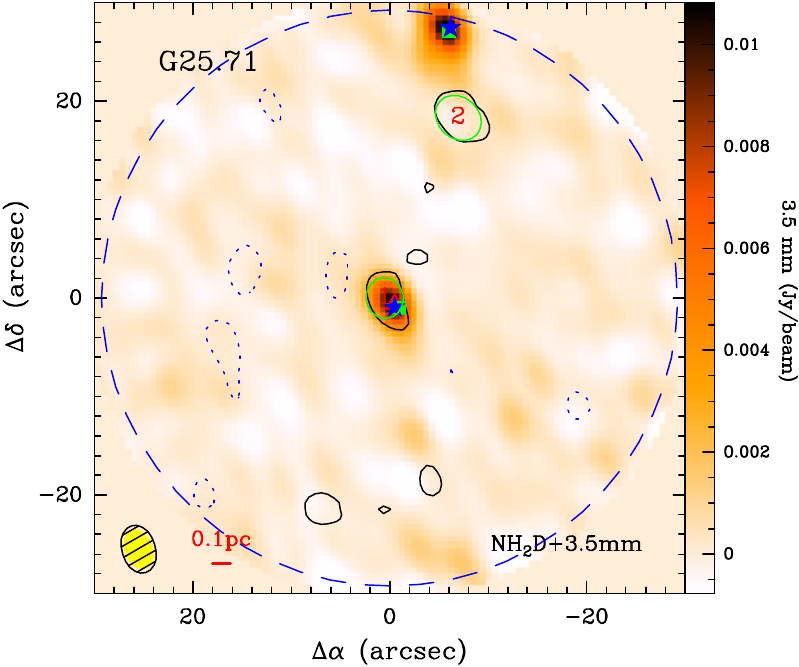}}
\caption{NH$_2$D integrated-intensity contours overlaid on a 3.5\,mm continuum with velocity range covering all the six HfS lines. The contour levels start at $-3\sigma$ in steps of $3\sigma$ for NH$_2$D with $\sigma_{\rm (a)-(h)} =$ 60.4, 60.9, 51.1, 53.2, 33.6, 23.4, 26.9, 50.4\,$\mjybkms$. The green ellipses with red numbers indicate the positions of extracted NH$_2$D cores. The symbols ``$\blacktriangle$'', ``$\blacksquare$'', and ``$\bigstar$'' indicate the positions of masers, \HII regions, and infrared sources, respectively. The synthesized beam sizes of each subfigure are indicated at the bottom-left corner. The dashed circle indicates the primary beam of the PdBI observations at 3.5\,mm.}
\label{Fig_nh2d_app}
\end{figure*}

\begin{figure*}
\centering
\subfigure[]{\includegraphics[width=0.33\textwidth,
angle=0]{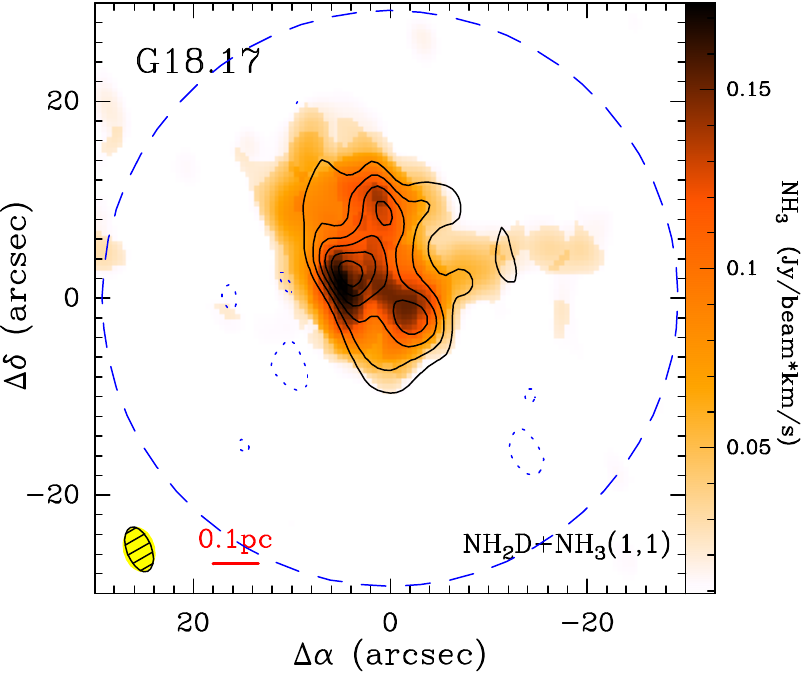}}
\subfigure[]{\includegraphics[width=0.33\textwidth,
angle=0]{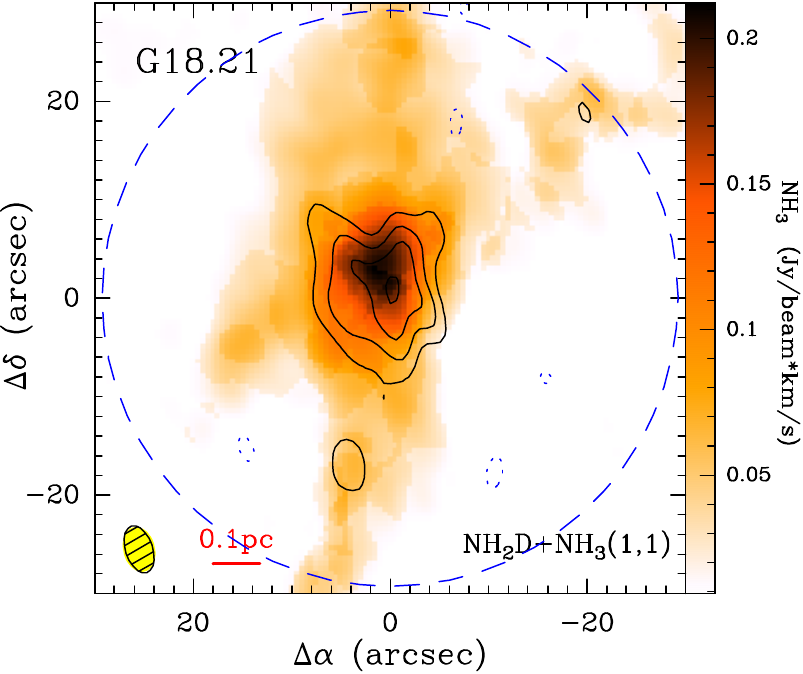}}
\subfigure[]{\includegraphics[width=0.33\textwidth,
angle=0]{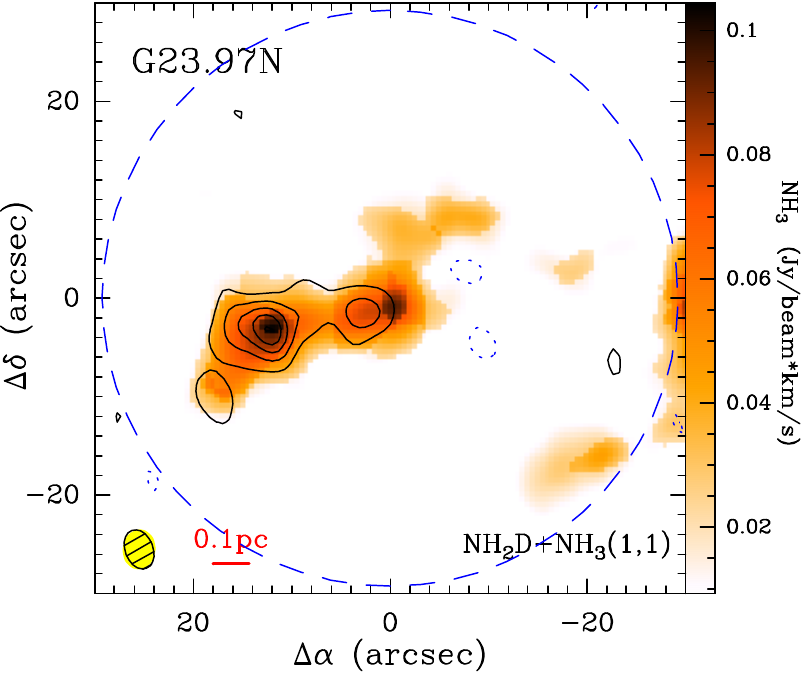}}
\subfigure[]{\includegraphics[width=0.33\textwidth,
angle=0]{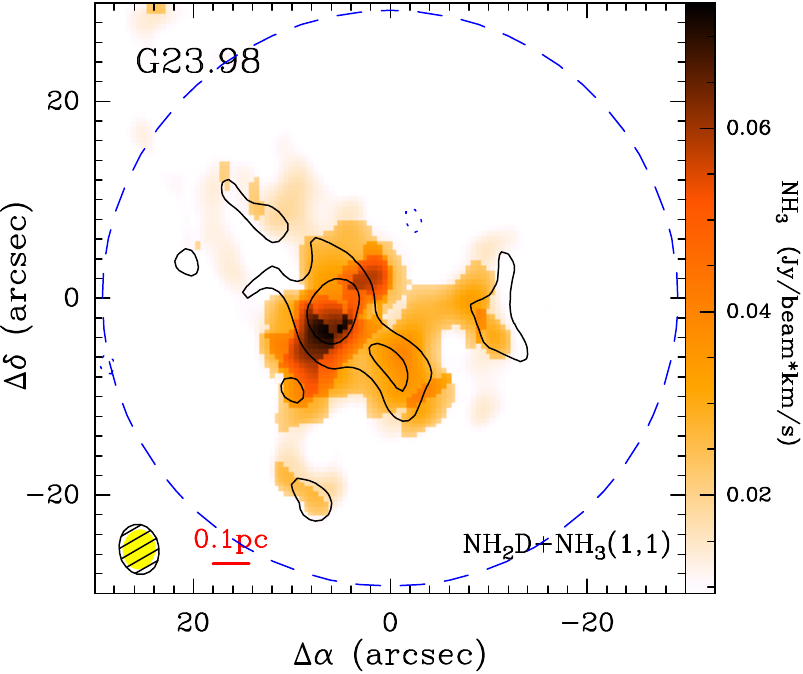}}
\subfigure[]{\includegraphics[width=0.33\textwidth,
angle=0]{figures/nh2d-nh3/g2344_bcd-eps-converted-to.pdf}}
\subfigure[]{\includegraphics[width=0.33\textwidth,
angle=0]{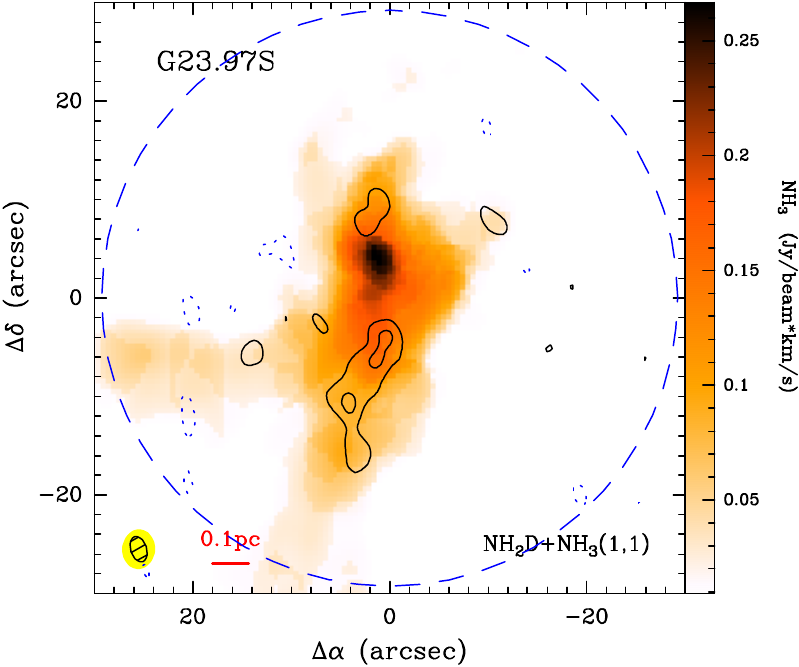}}
\subfigure[]{\includegraphics[width=0.33\textwidth,
angle=0]{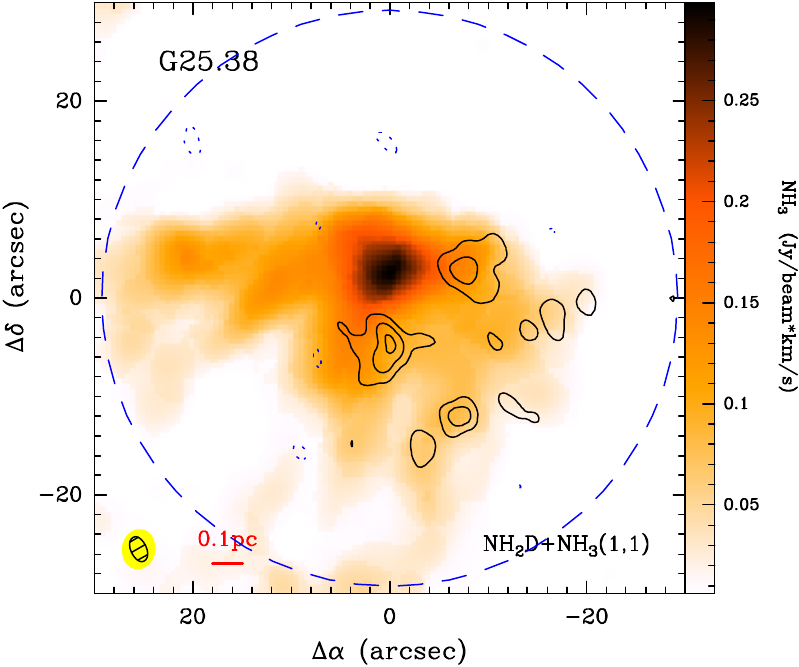}}
\subfigure[]{\includegraphics[width=0.33\textwidth,
angle=0]{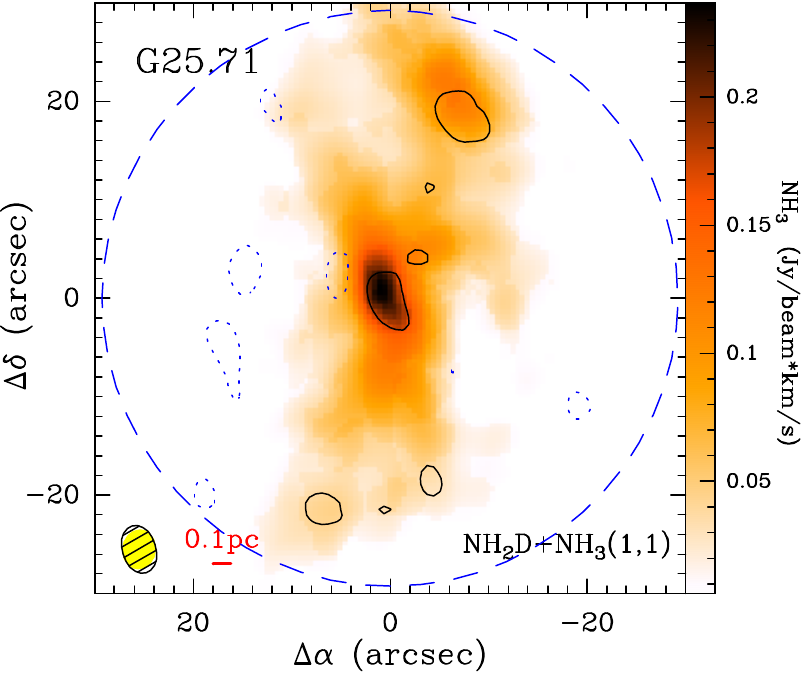}}
\caption{NH$_2$D integrated-intensity contours overlaid on an NH$_3$ (1,\,1) integrated-intensity image with velocity range covering all the six HfS lines. The contour levels start at $-3\sigma$ in steps of $3\sigma$ for NH$_2$D with $\sigma_{\rm (a)-(h)} =$ 60.4, 60.9, 51.1, 53.2, 33.6, 23.4, 26.9, 50.4\,$\mjybkms$. The red numbers indicate the positions of extracted NH$_2$D cores. The synthesized beam sizes of each subfigure are indicated at the bottom-left corner. The dashed circle indicates the primary beam of the PdBI observations at 3.5\,mm.}
\label{Fig_nh2d_nh3_app}
\end{figure*}

\begin{figure*}
\centering
\subfigure[]{\includegraphics[width=0.33\textwidth,
angle=0]{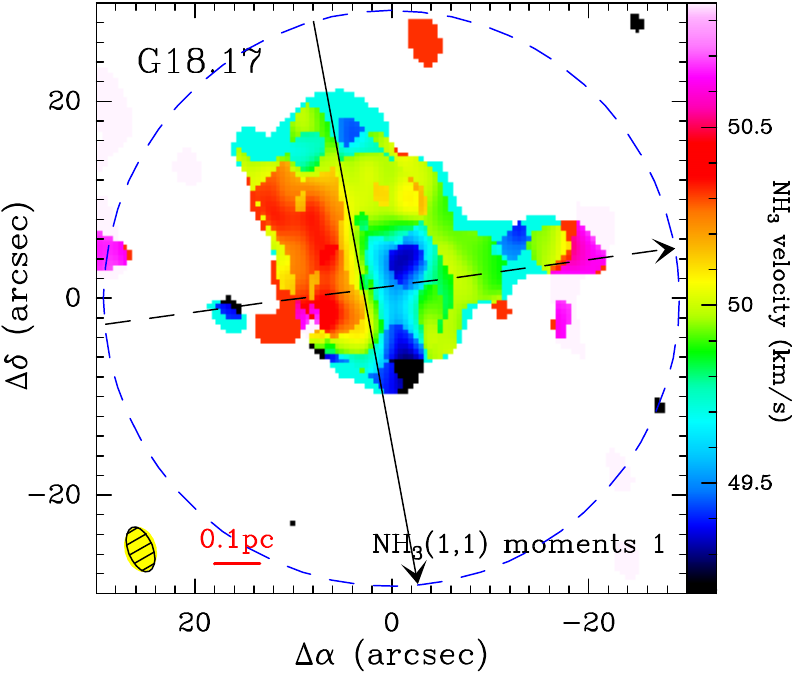}}
\subfigure[]{\includegraphics[width=0.33\textwidth,
angle=0]{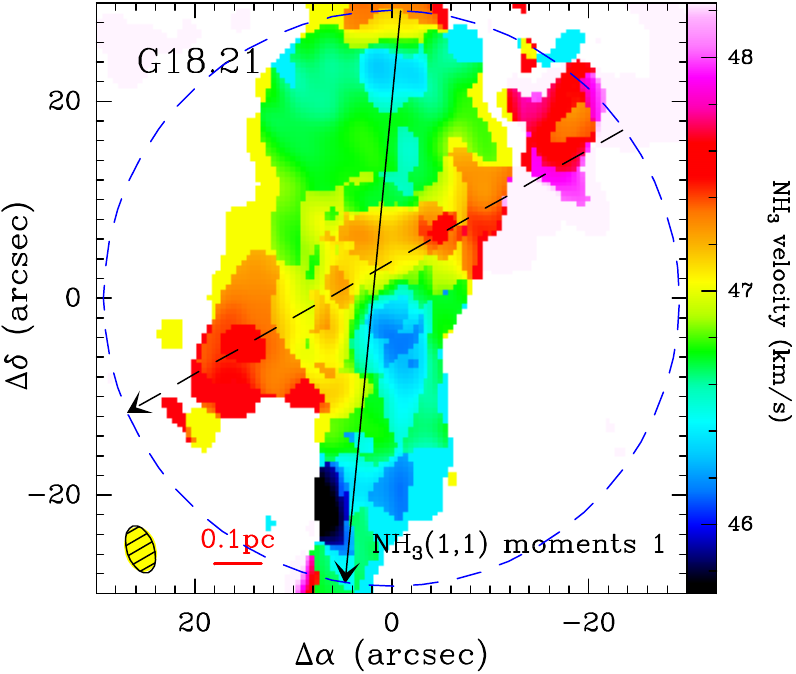}}
\subfigure[]{\includegraphics[width=0.33\textwidth,
angle=0]{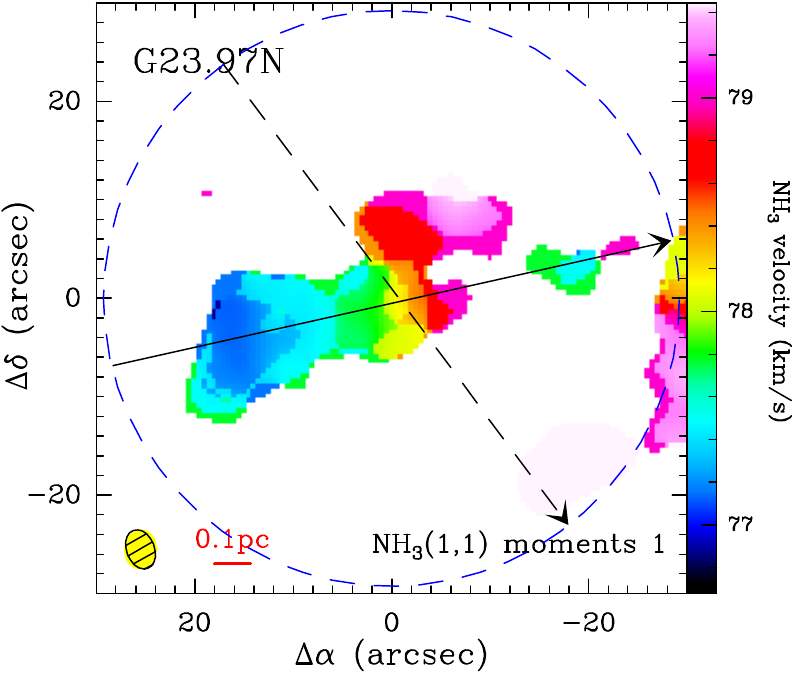}}
\subfigure[]{\includegraphics[width=0.33\textwidth,
angle=0]{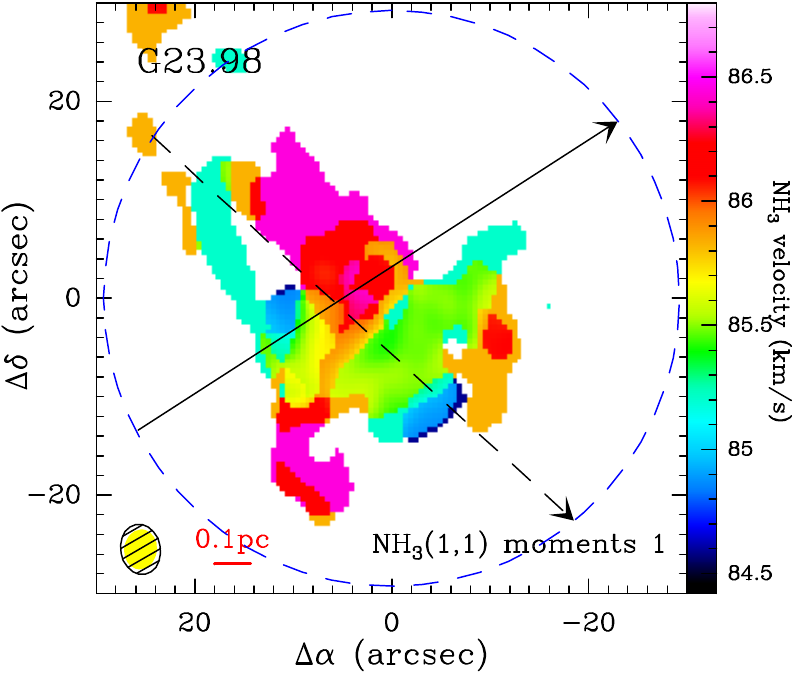}}
\subfigure[]{\includegraphics[width=0.33\textwidth,
angle=0]{figures/nh2d-moment1/g2344-11-eps-converted-to.pdf}}
\subfigure[]{\includegraphics[width=0.33\textwidth,
angle=0]{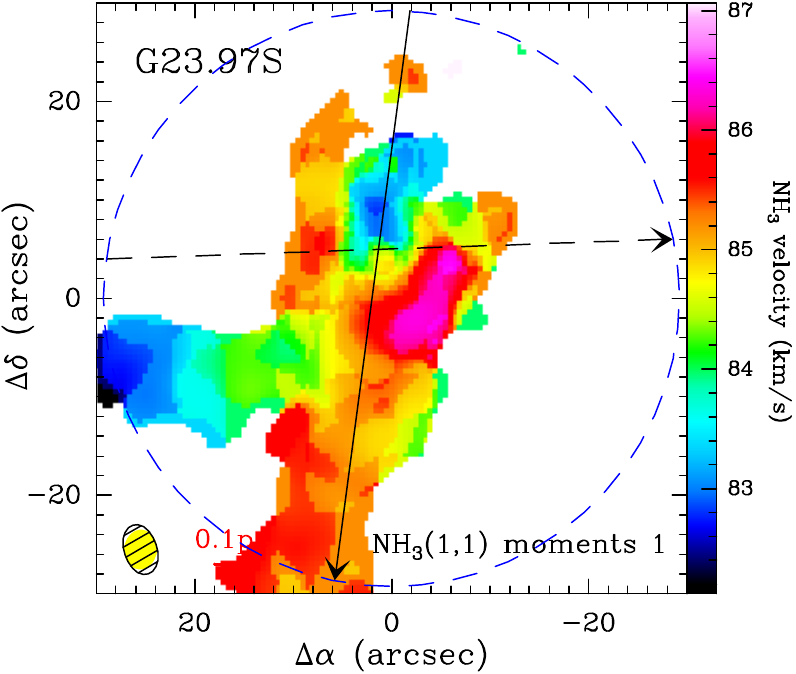}}
\subfigure[]{\includegraphics[width=0.33\textwidth,
angle=0]{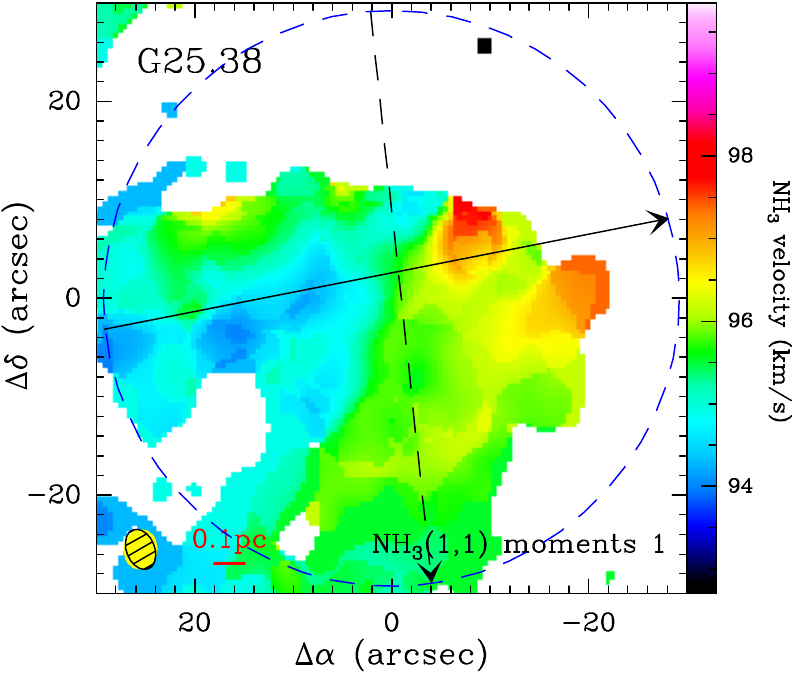}}
\subfigure[]{\includegraphics[width=0.33\textwidth,
angle=0]{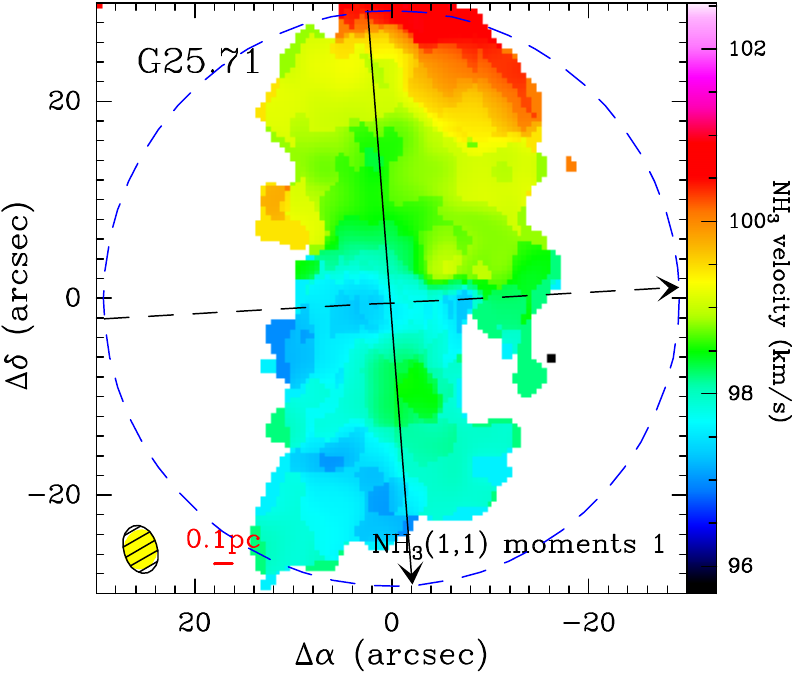}}
\caption{Velocity distribution (moment\,1) of NH$_3$ (1,\,1) line overlaid with integrated-intensity contours (moment\,0) of NH$_2$D line with velocity range covering all the six HfS lines. The contour levels start at $-3\sigma$ in steps of $3\sigma$ for NH$_2$D with $\sigma_{\rm (a)-(h)} =$ 60.3, 60.9, 51.1, 53.2, 84.1, 48.3, 49.0, 50.4 $\mjybkms$. The synthesized beam sizes of each subfigure are indicated at the bottom-left corner. The dashed circle indicates the primary beam of the PdBI observations at 3.5\,mm. The lines with arrows show the cutting direction in Figure\,\ref{Fig_pv_app}.}
\label{Fig_mom1_app}
\end{figure*}

\begin{figure*}
\centering
\subfigure[]{\includegraphics[width=0.33\textwidth,
angle=0]{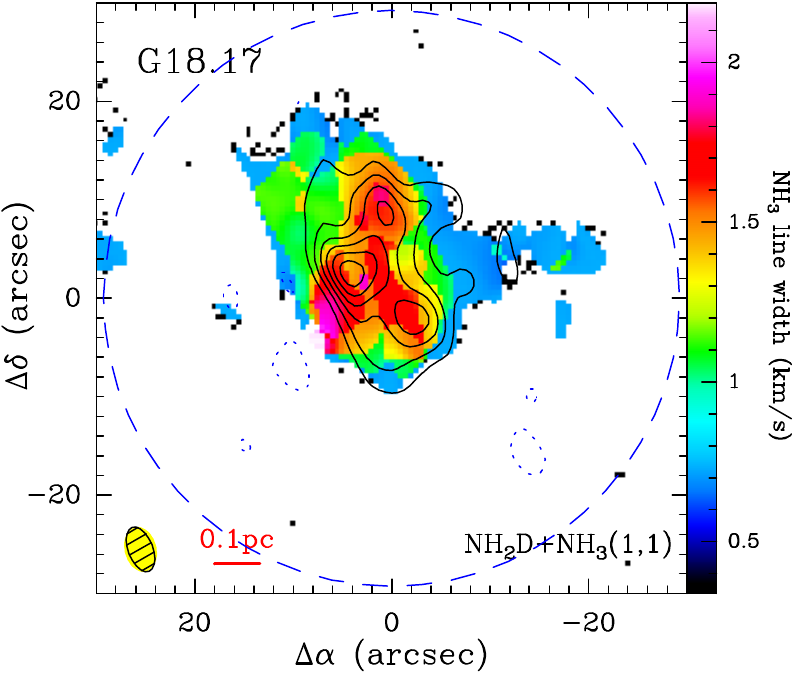}}
\subfigure[]{\includegraphics[width=0.33\textwidth,
angle=0]{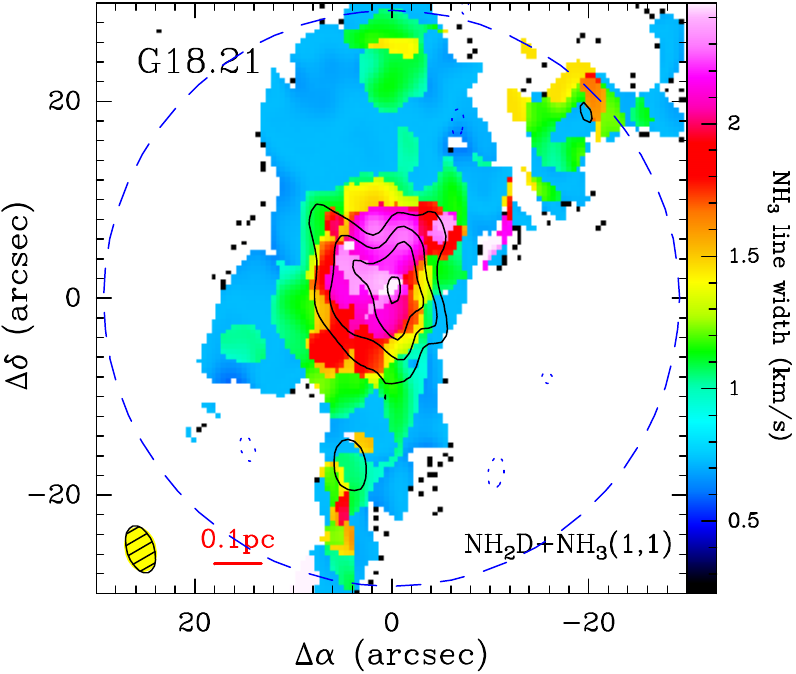}}
\subfigure[]{\includegraphics[width=0.33\textwidth,
angle=0]{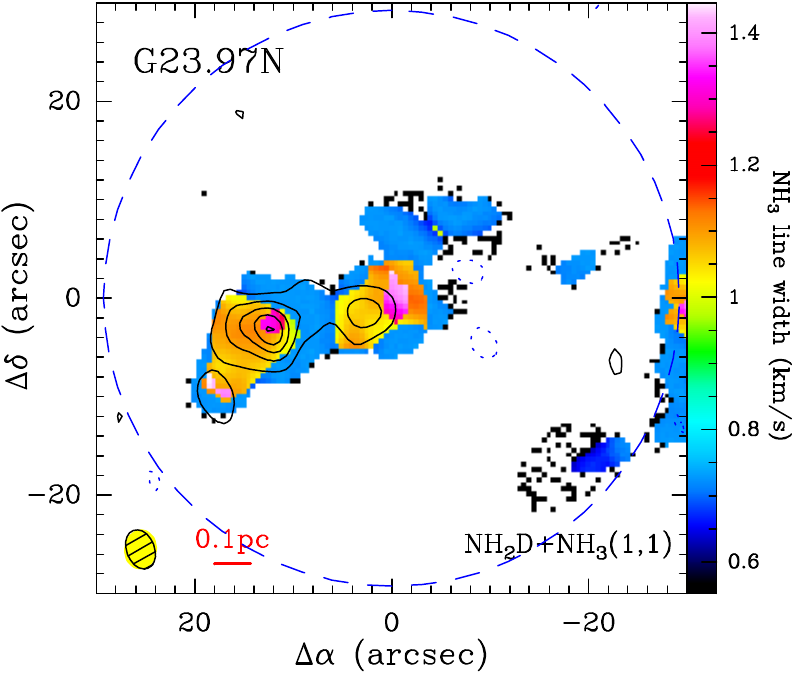}}
\subfigure[]{\includegraphics[width=0.33\textwidth,
angle=0]{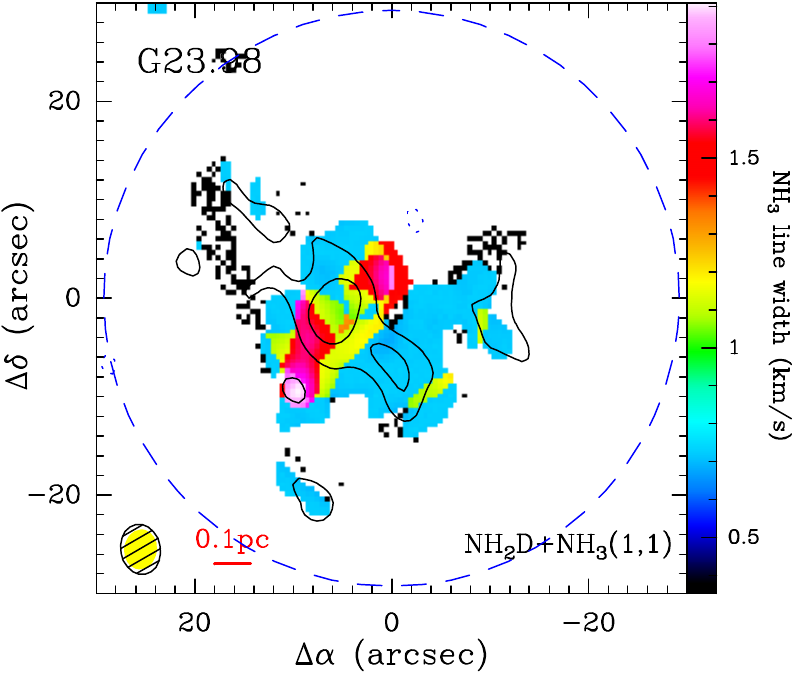}}
\subfigure[]{\includegraphics[width=0.33\textwidth,
angle=0]{figures/nh2d-moment2/g2344-11-eps-converted-to.pdf}}
\subfigure[]{\includegraphics[width=0.33\textwidth,
angle=0]{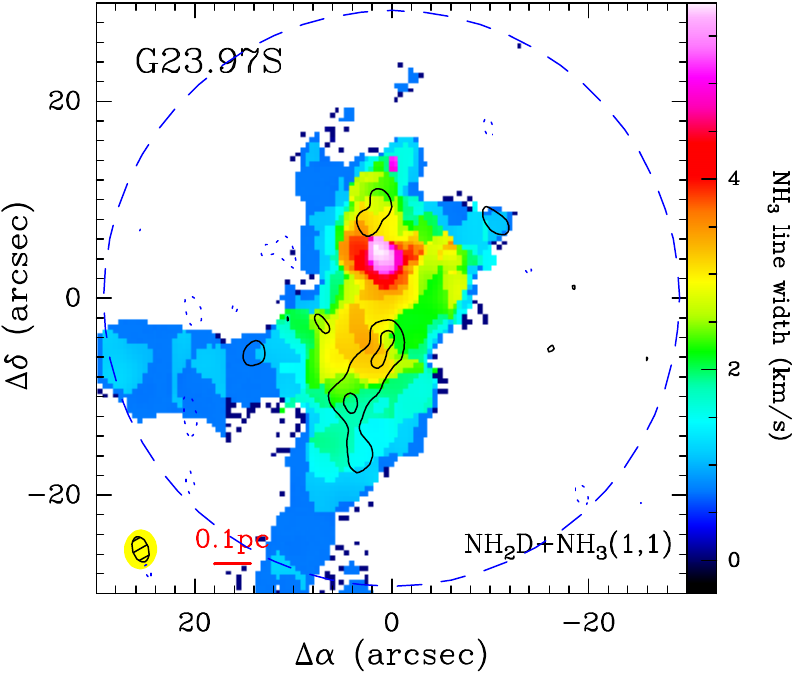}}
\subfigure[]{\includegraphics[width=0.33\textwidth,
angle=0]{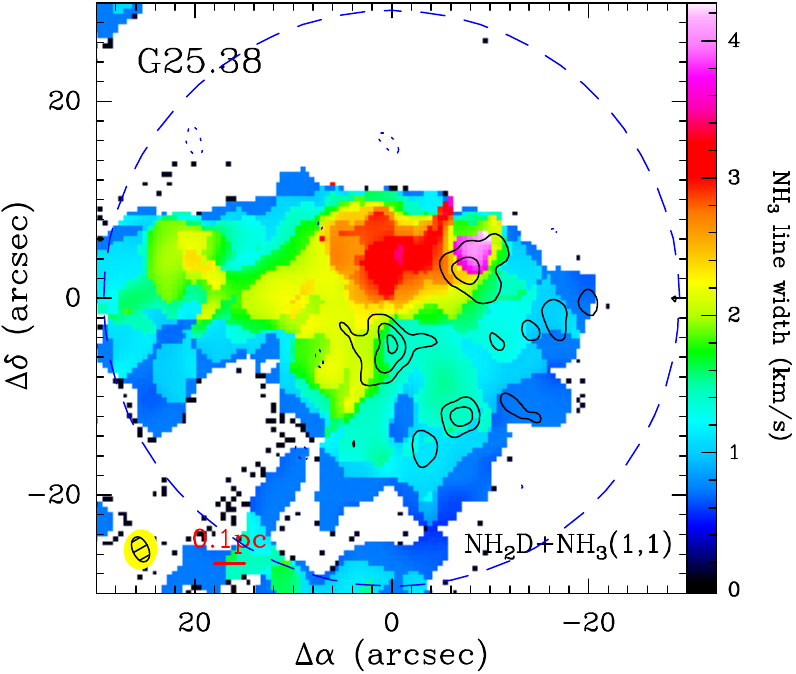}}
\subfigure[]{\includegraphics[width=0.33\textwidth,
angle=0]{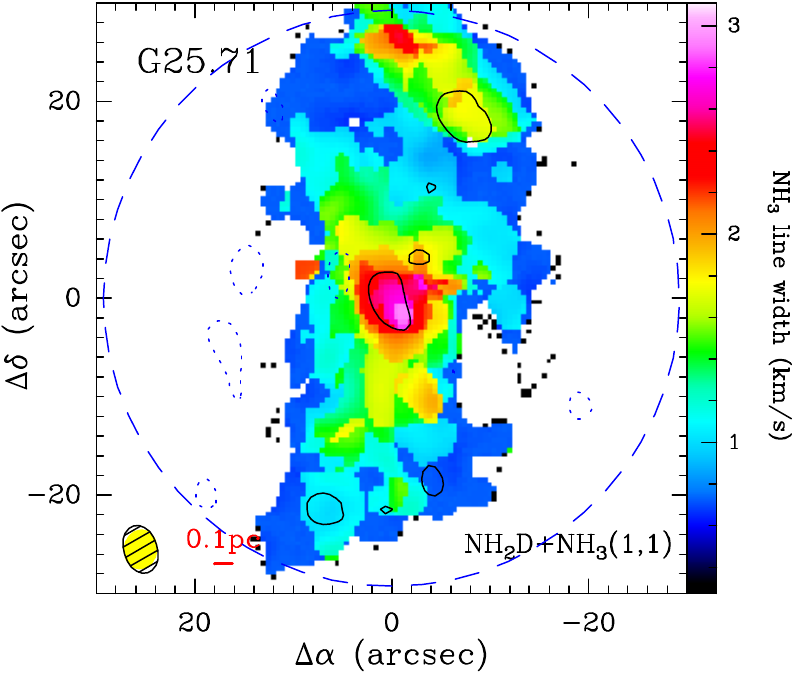}}
\caption{Line width distribution (moment\,2) of NH$_3$ (1,\,1) line overlaid with integrated-intensity contours (moment\,0) of NH$_2$D line with velocity range covering all the six HfS lines. The contour levels start at $-3\sigma$ in steps of $3\sigma$ for NH$_2$D with $\sigma_{\rm (a)-(h)} =$ 60.3, 60.9, 51.1, 53.2, 84.1, 48.3, 49.0, 50.4\,$\mjybkms$. The synthesized beam sizes of each subfigure are indicated at the bottom-left corner. The dashed circle indicates the primary beam of the PdBI observations at 3.5\,mm.}
\label{Fig_mom2_app}
\end{figure*}

\begin{figure*}
\centering
\subfigure[]{\includegraphics[width=0.30\textwidth,
angle=0]{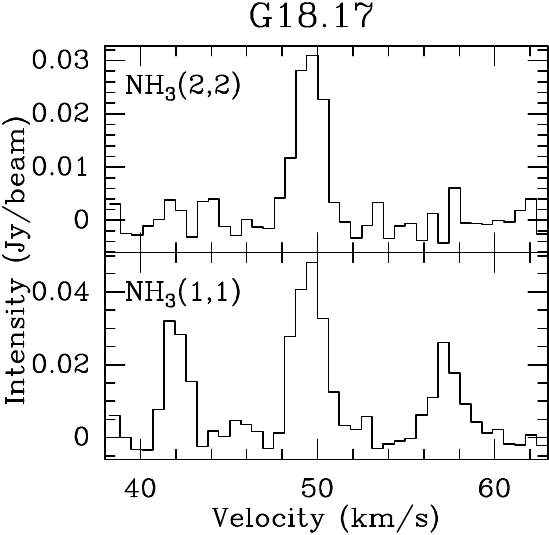}}
\subfigure[]{\includegraphics[width=0.30\textwidth,
angle=0]{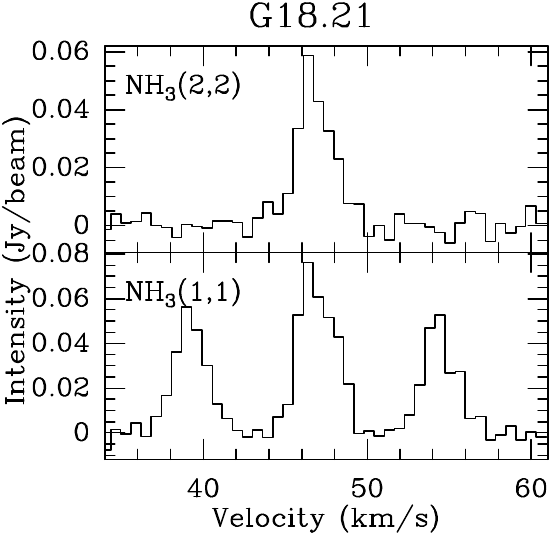}}
\subfigure[]{\includegraphics[width=0.30\textwidth,
angle=0]{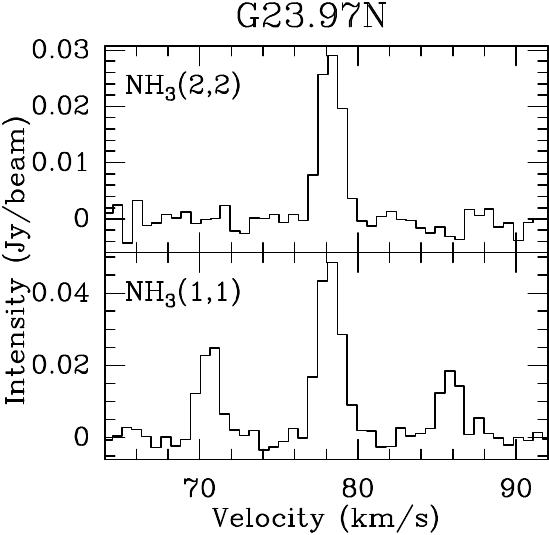}}
\subfigure[]{\includegraphics[width=0.30\textwidth,
angle=0]{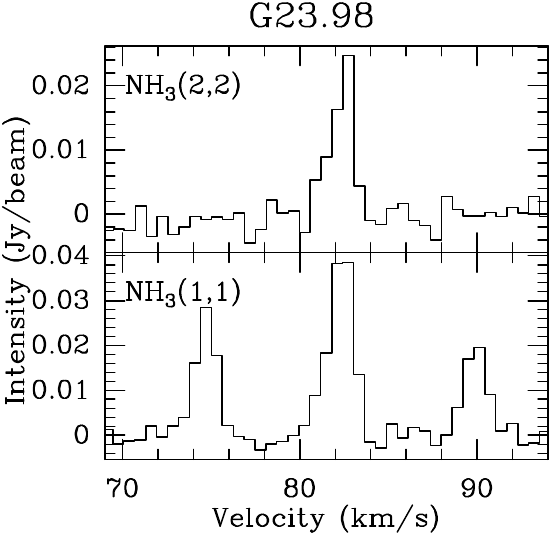}}
\subfigure[]{\includegraphics[width=0.30\textwidth,
angle=0]{figures/nh3-sum/g2344-spec-eps-converted-to.pdf}}
\subfigure[]{\includegraphics[width=0.30\textwidth,
angle=0]{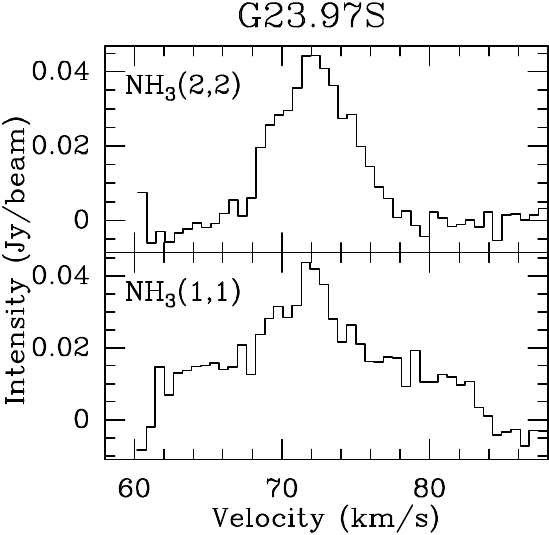}}
\subfigure[]{\includegraphics[width=0.30\textwidth,
angle=0]{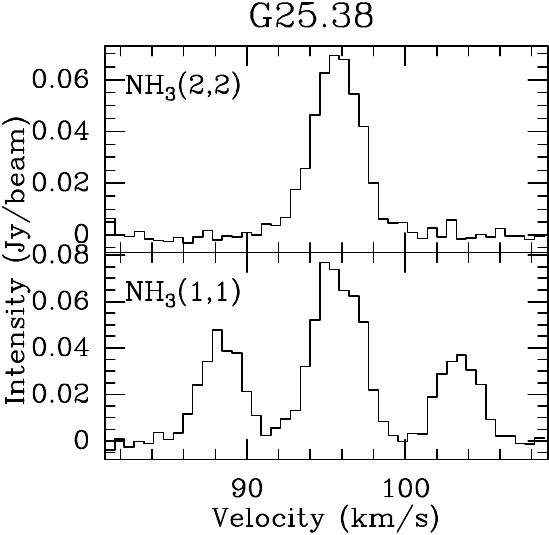}}
\subfigure[]{\includegraphics[width=0.30\textwidth,
angle=0]{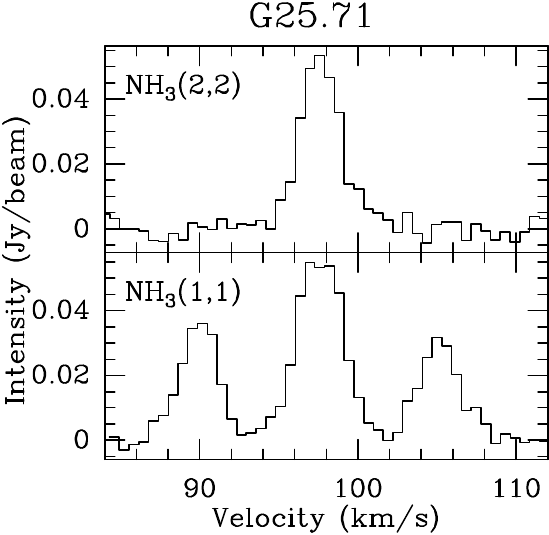}}
\caption{Spectra of NH$_3$ (1,\,1) and (2,\,2) obtained toward a single and the brightest pixel at 3.5\,mm continuum. The coordinates of the spectra are indicated at each corresponding lower panel of Figure\,\ref{Fig_pv_app}.}
\label{Fig_spectra_cont_app}
\end{figure*}

\begin{figure*}
\centering
\subfigure[]{\includegraphics[width=0.33\textwidth,
angle=0]{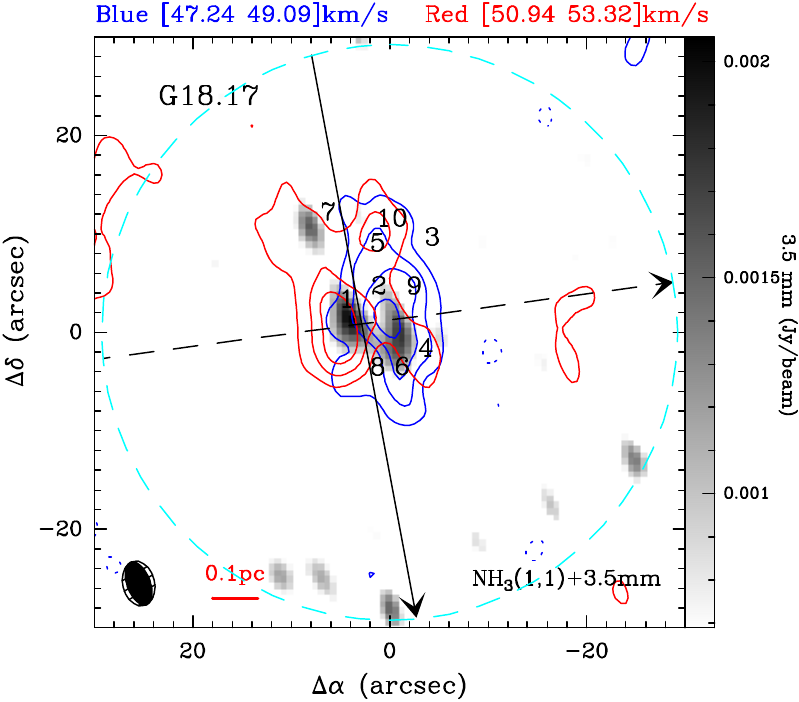}}
\subfigure[]{\includegraphics[width=0.33\textwidth,
angle=0]{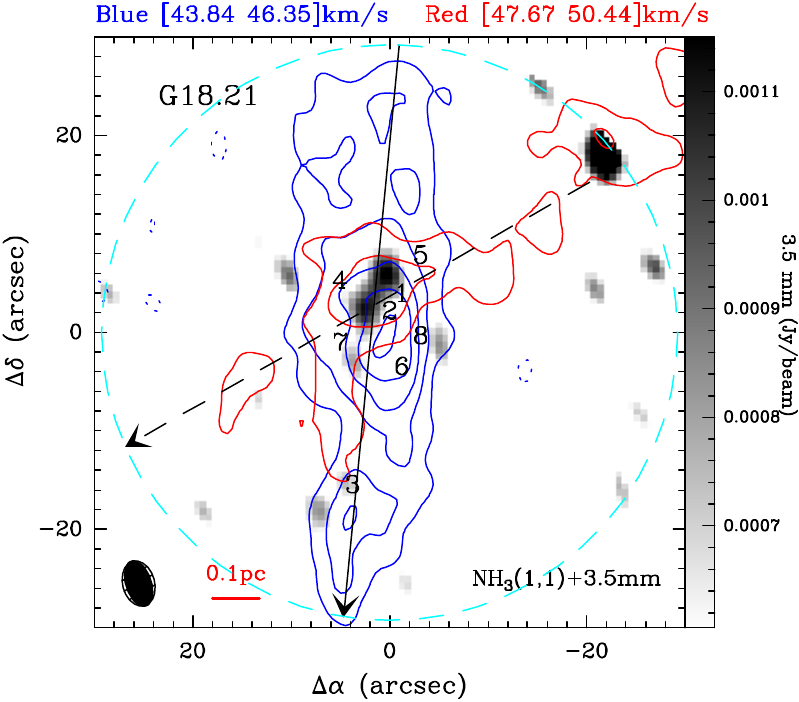}}
\subfigure[]{\includegraphics[width=0.33\textwidth,
angle=0]{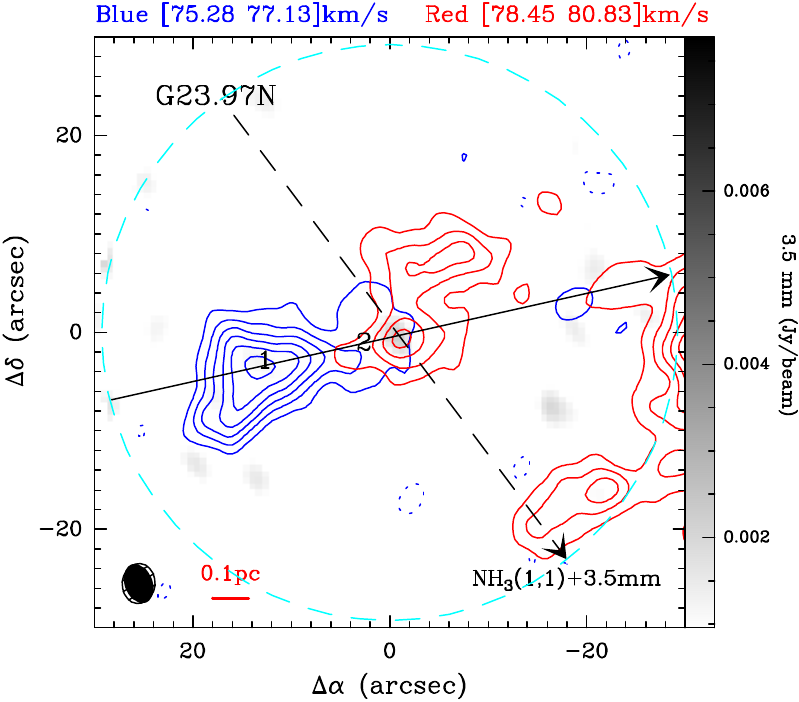}}
\subfigure[]{\includegraphics[width=0.33\textwidth,
angle=0]{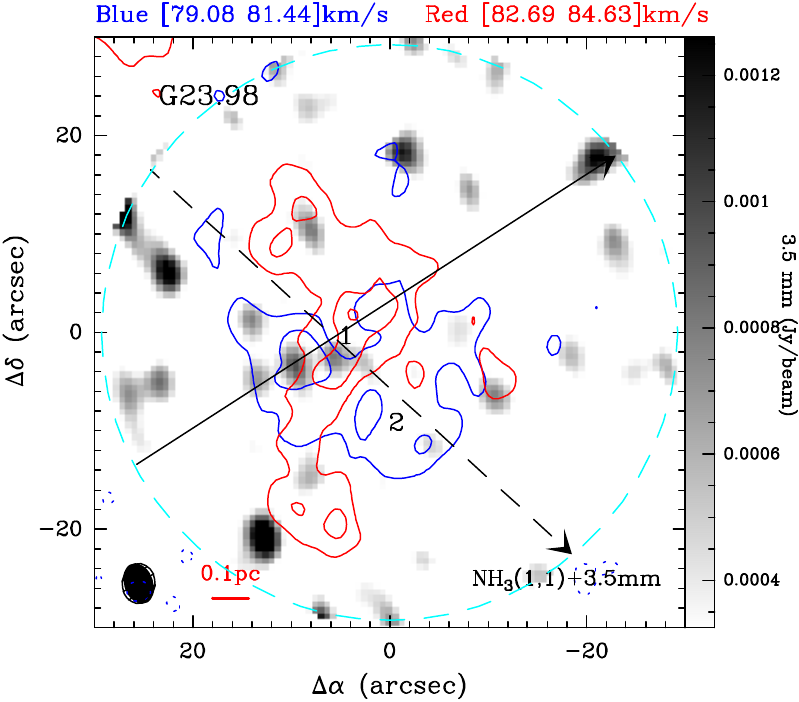}}
\subfigure[]{\includegraphics[width=0.33\textwidth,
angle=0]{figures/outflow/g2344-eps-converted-to.pdf}}
\subfigure[]{\includegraphics[width=0.33\textwidth,
angle=0]{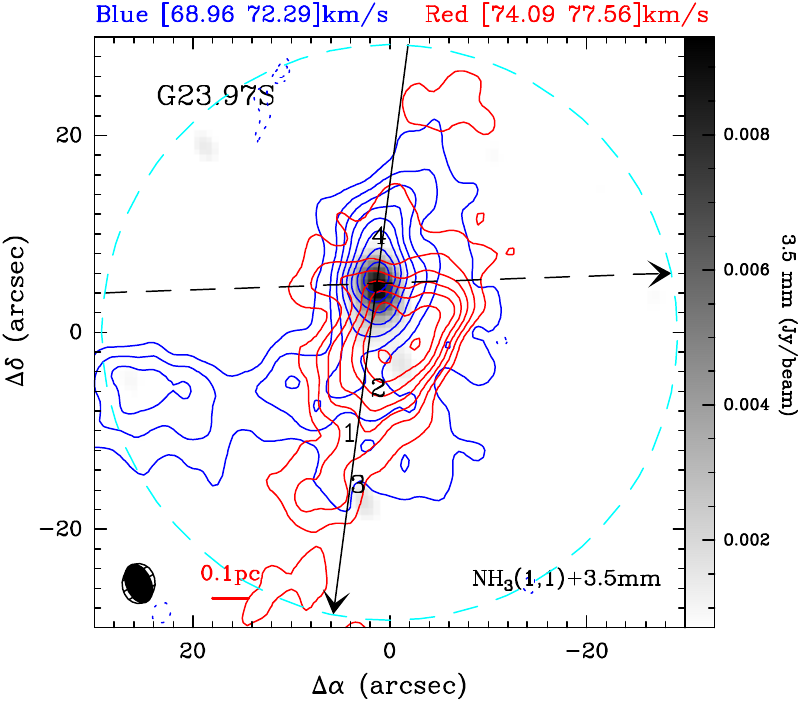}}
\subfigure[]{\includegraphics[width=0.33\textwidth,
angle=0]{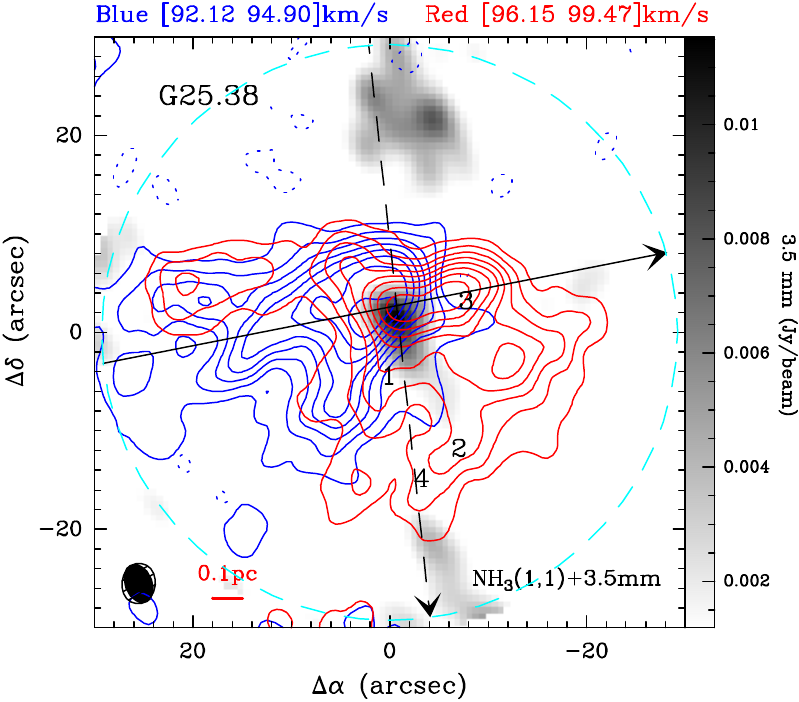}}
\subfigure[]{\includegraphics[width=0.33\textwidth,
angle=0]{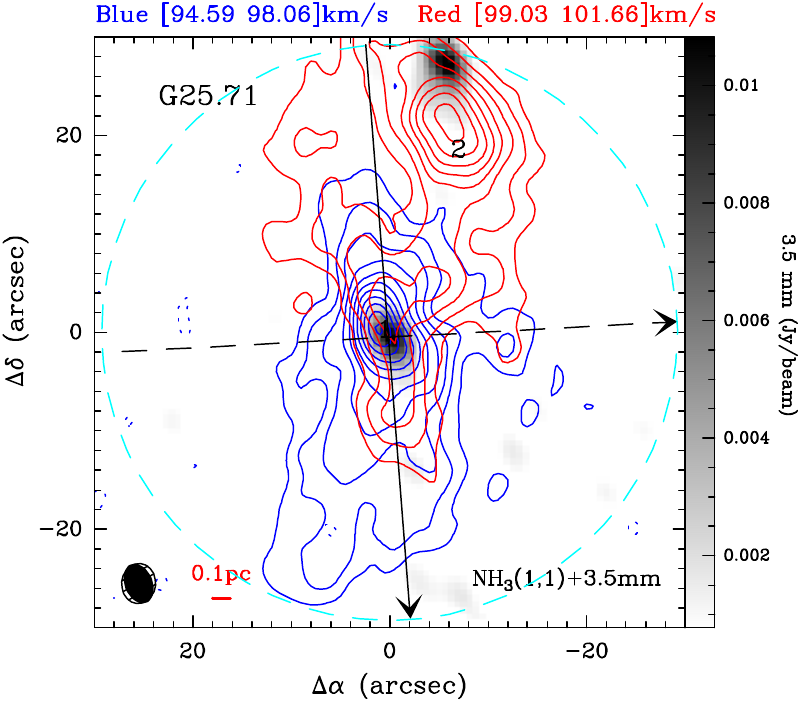}}
\caption{Blueshifted and redshifted NH$_3$ (1,\,1) integrated-intensity contours overlaid on a 3.5\,mm continuum. The blue and red contours are the blueshifted and redshifted velocity components, respectively. The blue contour levels start at $-3\sigma$ in steps of $3\sigma$ for NH$_3$ (1,\,1) with $\sigma_{\rm (a)-(h)} =$ 5.0, 6.8, 3.2, 4.9, 7.2, 5.5, 5.3, 5.6\,$\mjybkms$, and the red ones with $\sigma_{\rm (a)-(h)} =$ 5.7, 9.1, 4.5, 3.7, 4.8, 4.8, 5.2, 5.0\,$\mjybkms$. The black numbers indicate the positions of extracted NH$_2$D cores. The synthesized beam sizes of each subfigure are indicated at the bottom-left corner. The dashed circle indicates the primary beam of the PdBI observations at 3.5\,mm. The lines with arrows show the position-velocity cutting direction in Figure\,\ref{Fig_pv_app}.}
\label{Fig_outflow_app}
\end{figure*}

\begin{figure*}
\centering
\subfigure[]{\includegraphics[width=0.245\textwidth,
angle=0]{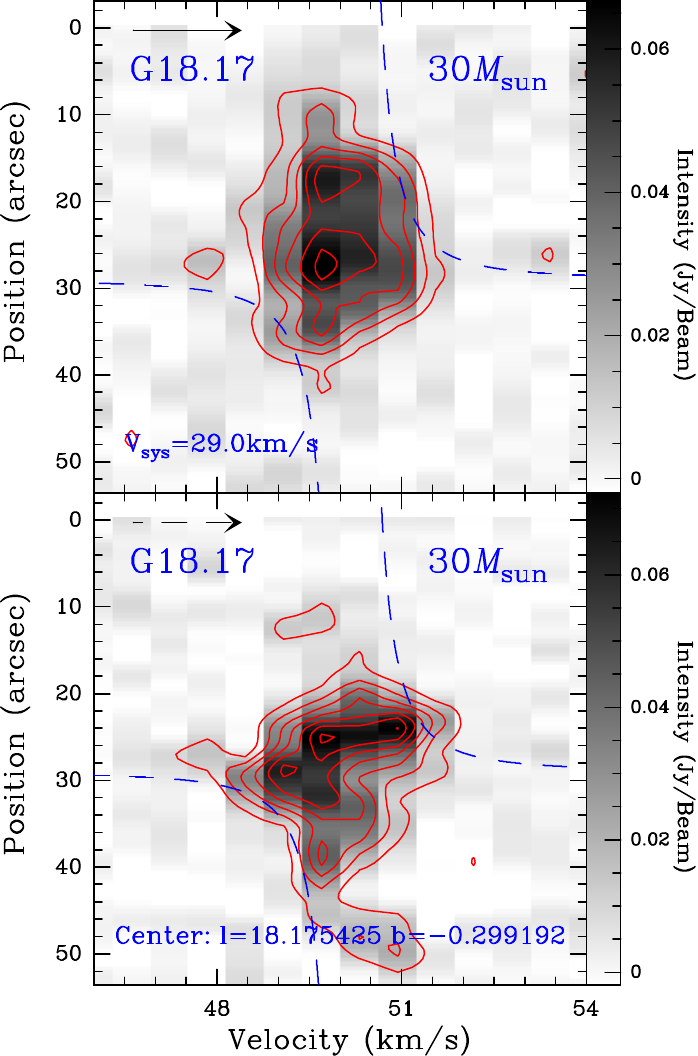}}
\subfigure[]{\includegraphics[width=0.245\textwidth,
angle=0]{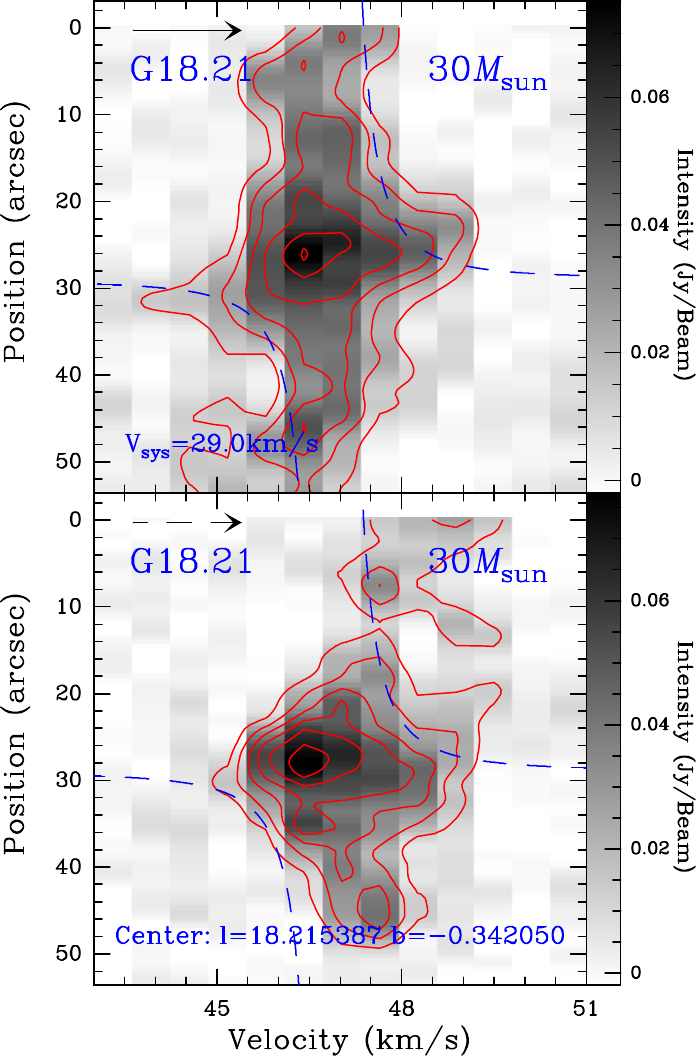}}
\subfigure[]{\includegraphics[width=0.245\textwidth,
angle=0]{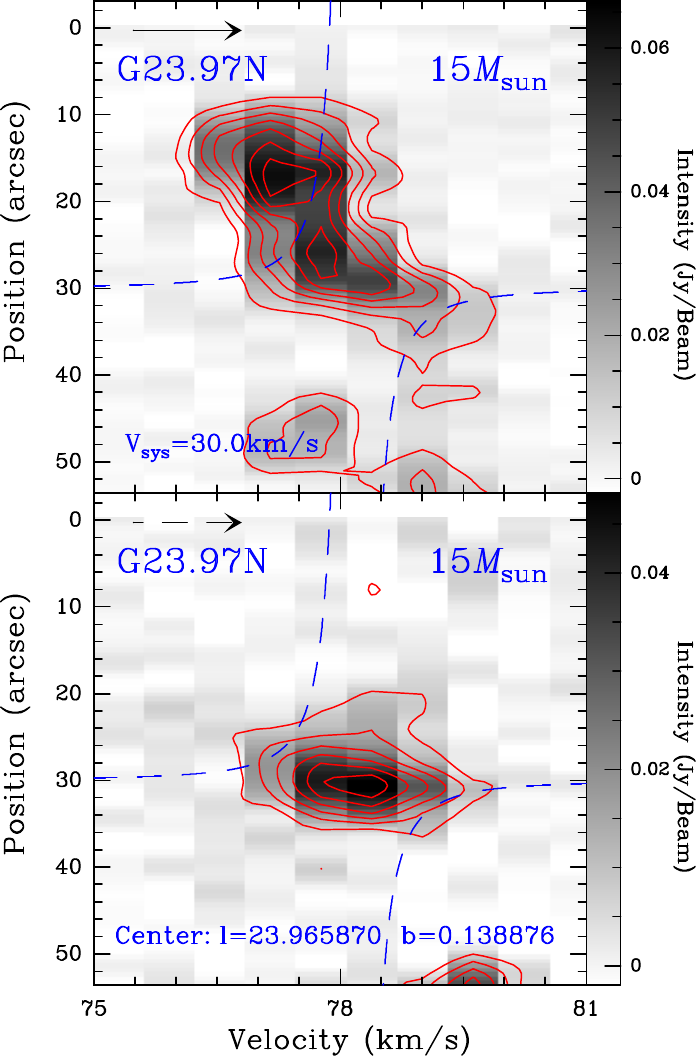}}
\subfigure[]{\includegraphics[width=0.245\textwidth,
angle=0]{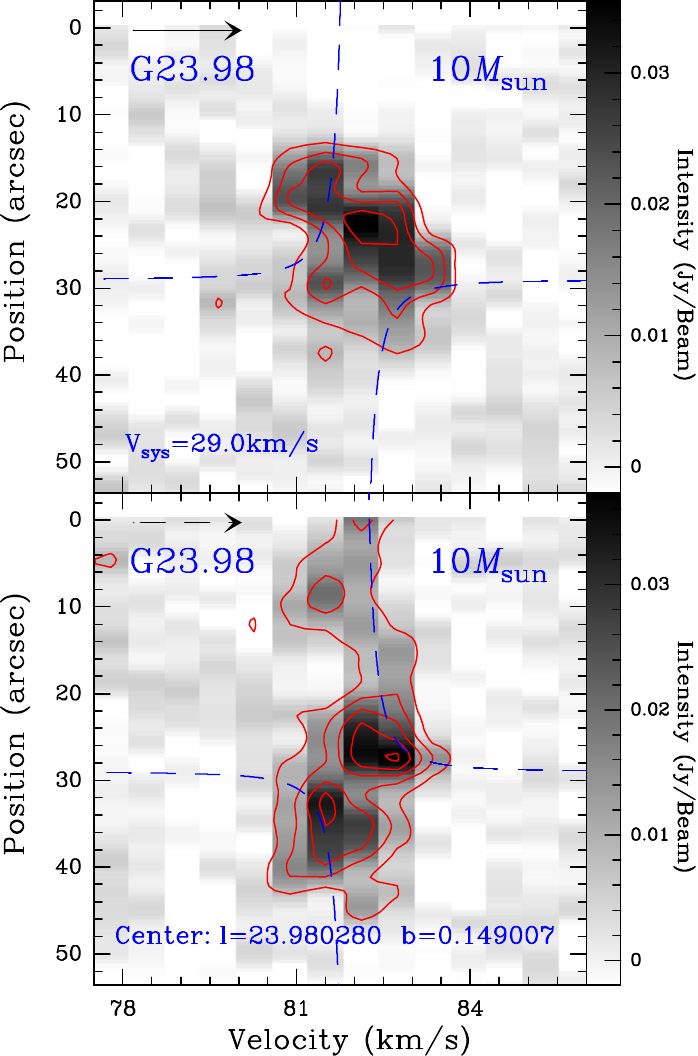}}
\subfigure[]{\includegraphics[width=0.245\textwidth,
angle=0]{figures/pv/g2344-eps-converted-to.pdf}}
\subfigure[]{\includegraphics[width=0.245\textwidth,
angle=0]{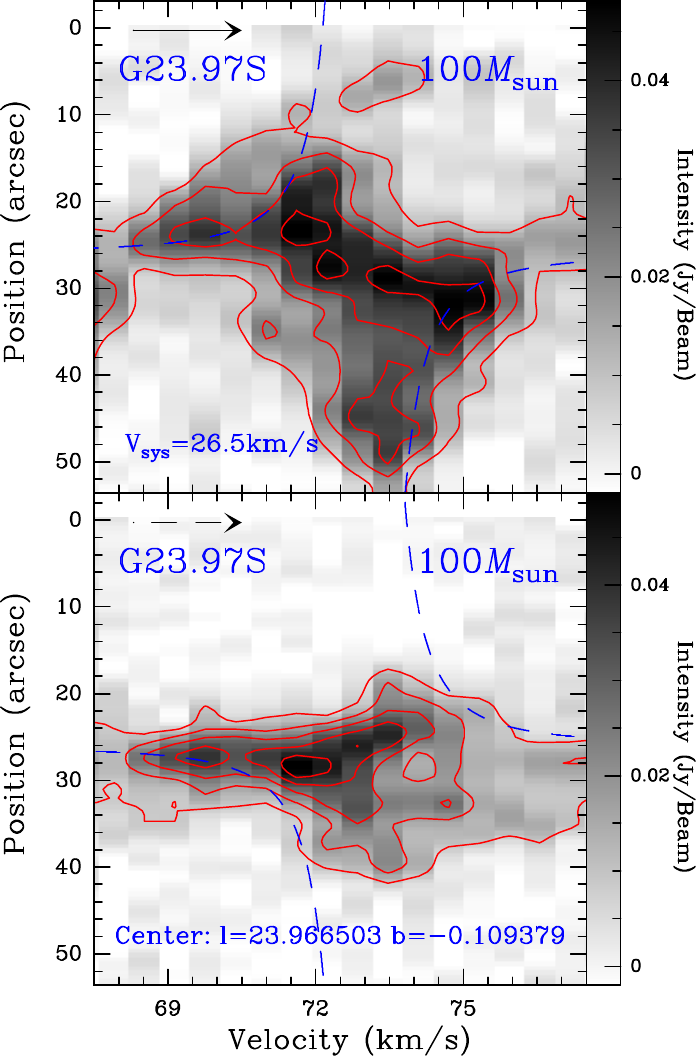}}
\subfigure[]{\includegraphics[width=0.245\textwidth,
angle=0]{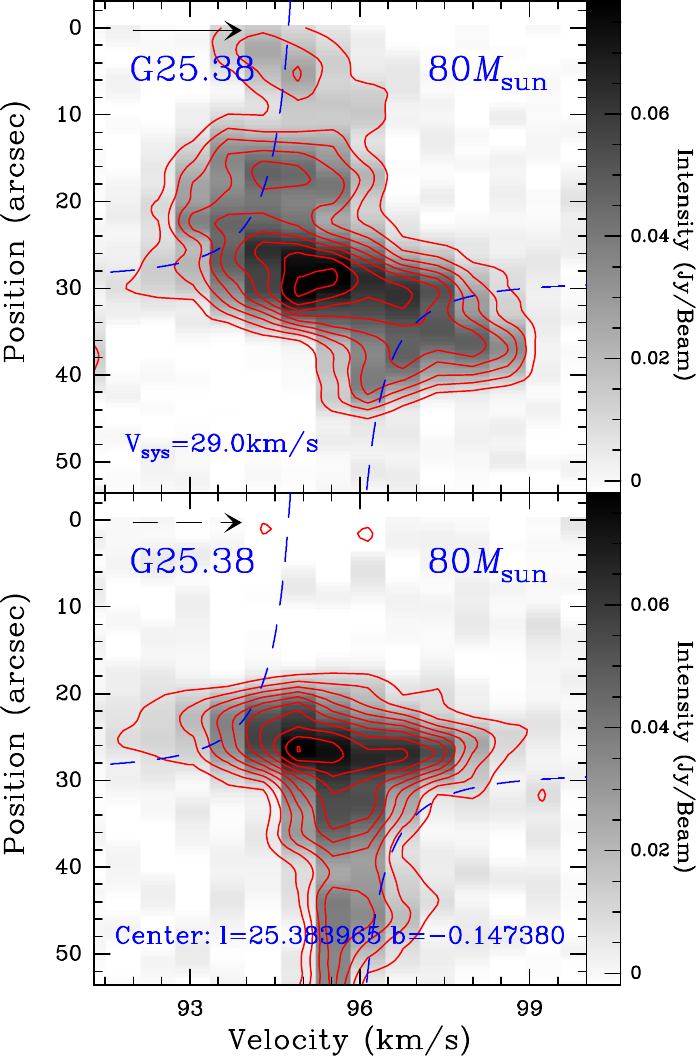}}
\subfigure[]{\includegraphics[width=0.245\textwidth,
angle=0]{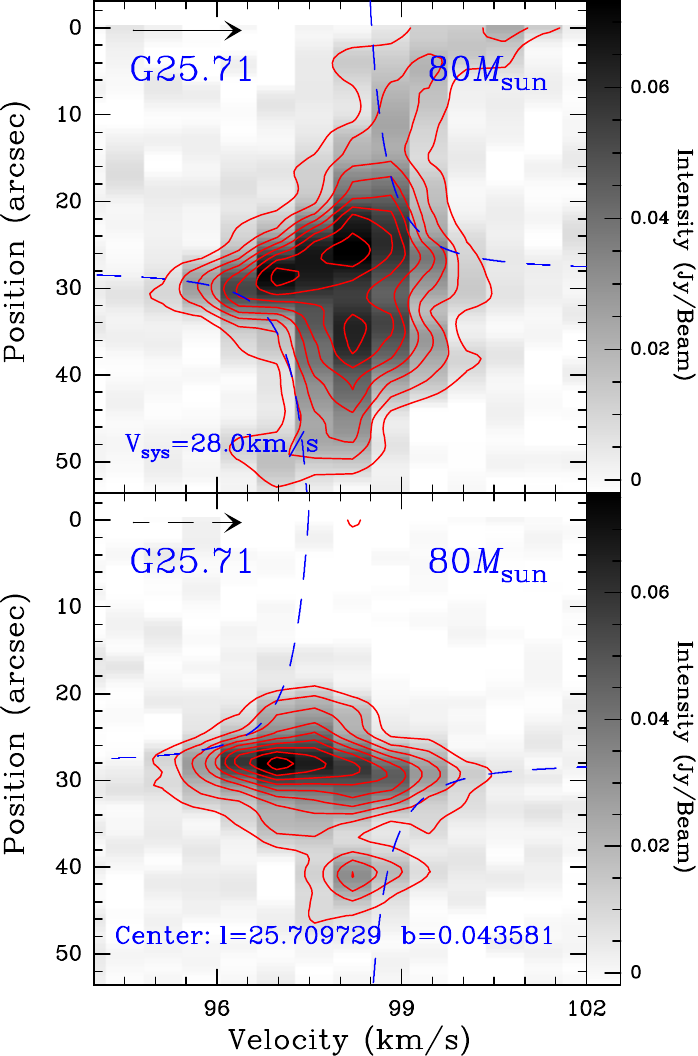}}
\caption{Position-velocity diagrams of the main line of NH$_3$ (1,\,1) HfS along the position-velocity slice indicated with solid and dashed lines in Figure\,\ref{Fig_mom1_app} (see also Figure\,\ref{Fig_outflow_app}). The arrows show the position-velocity cutting direction. Contour levels start at $3\sigma$ level and increase in steps of $3\sigma$ with $\sigma_{\rm (a)-(h)}$ = $\sim$3.4, $\sim$4.4, $\sim$2.5, $\sim$2.5, $\sim$3.3, $\sim$3.4, $\sim$2.7, and $\sim$2.8\,$\mjyb$ for source G18.17, G18.21, ..., G25.71, respectively. Blue dashed lines show a possible rotating toroids curve. The central mass, the central position, and the systemic velocity are indicated in each subfigure.}
\label{Fig_pv_app}
\end{figure*}

\end{document}